\documentclass[aps,reprint,twocolumn,superscriptaddress,nofootinbib,longbibliography]{revtex4-1}

\usepackage{amsmath, amssymb, bm, braket, dsfont, times, amsthm}
\usepackage{graphicx,xfrac,color}
\usepackage[breaklinks,colorlinks,allcolors=blue]{hyperref}
\usepackage{framed}
\definecolor{shadecolor}{gray}{0.95}

\usepackage[normalem]{ulem}

\makeatletter
\def\l@subsection#1#2{}
\def\l@subsubsection#1#2{}
\makeatother

\begin{document}

\title{Gapless topological phases and symmetry-enriched quantum criticality}
\author{Ruben Verresen}
\affiliation{Department of Physics, Harvard University, Cambridge, MA 02138, USA}
\author{Ryan Thorngren}
\affiliation{Department of Physics, Harvard University, Cambridge, MA 02138, USA}
\affiliation{Center of Mathematical Sciences and Applications, Harvard University, Cambridge, MA 02138, USA}
\affiliation{Department of Physics, Massachusetts Institute of Technology, Cambridge, MA 02139, US}
\author{Nick G. Jones}
\affiliation{Mathematical Institute, University of Oxford, Oxford, OX2 6GG, UK}
\affiliation{The Heilbronn Institute for Mathematical Research, Bristol, UK}
\author{Frank Pollmann}
\affiliation{Department of Physics, T42, Technische Universit\"at M\"unchen, 85748 Garching, Germany}
\affiliation{Munich Center for Quantum Science and Technology (MCQST), 80799 Munich, Germany}
\date{\today}                                       

\begin{abstract}
We introduce topological invariants for gapless systems and study the associated boundary phenomena. More generally, the symmetry properties of the low-energy conformal field theory (CFT) provide discrete invariants, establishing the notion of \emph{symmetry-enriched quantum criticality}. The charges of \emph{nonlocal} scaling operators, or more generally of symmetry defects, are topological and imply the presence of localized edge modes. 
We primarily focus on the $1+1d$ case where the edge has a topological degeneracy, whose finite-size splitting can be exponential or algebraic in system size depending on the involvement of additional gapped sectors. An example of the former is given by tuning the spin-1 Heisenberg chain to a symmetry-breaking Ising phase. An example of the latter arises between the gapped Ising and cluster phases: this symmetry-enriched Ising CFT has an edge mode with finite-size splitting $\sim 1/L^{14}$. In addition to such new cases, our formalism unifies various examples previously studied in the literature. 
Similar to gapped symmetry-protected topological phases, a given CFT can split into several distinct symmetry-enriched CFTs. This raises the question of classification, to which we give a partial answer---including a complete characterization of symmetry-enriched $1+1d$ Ising CFTs. Non-trivial topological invariants can also be constructed in higher dimensions, which we illustrate for a symmetry-enriched $2+1d$ CFT without gapped sectors.
\end{abstract}

\maketitle

\tableofcontents

\section{Introduction}
Topological phases of quantum matter are fascinating emergent phenomena, commonly characterized by nonlocal order parameters in the bulk and exotic behavior at the boundary.
A subclass of gapped topological phases are the so-called \emph{symmetry-protected topological} (SPT) phases, which are only nontrivial in the presence of certain global symmetries \cite{Fidkowski11class,Turner11class,Chen10,Schuch11,Chen13,Kapustin14,Kapustin15,Senthil15}. In one dimension, SPT phases have ground state degeneracies associated to their boundaries, called \emph{zero-energy edge modes} \cite{Affleck88,Kennedy90}. These edge modes are often exponentially localized with the same length scale as the bulk correlation length, seemingly suggesting that the bulk gap is essential.

However, in recent years, studies have shown the existence of critical chains that nevertheless host topologically protected edge modes, with either algebraic or exponential finite-size splitting
\cite{Kestner11,Cheng11,Fidkowski11longrange,Sau11,Ruhman12,Grover12,Kraus13,Ortiz14,Keselman15,Ruhman15,Kainaris15,Iemini15,Lang15,Ortiz16,Montorsi17,Wang17,Ruhman17,Scaffidi17,Guther17,Kainaris17,Jiang18,Zhang18,Verresen18,Parker18,Keselman18,Chen18}.
Our work provides a unifying framework for this phenomenon, placing it into the more general context of \emph{symmetry-enriched quantum criticality}. The idea is simple: a given universality class can split into various \emph{distinct} classes when additional symmetries are imposed. These can be distinguished by symmetry properties of either local or nonlocal operators. The latter nonlocal case serves as a \emph{topological invariant} and can imply emergent edge modes!

Our work unifies in two respects. First, as already mentioned, it offers a framework that incorporates previously studied examples, giving a common explanation for the observed edge modes at criticality---whilst also introducing qualitatively novel cases. Second, the invariants we introduce are direct generalizations of those studied in the gapped case, protecting edge modes in both settings. In particular, gapped SPT and spontaneously symmetry-breaking (SSB) phases correspond to the special cases where the bulk universality class is chosen to be trivial (i.e., there are no low-energy degrees of freedom). This puts in perspective how much is left to be explored: one can repeat the study and classification for any choice of universality class. In this work, we focus on universality classes described by conformal field theories (CFTs).

\begin{figure}
\includegraphics[scale=0.97,trim=5 5 4 3,clip]{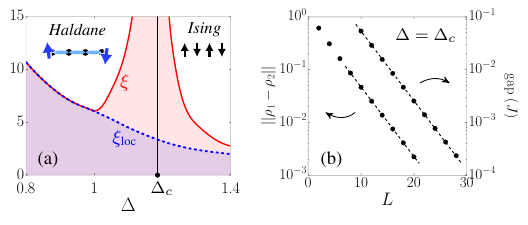}
\caption{\textbf{Persistent edge mode in a spin-1 XXZ chain at criticality.} (a) The bulk correlation length $\xi$ blows up, whereas the edge mode localization length $\xi_\textrm{loc}$ stays finite. (b) The latter can be measured in two ways: the two ground states $|\uparrow_l \downarrow_r \rangle \pm |\downarrow_l \uparrow_r \rangle$ have a splitting which is exponentially small in system size $L$, or, equivalently, the two states differ by an operator which is exponentially localized near the edges ($L$ is the region we trace out near an edge before calculating the distance of the two density matrices). The dashed lines are proportional to $\exp(-L/\xi_\textrm{loc})$ with $\xi_\textrm{loc}\approx 3.3$. \label{fig:xiloc}}
\end{figure}

Examples of symmetry-enriched criticality are hiding in plain sight.
These tend to occur at phase transitions where all neighboring gapped phases are non-trivial (either SPT or SSB). For example, a paradigmatic SPT phase is the Haldane phase, realized in the spin-1 Heisenberg chain \cite{Haldane83,Kennedy90,Kennedy92,Gu09,Pollmann10,Chen10}. By introducing an easy-axis anisotropy, this can be driven to an Ising phase:
\begin{equation}
H_\textrm{XXZ} = J \sum_n \left( S^x_n S^x_{n+1} + S^y_n S^y_{n+1} + \Delta S^z_n S^z_{n+1} \right). \label{eq:XXZ}
\end{equation}
These two gapped phases are separated by an Ising critical point at $\Delta_c \approx 1.1856$ \cite{Chen03}. We point out that, although the bulk correlation length diverges, a twofold degeneracy with open boundaries remains exponentially localized, shown in Fig.~\ref{fig:xiloc}. We will show that this Ising CFT is topologically \emph{enriched} by $\pi$-rotations around the principal axes (a $\mathbb Z_2 \times \mathbb Z_2$ group), explaining the observed edge modes.

The concepts developed in this work apply to symmetry-enriched CFTs with a general on-site symmetry group $G$; we refer to them also as $G$-enriched CFTs, or $G$-CFTs for short. This is not only relevant for condensed matter systems---such as the example described above---but also to high-energy physics; in particular, $G$-CFTs can be related to discrete torsion of orbifold CFTs, arising in string theory \cite{Vafa86,Vafa95,Douglas98,Sharpe03}, although edge modes had not yet been pointed out in that context. While we introduce our framework for general symmetry groups $G$ and discuss a variety of examples, for clarity we will illustrate these concepts in detail for two particular symmetry groups: the unitary group $\mathbb Z_2 \times \mathbb Z_2$ and the anti-unitary $\mathbb Z_2 \times \mathbb Z_2^T$ (where complex conjugation is defined relative to an on-site basis). In particular, for the Ising CFT, the former (latter) symmetry group gives rise to edge modes whose finite-size splitting is exponentially (algebraically) small in system size. For the algebraic splitting, we show that the power is remarkably high, $\sim 1/L^{14}$. This is derived by studying the symmetry properties of descendant operators in the CFT. To emphasize the generality of our approach, we present examples for other symmetry groups, including $U(1) \times \mathbb Z_2 \times \mathbb Z_2^T$, $U(1) \rtimes \mathbb Z_2^T$ and $\mathbb Z_3 \times \mathbb Z_3$. We also discuss the classification of such symmetry-enriched CFTs, providing a complete answer in the case of the Ising CFT for a general symmetry group $G$. 
The topological invariants we introduce can also be generalized to higher dimensions, which we illustrate at the end of this work.

The remainder of this paper is organized as follows. Since the essential concepts can be readily understood, we give an overview in Section~\ref{sec:overview}. This provides a self-contained summary for the reader in a hurry, and can serve as a guide to the other sections. In particular, it introduces and motivates the notion of \emph{symmetry flux}: this is the key player throughout this work and allows us to define topological invariants at criticality. Moreover, Section~\ref{sec:overview} gives a sense of how this invariant protects edge modes at criticality. Symmetry fluxes and edge modes are discussed more systematically in Sections~\ref{sec:symflux} and \ref{sec:edge}, respectively. Section~\ref{sec:classification} concerns the classification of $G$-enriched CFTs. A $\mathbb Z_3\times \mathbb Z_3$ and a fermionic $\mathbb Z_2^f \times \mathbb Z_2^T$ example are presented in Section~\ref{sec:other}, and in Section~\ref{sec:lit} we show how previous works fit into our framework. Finally, Section~\ref{sec:higher} discusses generalizations to higher dimensions.

\section{A conceptual overview \label{sec:overview}}

This section functions as a primer, motivating concepts introduced in greater generality in subsequent sections. The reader seeking a taste of symmetry-enriched criticality may choose to read only this section. Alternatively, readers preferring generality and not requiring motivational examples may skip this section altogether and continue to Section~\ref{sec:symflux}.

We will first review the known order parameters for gapped phases in 1D, which are either local (SSB) or nonlocal (SPT). In Section~\ref{subsec:Z2Z2T} we will show that, similarly, critical points can also host local and nonlocal order parameters. Non-trivial nonlocal order parameters can be related to the existence of edge modes, discussed in Section~\ref{ssec:edgemodesymflux}. Comments on the classification of such symmetry-enriched critical points, fermionic versions and higher-dimensional generalizations can be found in Sections~\ref{ssec:classoverview}, \ref{ssec:fermionicoverview} and \ref{subsec:twist}. These subsections serve as pointers to the main sections of this paper.

In order to present some of the key concepts of this work in a simple manner, in this overview we will fix the symmetry group $G = \mathbb Z_2 \times \mathbb Z_2^T$ as an illustrative example ($\mathbb Z_2^T$ denotes the group generated by an anti-unitary symmetry that squares to the identity).
Other symmetry groups encountered in this work include $\mathbb Z_2 \times \mathbb Z_2$ (Section~\ref{sec:symflux}), $\mathbb Z_3 \times \mathbb Z_3$ and $\mathbb Z_2^f \times \mathbb Z_2^T$ (Section~\ref{sec:other}), $U(1) \times \mathbb Z_2 \times \mathbb Z_2^T$, $U(1) \rtimes \mathbb Z_2^T$ (Section~\ref{sec:lit}) and $\mathbb Z_2 \times \mathbb Z_2 \times \mathbb Z_2$ (Section~\ref{sec:higher}). 
Below, we consider spin-$1/2$ chains, denoting the Pauli matrices by $X$, $Y$ and $Z$, and representing the symmetry group $\mathbb Z_2 \times \mathbb Z_2^T$ by the spin-flip $P \equiv \prod_n X_n$ and an anti-unitary symmetry given by complex conjugation in this basis, $T  \equiv K$.

\subsection{Local and nonlocal order parameters for gapped phases \label{subsec:gappedreview}}

Naturally, phases that spontaneously break a symmetry can be diagnosed by long-range order of some local operator with a non-trivial charge (i.e., the operator does not commute with the symmetry). The standard example is the Ising chain $H_{\rm Ising} = - \sum_n Z_n Z_{n+1}$, where the ground state has $\lim_{|n-m| \to \infty} \langle Z_m Z_n \rangle \neq 0$. Since $Z_n$ has charge $-1$ under the $\mathbb Z_2$ Ising symmetry $P$, this is distinct from the trivial phase, which can have long-range correlations only for operators with charge $+1$. Moreover, a physical state (i.e., without long-range cat-state-like entanglement) will obey clustering: $\lim_{|n-m| \to \infty} \langle Z_m Z_n \rangle = \langle Z_m \rangle \langle Z_n \rangle$. We conclude that $\langle Z_n \rangle \neq 0$ which means the ground state is no longer invariant under the $\mathbb Z_2$ symmetry\footnote{Otherwise, $\langle Z_n \rangle = \langle Z_n P \rangle =  - \langle P Z_n \rangle = - \langle Z_n \rangle \Rightarrow \langle Z_n \rangle = 0$.}, implying a ground state degeneracy.

In the presence of $\mathbb Z_2$ symmetry there is only one symmetry-breaking `Ising' phase. If we additionally impose complex conjugation symmetry $\mathbb Z_2^T$, then this single Ising phase subdivides into two distinct phases. Indeed, consider $H_{\rm Ising}' = - \sum_n Y_n Y_{n+1}$. If we only preserve $\mathbb Z_2$, one can smoothly connect $H_{\rm Ising}$ to $H_{\rm Ising}'$ without closing a gap by rotating $Y_n$ into $Z_n$. However, the Ising order parameter $Y_n$ is imaginary, whereas $Z_n$ is real. Hence, these two Hamiltonians \emph{cannot} be connected whilst preserving $\mathbb Z_2 \times \mathbb Z_2^T$. They must be separated by a quantum phase transition\footnote{In fact, a direct phase transition between the two (e.g., the XX chain) would be an example of a deconfined quantum critical point in one dimension \cite{Senthil04,Jiang19,Roberts19}.}. We will see in Section~\ref{subsec:Z2Z2T} that if the system is gapless, the charges of local operators can distinguish phases even in the absence of symmetry-breaking.

An SPT phase, on the other hand, cannot be probed by any local observable. Instead, one can associate to any\footnote{This is simplest for on-site symmetries. For other cases, see Ref.~\cite{Pollmann12b}.} unbroken symmetry a string order parameter (this is derived using the concept of symmetry fractionalization \cite{Pollmann10,Turner11class} as discussed in Section~\ref{sec:symflux} and reviewed in Appendix~\ref{app:gappedsym}). For instance, the trivial paramagnet $H_{\rm triv} = -\sum_n X_n$ has long-range order $\lim_{|n-m| \to \infty} \langle X_m X_{m+1} \cdots X_{n-1} X_n \rangle \neq 0$ whereas the paradigmatic cluster SPT model \cite{Suzuki71,Raussendorf01,Keating04,Son11,Verresen17}
\begin{equation}
H_{\rm cluster} = - \sum_n Z_{n-1} X_n Z_{n+1}
\end{equation}
has long-range order\footnote{The simplest way of deriving this is by noting that it equals the product $\prod_{k=m}^n Z_{k-1} X_k Z_{k+1}$, and $Z_{k-1} X_k Z_{k+1}=+1$ in the ground state.} \begin{equation}
\lim_{|n-m| \to \infty} \langle Z_{m-1} Y_m X_{m+1} \cdots X_{n-1} Y_n Z_{n+1} \rangle \neq 0. \label{eq:clusterstring}
\end{equation}
We see that in both cases, the string is made out of the unbroken $\mathbb Z_2$ symmetry $P$. However, their endpoint operators are distinct: in the trivial (non-trivial) case it is real (imaginary). As before, these discretely distinct charges mean that the two Hamiltonians must be separated by a quantum phase transition.

In the SSB case, the long-range order of a \emph{local} operator together with the clustering property allowed us to derive a bulk degeneracy. This does not work for string-like objects since there is no natural way of factorizing the correlation function. However, if we consider the system with open boundaries, then it is well-defined to consider a string with a single endpoint (i.e., the other end disappears into the boundary)\footnote{More precisely, consider the long-range order with the endpoints of the strings close to the left and right boundary. Now act with the global symmetry $\prod_n X_n$ and apply clustering.}. It is for this reason that SPT phases come with protected zero-energy edge modes. In the case of the above cluster model, there is a protected qubit (a Kramers pair) at each edge, giving a global fourfold degeneracy with open boundaries \cite{Suzuki71,Raussendorf01,Keating04,Son11,Verresen17}.

More generally, the string order parameter \cite{Pollmann12b,Else13} for an on-site symmetry $\prod_n U_n$ has the form
\begin{equation}
\mathcal S_n = \cdots U_{n-2} U_{n-1} \mathcal O_n \label{eq:String}
\end{equation}
where $\mathcal O_n$ is some local operator such that $\langle \mathcal S^\dagger_m \mathcal S_n \rangle$ has long-range order, as in Eq.~\eqref{eq:clusterstring}.
In this work, we will use the term `symmetry flux' for such a string operator.
In principle, one could use this terminology more broadly for \emph{any} string operator of the form of Eq.~\eqref{eq:String}, since it acts like a source of flux for operators charged under the symmetry (see Section~\ref{subsec:twist}). We use this term in the more restrictive sense (i.e. for endpoint operators that give rise to long-range order) since it is the symmetry properties of its endpoint operator $\mathcal O_n$ which encode the projective representation (or, equivalently, the second group cohomology class) labeling the gapped phase \cite{Pollmann12b,Else13}.
(Relatedly, it has been appreciated how SPT phases can be diagnosed by what happens when you gauge the global symmetry, in which case these symmetry fluxes become local operators\footnote{Depending on the charge of these long-range ordered strings, they lead to distinct symmetry-breaking phases after gauging, and they incorporate the `correct' way of terminating a gauge string.} \cite{Levin12,Wang15}.)
In this work, we generalize these string order parameters (equivalently, symmetry fluxes) to cases where long-range order is replaced by `longest-range order' as we discuss now.

\subsection{Generalized order parameters for gapless phases \label{subsec:Z2Z2T}}

We saw that for gapped phases the key idea was that \emph{symmetry properties of both local and nonlocal operators allow us to define discrete invariants}.
We now show that this generalizes to gapless phases.

Let us note that we do not consider the trivial instance where the symmetry group of interest acts exclusively on gapped degrees of freedom (i.e., gapless modes would be uncharged). For instance,
if one stacks a gapped SPT phase on top of a critical chain such that all the symmetries protecting the former act trivially on the latter, then the SPT phase is automatically stable by virtue of, e.g., there not being any symmetry-allowed way of coupling the SPT edge mode operators to the critical bulk. We exclude such cases, instead focusing on situations where (part of) the protecting symmetry acts on the gapless low-energy theory---making the latter an essential piece of the physics at play. Remarkably, there are still local and nonlocal order parameters in this case.

The simplest example---with a local order parameter---is given by the two critical Ising chains:
\begin{equation}
\begin{array}{ccc}
H &= &- \sum_n (Z_n Z_{n+1} + X_n), \\
H'&= &- \sum_n (Y_n Y_{n+1} + X_n).
\end{array}
\end{equation}
These are at a phase transition between the trivial paramagnet and the above $H_{\rm Ising}$ and $H_{\rm Ising}'$, respectively.
Both are described by the Ising universality class, or, equivalently, the conformal field theory (CFT) with central charge $c=1/2$ \cite{DiFrancesco97}. This universality class has a \emph{unique local} operator $\sigma(x)$ with scaling dimension $\Delta_\sigma = 1/8$ (meaning that $\langle \sigma(x) \sigma(0) \rangle \sim 1/x^{2\Delta_\sigma}$) \cite{DiFrancesco97}.
The lattice operators that have overlap with this continuum field $\sigma$ are naturally given by the Ising order parameters of the nearby phases, i.e.\footnote{Throughout this paper, we will use the notation `$\textrm{CFT operator} \sim \textrm{lattice operator}$' to express which lattice operator contains which dominant scaling operator (as evidenced by, e.g., correlation functions).} $\sigma(x) \sim Z_n$ for $H$ and $\sigma(x) \sim Y_n$ for $H'$. We observe that these two operators transform \emph{differently} under complex conjugation $T$. We say that these two Ising CFTs are \emph{enriched} by the $\mathbb Z_2^T$ symmetry $T$, with the former CFT obeying $T \sigma T =+\sigma$ and the latter $T \sigma T = -\sigma$. Indeed, one can argue that this charge is always \emph{discrete}, i.e. $T \sigma T = \pm \sigma$, and that it is \emph{well-defined}, i.e., all choices of lattice operators that generate $\sigma$ in the low-energy effective theory have the same charge/sign \emph{if} $T$ is an unbroken symmetry\footnote{Both statements follow from that the fact that if $\mathcal O_1$ and $\mathcal O_2$ both generate $\sigma$, then $\langle \mathcal O_1 \mathcal O_2 \rangle \neq 0$ (following from the fusion rule $\sigma \times \sigma = 1 +\varepsilon$) which must be real if $T$ is unbroken.}. This discrete invariant cannot change as long as we stay within the Ising universality class.

What is the consequence of such an invariant? It means that any $G$-symmetric path (here $G = \mathbb Z_2 \times \mathbb Z_2^T$) of gapless Hamiltonians connecting $H$ and $H'$ must at some point go through a different universality class. For example, consider the interpolation $\lambda H + (1-\lambda) H'$ (with $0 \leq \lambda \leq 1$): this is everywhere in the Ising universality class \emph{except} at the halfway point $\lambda=1/2$, where the system passes through a multi-critical point  (with a dynamical critical exponent $z_\textrm{dyn}=2$). Alternatively, $\lambda H - (1-\lambda) H'$ passes through a Gaussian fixed point (central charge $c=1$) at $\lambda = 1/2$. It is at these non-Ising points that the property $T \sigma T = \pm \sigma$ changes. Of course, the $G$-symmetry of the path is key: $H$ and $H'$ are unitarily equivalent by a rotation around the $x$-axis, $e^{-i\alpha \sum_n X_n}$ ($0 \leq \alpha \leq \pi/4$), but this path violates complex conjugation symmetry $T$. This is similar to how we saw that the nearby gapped phases, $H_{\rm Ising}$ and $H_{\rm Ising}'$, are distinguished by the additional symmetry.

The above examples could be distinguished by symmetry properties of \emph{local} operators. A more interesting case is when two enriched critical points can only be distinguished by the symmetry properties of \emph{nonlocal} operators. An example is given by:
\begin{equation}
\begin{array}{ccl}
H &= &- \sum_n (Z_n Z_{n+1} + X_n), \\
H''&= &- \sum_n (Z_n Z_{n+1} + Z_{n-1}X_nZ_{n+1}). \label{eq:example}
\end{array}
\end{equation}
The former (latter) is a phase transition between the trivial phase and the Ising (cluster SPT) phase that we encountered above.
These two systems, $H$ and $H''$, cannot be distinguished by a local operator; in particular, both are described by the Ising universality class with $\sigma(x) \sim Z_n$ \cite{Jones19}, implying $T \sigma T = + \sigma$. In previous work, $H$ and $H''$ were shown to be topologically distinct by mapping them to free-fermion chains and then appealing to a winding number \cite{Verresen18,Jones19}. In this work, we go significantly beyond this by identifying a topological invariant which is well-defined in the presence of interactions. Indeed, the Ising CFT also has a \emph{nonlocal} operator $\mu(x)$ with scaling dimension $\Delta_\mu = 1/8$---this is related to the local $\sigma(x)$ under Kramers-Wannier duality. For the usual critical Ising chain, this is known to be the string operator $\mu(x) \sim \cdots X_{n-2} X_{n-1} X_n$ at any position $n$. For $H''$, however, one can show that $\mu(x) \sim \cdots X_{n-2} X_{n-1} Y_n Z_{n+1}$ \cite{Jones19}. (Note that in both cases, the nearby symmetry-preserving phase\footnote{This can be reached by perturbing with $\varepsilon(x) \sim - \mu(x) \mu(x+a) + \sigma(x) \sigma(x+a)$.} has long-range order $\lim_{|x-y| \to \infty} \langle \mu(x) \mu(y) \rangle \neq 0$---see e.g. Eq.~\eqref{eq:clusterstring}---as expected by Kramers-Wannier duality.) This suggests that the two models are two distinct $G$-enriched Ising CFTs distinguished by $T\mu T = \pm \mu$. To establish this, we need to ensure that the charge of $\mu$ is well-defined, i.e., that it is independent of our choice of operator on the lattice. This requires us to generalize the notion of symmetry flux or string order parameter (which we reviewed in the previous subsection) to gapless cases.

The symmetry flux of $P = \prod_n X_n$ is a string operator consisting of the same on-site unitaries as $P$, i.e., $\mathcal S_n = \big(\prod_{m<n} X_{m}\big) \mathcal O_n$, with a condition on the endpoint operator $\mathcal O_n$.
In the gapped case, we could choose the latter such that the string operator has long-range order.
In the gapless case, we generalize this condition by demanding that $\mathcal O_n$ is chosen such that $\langle \mathcal S^\dagger_m \mathcal S_n \rangle$ has slowest possible algebraic decay. For the Ising CFT, it is known that $\Delta=1/8$ is the smallest possible scaling dimension \cite{DiFrancesco97}. Hence, the above discussion about $\mu$ tells us that the symmetry flux of $P$ has the endpoint operator $\mathcal O_n = X_n$ for $H$ and $\mathcal O_n = Y_n Z_{n+1}$ for $H''$; note that their charges under $T$ are distinct.
One can argue---as we do in Section~\ref{sec:symflux}---that once one fixes the $\mathbb Z_2$ symmetry under consideration (here $P$), then the Ising universality class has a \emph{unique} symmetry flux, even in the presence of additional gapped degrees of freedom\footnote{E.g., consider a gapped SPT phase with long-range order in a string order parameter $\mathcal A_n$ which is odd under $T$. If this is stacked on top of a critical chain, then $\mu$ and $\mu \mathcal A$ have the \emph{same} scaling dimension $\Delta = 1/8$ yet \emph{distinct} charges. Fortunately, they are distinguished by their string, i.e., they are symmetry fluxes for \emph{distinct} symmetries.}! This means that we have a well-defined charge $T \mu T = \pm \mu$. We say that the above two models, $H$ and $H''$, realize two distinct $\mathbb Z_2 \times \mathbb Z_2^T$-enriched Ising CFTs. Moreover, since this is based on charges of nonlocal operators, we say that they are \emph{topologically distinct}.

We thus propose that distinct symmetry-enriched CFTs can be distinguished by how symmetry fluxes of $g \in G$ are charged under the other symmetries in $G$. For the models in Eq.~\eqref{eq:example}, the discrete invariant is the sign picked up when conjugating the symmetry flux of $P$ by $T$. A nice feature of this definition of symmetry flux is that in the gapped case it reduces to the usual string order parameter. Indeed, in that case, asking for the slowest possible decay is asking for long-range order.
Moreover, in the gapped symmetry-preserving case, one can argue that the symmetry flux is \emph{always} unique in one spatial dimension; this follows from the principle of symmetry fractionalization \cite{Fidkowski11class,Turner11class}. Indeed, the symmetry properties of its endpoint operator $\mathcal O_n$ encode the projective representation (or, equivalently, the second group cohomology class) labeling the gapped phase.
In the gapless case, the uniqueness or degeneracy of the symmetry flux depends on the particular CFT, as we discuss in detail in Section~\ref{sec:symflux}.

\subsection{Edge modes from charged symmetry fluxes \label{ssec:edgemodesymflux}}

If a symmetry flux has a non-trivial charge under another symmetry, then this can be linked to ground state degeneracies in the presence of open boundary conditions.
This is well-known for the case with a bulk gap \cite{Pollmann12b} and in this work we show that it extends to the gapless case.
We argue this more generally in Section~\ref{sec:classification}; here we illustrate this for the lattice model $H''$ in Eq.~\eqref{eq:example}, taking a half-infinite system with sites $n=1,2,\dots$. We show that the boundary of this critical chain spontaneously magnetizes (see Ref.~\onlinecite{Scaffidi17} for a related system with $G = \mathbb Z_2 \times \mathbb Z_2$). Indeed, on the lattice we see that $Z_1$ commutes with $H''$; the spontaneous edge magnetization $Z_1 = \pm 1$ thus gives a twofold degeneracy.
To see that this magnetization is not a mere artifact of our fine-tuned model, we can study its stability in the Ising CFT starting with $\langle \sigma(x\approx 0) \rangle \neq 0$ (with the boundary at $x=0$). If we add the $P$-symmetry flux $\mu(0)$ to the Hamiltonian, this would connect the two ordering directions and hence destabilize them. This is what would happen for the usual Ising chain $H$, where the symmetry flux is condensed near the boundary, $\langle \mu(0) \rangle \neq 0$, giving us a symmetry-preserving boundary condition, $\langle \sigma(0) \rangle =0$ \cite{Watts01}. However, if $T \mu T = -\mu$, the $\mathbb Z_2^T$ symmetry prevents us from adding this perturbation, and, remarkably, the edge magnetization is stable! Indeed, in Section~\ref{sec:edge} we show that all symmetry-\emph{allowed} perturbations correspond to operators with scaling dimension greater than one---implying that they are irrelevant for the boundary RG flow. In summary, the zero-dimensional edge spontaneously breaks $P$ symmetry, stabilized by the $\mathbb Z_2 \times \mathbb Z_2^T$-enriched bulk CFT.

It is important to note that this degeneracy crucially relies on the presence of an edge. The key reason for this is that, while $\mu(x)$ is \emph{nonlocal} in the bulk, it is \emph{local} near a boundary, i.e., its string can terminate\footnote{Note that terminating its string in the bulk would effectively lead to a \emph{two}-point operator $\mu(x) \mu(y)$.}; see Section~\ref{sec:edge} for how this enters the general argument. Indeed, $H''$ has a unique ground state with periodic boundary conditions. In fact, in that case $H$ and $H''$ are related by the unitary transformation $U = \prod_n (CZ)_{n,n+1}$, where $CZ$ is the control-$Z$ gate---and it is well-established that the critical Ising chain $H$ has a unique ground state. This unitary transformation also relates the trivial \emph{gapped} phase to the non-trivial \emph{gapped} cluster phase \cite{Raussendorf01}.

Thus far, we have focused on a single end, giving the complete story for a half-infinite system. For a finite system of length $L$, we need to consider the finite-size splitting of the symmetry-preserving states which entangle both edges:
\begin{equation}
|\uparrow_l \uparrow_r \rangle \pm |\downarrow_l \downarrow_r \rangle \quad \textrm{and} \quad |\uparrow_l \downarrow_r\rangle \pm |\downarrow_l \uparrow_r \rangle. \label{eq:overviewstates}
\end{equation}
In general, to analytically determine such finite-size splitting, it is useful to start in the scale-invariant RG fixed point limit and then consider what additional perturbations are necessary to distinguish the states. For example, if the system is gapped, then all four states in Eq.~\eqref{eq:overviewstates} are degenerate in the fixed point limit. The only way a local perturbation can then couple the two edges is at $L^\textrm{th}$ order in perturbation theory. Indeed, for gapped SPT phases, the finite-size splitting is exponentially small in system size \cite{Verresen17}. For critical systems, however, the RG fixed point limit is richer, being described by a CFT. In particular, it is known \cite{Cardy86} that the two anti-ferromagnetic states in Eq.~\eqref{eq:overviewstates} are split from the ferromagnetic ones at the energy scale\footnote{This is the \emph{only} possible nonzero energy scale at a fixed point limit.} $\sim 1/L$, the same as the finite-size \emph{bulk} gap. This is due to the spontaneous boundary magnetizations sensing their (mis)alignment through the critical bulk. The remaining two states, $|\uparrow_l \uparrow_r \rangle \pm |\downarrow_l \downarrow_r \rangle$, are exactly degenerate within the CFT. They can be split by perturbing away from the fixed point limit by adding RG-irrelevant perturbations. The splitting can already occur at second order in perturbation theory (intuitively, each edge has to couple to the critical bulk). Nevertheless, we show that the dominant contribution has a surprisingly large power $\sim 1/L^{14}$, caused by the so-called \emph{seventh descendant} of $\mu$; see Section~\ref{sec:edge} for a derivation.

The above twofold degeneracy with open boundaries is the generic result for a topologically non-trivial symmetry-enriched Ising CFT. The nature of its finite-size splitting depends on the protecting symmetry. In the above case, where $\mu$ is odd under an anti-unitary symmetry, we found an algebraic splitting $\sim 1/L^{14}$. In other scenarios, $\mu$ may be odd under a symmetry associated to additional gapped degrees of freedom, which by the same perturbative argument would lead to an exponentially small finite-size splitting. The latter case is in agreement with the observations in Ref.~\onlinecite{Scaffidi17}, and is also what we observe in Fig.~\ref{fig:xiloc}. Both scenarios are discussed in detail in Section~\ref{sec:edge}.

Since algebraically-localized edge modes are not common in the current literature (exceptions being systems with long-range interactions \cite{Vodola14,Vodola15,Patrick17,Jaeger20}), one might wonder to what extent they are, in fact, localized. The key feature that makes an edge mode meaningful is normalizability. Indeed, this is what allows one to make statements of the type ``$90\%$ of the weight is contained in the first five sites''. Algebraic modes can still be normalizable. For instance, suppose one has a Majorana edge mode of the form $\gamma_L \sim \sum_n \frac{1}{n^\alpha} \gamma_n$ (where $\gamma_n$ is a hermitian Majorana mode on site $n$, satisfying $\{\gamma_n,\gamma_m\}=2\delta_{n,m}$), by which we mean\footnote{In the non-interacting case, a stronger property will hold: an edge mode means that $[\gamma_L,H]=0$ such that all energy levels are degenerate. This is not generic for interacting systems.} that if $|\psi\rangle$ is a ground state, then so is $\gamma_L|\psi\rangle$. Then $\gamma_L^2 = \frac{1}{2}\{ \gamma_L, \gamma_L \} \sim \sum_n \frac{1}{n^{2\alpha}}$, which is finite if $\alpha > 1/2$, i.e., the operator is normalizable. Relatedly, an algebraic edge mode has a localization length. One way of understanding this is by noting that their localization is determined by perturbations which are irrelevant in the renormalization group flow---and such operators naturally have a length scale.
To the best of our knowledge, this work contains the first example of an algebraically-localized edge mode that cannot be destroyed whilst preserving the symmetry and universality class.

\subsection{Classifying \texorpdfstring{$G$}{G}-CFTs \label{ssec:classoverview}}

In Section~\ref{subsec:Z2Z2T}, we defined discrete invariants. Two natural questions arise: given a universality class, how many distinct invariants can one realize; and are these invariants complete? More precisely, if all local operators and symmetry fluxes have the same symmetry properties, can the models be smoothly connected whilst preserving their universality class? We explore this in Section~\ref{sec:classification}, arguing that these invariants are indeed complete for the Ising CFT. In particular, we discuss the case of the symmetry group $G = \mathbb Z_2 \times \mathbb Z_2$ for illustrative purposes. We first recall the six distinct \emph{gapped} phases in this symmetry class, after which we study the universality classes that naturally arise at the direct transitions between these phases. In particular, we find nine distinct $\mathbb Z_2 \times \mathbb Z_2$-enriched Ising ($c=1/2$) universality classes, whereas we show that \emph{all} $\mathbb Z_2\times \mathbb Z_2$-enriched Gaussian ($c=1$) transitions can be connected (if they have minimal codimension, as explained in Section~\ref{sec:classification}). For the $c=1$ case, we construct exactly-solvable models that allow us to connect symmetry-enriched CFTs which have seemingly distinct symmetry properties. This is possible since---unlike for the Ising CFT---there is a \emph{connected} family of different $c=1$ universality classes along which scaling dimensions can cross, such that the symmetry properties of symmetry fluxes need not be invariant for $c=1$. However, as we discuss in Sections~\ref{sec:classification} and \ref{sec:lit}, there are still non-trivial $G$-CFTs for $c\geq 1$.

\subsection{Majorana edge modes at criticality \label{ssec:fermionicoverview}}

The edge mode encountered in the spin chain $H''$ in Eq.~\eqref{eq:example} is rather unusual from the gapped perspective. Firstly, the ground state is unique with periodic boundary conditions, and \emph{twofold}\footnote{We do not count the states whose finite-size splitting scales as the bulk gap $\sim 1/L$.} degenerate with open boundaries. There is no gapped bosonic SPT phase with this property. Moreover, while the ground states $|\uparrow_l \uparrow_r\rangle \pm |\downarrow_l \downarrow_r\rangle$ can be toggled by a local edge mode operator $Z_1 \sim \sigma(0)$, such cat states are unstable and the system would collapse into $|\uparrow_l \uparrow_r\rangle$ or $|\downarrow_l \downarrow_r\rangle$ (i.e., the zero-dimensional edges exhibits spontaneous symmetry breaking, which is not possible in the absence of a bulk). In this collapsed basis, one would need an extensive operator $P$ to toggle between them.

In Section~\ref{sec:ferm} we show that under a Jordan-Wigner transformation, the above becomes a fermionic example where the edge mode is more conventional: the bulk is a free Majorana $c=\frac{1}{2}$ CFT, and each boundary hosts a localized zero-energy Majorana edge mode. This has the same ground state degeneracy as the usual gapped Kitaev chain \cite{Kitaev01}. Moreover, similar to the latter, there is an edge Majorana operator that toggles between the two (stable) ground states. The system is characterized by the symmetry flux of fermionic parity symmetry being odd under spinless time-reversal symmetry.

This critical Majorana chain arises as a phase transition from the gapped phase with two Majorana modes (per edge) protected by spinless time-reversal, to the Kitaev chain phase with one Majorana mode (per edge)---at the transition, one edge mode delocalizes, becoming the bulk critical mode, whereas the other mode remains localized \cite{Verresen18}.
Mapping this back to the spin chain language, starting from the gapped cluster phase, two of the four degenerate ground states with open boundary conditions have a splitting that is determined by the \emph{bulk} correlation length: these become delocalized at the critical point toward the Ising phase, with only the above twofold degeneracy remaining. Entering the gapped Ising phase, we have an edge magnetization which gradually merges with the bulk magnetization as we go deeper into the phase; see also Ref.~\cite{Parker18b} where an analogous case with gapped degrees of freedom is discussed.

\subsection{A unified language}

There is already a considerable body of work on critical one-dimensional systems with topological edge modes \cite{Kestner11,Cheng11,Fidkowski11longrange,Sau11,Ruhman12,Grover12,Kraus13,Ortiz14,Keselman15,Ruhman15,Kainaris15,Iemini15,Lang15,Ortiz16,Montorsi17,Wang17,Ruhman17,Scaffidi17,Guther17,Kainaris17,Jiang18,Zhang18,Verresen18,Parker18,Keselman18,Chen18}. Section~\ref{sec:lit} is devoted to demonstrating how our formalism allows us to unify previous works. We illustrate this for Refs.~\cite{Kestner11,Grover12,Keselman15,Scaffidi17,Verresen18}, showing how the models introduced therein can be interpreted as $G$-enriched Ising or Gaussian CFTs. This automatically identifies novel discrete bulk invariants for these systems in terms of their symmetry fluxes, ensuring the presence of protected edge modes.

\subsection{Generalization to arbitrary dimensions: twisted sectors \label{subsec:twist}}

While the present work focuses on one spatial dimension, these concepts can be generalized to higher dimensions, which is the focus of Section~\ref{sec:higher}.
To this end, we first reformulate the topological invariant in the one-dimensional case.
Any string operator associated to an on-site symmetry $U$ can be interpreted as creating a flux for that symmetry.
To see this, one can imagine inserting this operator in spacetime, then any operator charged under $U$ will pick up a phase factor when encircling the endpoint of the string---the defining characteristic of a flux.
More precisely, in Section~\ref{sec:higher} we explain how the operator-state correspondence relates the symmetry flux operator defined above (i.e., the slowest-decaying string operator) to a Hamiltonian in the presence of an external flux (i.e., certain terms in the Hamiltonian have been `twisted' by phase factors).
In particular, the charge of the endpoint operator of the symmetry flux naturally coincides with the charge attached to the external flux.
This gives a different perspective on the topological invariant for both gapped and gapless models.
In the gapped case, this coincides with a well-known approach for detecting and classifying SPT phases in terms of the response functions of external flux insertions \cite{Levin12,Hung13,Chen13,Barkeshli13,Wen14,Cheng14,Zaletel14,Teo14,Else14,Kapustin14,Kapustin14b,Wang15,Tarantino16,Tiwari18}.
For a simple illustration of detecting the SPT invariant for the cluster chain by inserting fluxes (i.e., twisting certain terms), see Eqs.~\eqref{eq:Scaffidi_twistA} and \eqref{eq:clustercharge} in Section~\ref{sec:higher}.

We show how this approach of encoding SPT invariants in the charges of external fluxes can generalize to the gapless case in general dimensions. In particular, we consider two-dimensional systems where the Hamiltonian has been twisted along a one-dimensional line (i.e., flux threads through the system). If the corresponding ground state is unique and its charge is distinct from the untwisted case, it provides a bulk topological invariant even in the gapless case, which can moreover be linked to edge modes. This idea is broadly applicable, and we illustrate it in detail for several copies of the 2+1D Ising CFT with a non-trivial topological invariant protected by $\mathbb Z_2^3$ symmetry; this constitutes an example which does not rely on any gapped sector.

\section{Symmetry fluxes and topological invariants \label{sec:symflux}}

In this section, we explain how to define discrete topological invariants for both gapped and gapless systems in one spatial dimension (higher-dimensional generalizations are discussed in Section~\ref{sec:higher}). We do this by introducing the notion of symmetry fluxes; their charges will be the invariants. In Section~\ref{subsec:symfluxHeisenberg}, we illustrate how this indeed associates a topological invariant to the Ising critical point of the spin-$1$ XXZ chain encountered in Fig.~\ref{fig:xiloc}. These invariants constrain the possible structure of phase diagrams, as we illustrate in Section~\ref{subsec:phasediagram}. This section is devoted to \emph{bulk} properties; the related phenomenon of topological edge modes is discussed in Section~\ref{sec:edge}.

\subsection{Defining symmetry fluxes and their charges \label{subsec:symfluxdef}}

Consider a symmetry element $g \in G$, where $G$ is the symmetry group of the Hamiltonian. If this is represented on the lattice by an \emph{unbroken} on-site unitary operator $U^g = \prod_n U^g_n$, we can associate to it a \emph{symmetry flux}. This is defined to be a (half-infinite) string operator of the form $\mathcal S^g_n \equiv \cdots U^g_{n-3} U^g_{n-2} U^g_{n-1} \mathcal O^g_n$ where the \emph{local} endpoint operator $\mathcal O^g_n$ is chosen such that the resulting correlator,
\begin{equation}
\langle \mathcal S_{m}^{g\dagger} \; \mathcal S^g_n \rangle = \langle \mathcal O_m^{g\dagger} U^g_m U^g_{m+1} \cdots U^g_{n-1} \mathcal O^g_n \rangle, \label{eq:correlator}
\end{equation}
has the slowest possible decay as a function of $|n-m|$.
In practice, the correlator in Eq.~\eqref{eq:correlator} has two possible functional forms, depending on whether or not $U^g$ is associated to gapped degrees of freedom---meaning that all particles charged under this symmetry are massive. If this is the case, its symmetry flux has long-range order (i.e., it tends to a finite positive value). Otherwise, there is algebraic decay, $\langle \mathcal S^{g\dagger}_{m} \mathcal S^g_n \rangle \sim 1/|n-m|^{2\Delta_g}$. (Note that any decay faster than algebraic would imply symmetry breaking.) The exponent $\Delta_g$ is called the scaling dimension of $\mathcal S_g$ and is by definition as small as possible; long-range order can be seen as the special case $\Delta_g=0$. The universality class of the system determines the value of $\Delta_g$.

An obvious and important question is whether---for a given model---the above definition specifies a \emph{unique} symmetry flux.
Note that we want to avoid trivial overcounting: multiplying any symmetry flux by, e.g., a phase factor $e^{ikn}$ will again give a symmetry flux. For this reason, we only consider non-oscillatory symmetry fluxes, i.e., $\lim_{|n-m|\to \infty} |n-m|^{2\Delta_g} \langle \mathcal S_m^{g\dagger} \mathcal S^g_n\rangle$ should be a well-defined finite value. Moreover, if $\mathcal S^g$ and $\tilde{\mathcal S}^g$ differ only by sub-leading correlations, then we do not want to count them as being distinct. In other words, we define an equivalence class where $\mathcal S^g$ and $\tilde{\mathcal S}^g$ are in the same class if and only if $\mathcal S^g - \tilde{\mathcal S}^g$ is \emph{not} a symmetry flux (i.e., its two-point correlator decays \emph{faster} than $\sim 1/|n-m|^{2\Delta}$).
We then naturally have a \emph{vector space} spanned by the (classes of) symmetry fluxes of $g\in G$, and its \emph{dimension} $D_g$ is a property of the universality class, as we will soon discuss. For instance, we will see that this space is one-dimensional for gapped systems and for the Ising universality class, which simplifies things considerably. We now discuss the action of the symmetry group $G$ on this vector space of symmetry fluxes.

\subsubsection{Charges of symmetry fluxes \label{subsubsec:charge}}

The group $G$ has a natural action on symmetry fluxes via conjugation:
\begin{equation}
U^h \mathcal S^g U^{h\dagger} = \cdots U_{n-2}^{hgh^{-1}} U_{n-1}^{hgh^{-1}} \left( U^h \mathcal O^g_n U^{h\dagger} \right).
\end{equation}
It is easy to see\footnote{Since $U^h$ is a symmetry, this string order still has scaling dimension $\Delta_g$. This shows that $\Delta_{hgh^{-1}} \leq \Delta_g$. By similarly starting from a symmetry flux of $hgh^{-1}$, we derive $\Delta_{hgh^{-1}} \geq \Delta_g$. Hence, $\Delta_{hgh^{-1}} = \Delta_g$.} that this is a symmetry flux for $hgh^{-1}$. It is hence natural to take $h$ to be an element of the subgroup of all elements commuting with $g$---the \emph{stabilizer} $C(g)$---which implies that $U^h \mathcal S^g U^{h\dagger}$ is again a symmetry flux for $g$.

Let us consider the case where the space of symmetry fluxes of $g$ is one-dimensional. Then the new flux $U^h \mathcal S^g U^{h\dagger}$ and $\mathcal S^g$ must be linearly dependent. We define the relative phase to be \emph{the charge of the symmetry flux of $g$ under $h$}. In general, the charge $\chi_g(h) \in U(1)$ is given by the prefactor in $U^h \mathcal S^g U^{h\dagger} = \chi_g(h) \mathcal S^g + \mathcal S^g_\textrm{sub}$, where $\mathcal S^g_\textrm{sub}$ is the subdominant (i.e., faster-decaying) piece. This is well-defined, i.e., independent of the choice of symmetry flux (within a given class), since this choice would only affect the subdominant piece. However, in practice one chooses the endpoint operator $\mathcal O^g$ to transform nicely under $C(g)$, in which case we can directly read off the charge from $U^h \mathcal S^g U^{h\dagger} = \chi_g(h) \mathcal S^g$. If $G$ is abelian, it can be shown that these charges are classified by $H^2(G,U(1))$; see Appendix~\ref{app:symflux} for details.

The charges under anti-unitary symmetries are slightly more subtle, since they seemingly change if we redefine $\mathcal S^g \to i \mathcal S^g$.
Here we show how to properly define such a charge for a symmetry flux of a unitary symmetry $g$ of order two (i.e., $(U^g)^2=1$), which will be enough for all the anti-unitary cases studied in this work. The key point is that a unitary that squares to unity must necessarily be hermitian: $U^g=U^{g\dagger}$. Using this, one can show that one can always choose the endpoint operator $\mathcal O^g$ of its symmetry flux to be hermitian\footnote{Indeed, $\mathcal O^g = (A+iB)/2$ with hermitian operators $A = \mathcal O^g + \left(\mathcal O^g\right)^\dagger$ and $B = i( \left(\mathcal O^g\right)^\dagger - \mathcal O^g )$. If neither $A$ nor $B$ give rise to symmetry fluxes, then also their linear combination $\mathcal O^g$ cannot.}, in which case its charge under an anti-unitary symmetry is well-defined\footnote{Alternatively, one can combine the anti-unitary symmetry with taking the dagger to get an effective unitary action.}.

This framework of symmetry fluxes applies to finite and continuous groups alike. However, in practice it is often sufficient to consider charge of discrete subgroups. This is already well-known for gapped SPT phases (e.g., the Haldane phase can be said to be protected by the $SO(3)$ group of spin rotations, but just as well by its $\mathbb Z_2 \times \mathbb Z_2$ subgroup of $\pi$-rotations \cite{Kennedy92}). The reason for this is simple to understand: the symmetry flux is well-defined for any element $g$ of a continuous group, but usually the charges under its stabilizer $C(g)$ are trivial. For instance, for a generic rotation in $SO(3)$, the only elements that commute with it are other rotations along the \emph{same} axis, and one can argue that these always lead to trivial charges. It is only at special high-symmetry points of $SO(3)$, namely at $\pi$-rotations, where the stabilizer contains non-trivial elements (such as $\pi$-rotations along orthogonal axes).

The above showed that charges are straightforwardly defined if the symmetry flux is unique, directly giving access to discrete invariants. We now show that this uniqueness is indeed guaranteed if, for example, the bulk is gapped or described by the Ising universality class. We also touch upon the more general case where the space of symmetry fluxes is higher-dimensional; in this case one can also obtain discrete invariants.

\subsubsection{Symmetry fluxes for gapped phases \label{subsubsec:gapped}}

In the gapped case, the symmetry flux for any unbroken symmetry $g \in G$ is \emph{unique}, i.e., any two symmetry fluxes are linearly dependent. This follows from the concept of \emph{symmetry fractionalization} (see Appendix~\ref{app:gappedsym} for a proof). Indeed, in this case the above definition of the symmetry flux coincides with the well-known notion of a string order parameter characterizing the phase \cite{Pollmann12b,Else13}. Note that this uniqueness applies even to \emph{critical} systems as long as $U^g$ acts non-trivially on \emph{gapped} degrees of freedom only, at least if we restrict ourselves to cases without bulk degeneracies\footnote{From the perspective of conformal field theory, this means that we assume that there is only a single local operator of scaling dimension zero.}. The uniqueness of this symmetry flux implies we have well-defined charges $\chi_g(h)$.

It is known that gapped phases are classified by topologically distinct projective representations of $G$ which are labeled by a so-called cocycle $\omega \in H^2(G,U(1))$ \cite{Chen10,Fidkowski11class,Turner11class}. The above charges $\chi_g(h)$ can be expressed in terms of this cocycle (see Appendix~\ref{app:gappedsym}). The converse is also true for, e.g., abelian groups $G$: in Appendix~\ref{app:gappedsym} we prove that the cocycle can be reconstructed from knowing the charges. This is not true for arbitrary groups $G$. Nevertheless, in practice knowing the charges is often equivalent to knowing the projective representations (the simplest known counter-example involves a group of 128 elements \cite{Pollmann12b}). 

\subsubsection{Symmetry fluxes for the Ising universality class \label{subsubsec:ising}}

We now show that if one is given a $G$-symmetric lattice model where some degrees of freedom are described by the Ising universality class and other degrees of freedom (if present) are gapped, then any $g \in G$ has a \emph{unique} symmetry flux $\mathcal S^g$. Firstly, the Ising CFT is known to have an emergent $\mathbb Z_2$ symmetry (i.e., this makes no reference to what the symmetries of the lattice model might be) and it is known that this has a \emph{unique}\footnote{This is because the partition function in the sector twisted by the emergent $\mathbb Z_2$ symmetry has a unique ground state; see Section~\ref{sec:higher}.} symmetry flux $\mu$ which moreover has scaling dimension $\Delta_\mu = 1/8$ \cite{DiFrancesco97}. Secondly, this is the \emph{only} unitary on-site symmetry of the Ising CFT \cite{Ruelle98}, hence any lattice symmetry which acts non-trivially on the local degrees of freedom of the CFT must coincide---for the gapless part of the spectrum---with the emergent $\mathbb Z_2$ symmetry, thereby inheriting the uniqueness of its symmetry flux. A part of the symmetry can of course also act non-trivially on additional \emph{gapped} degrees of freedom, but as argued in Section~\ref{subsubsec:gapped}, this does not affect the conclusion of uniqueness.

In summary, for any $g \in G$ we have a unique symmetry flux. Its scaling dimension is $\Delta_g = 1/8$ if $g$ acts non-trivially on the CFT, otherwise $\Delta_g = 0$. As described in Section~\ref{subsubsec:charge}, this uniqueness gives us well-defined charges $\chi_g(h) \in U(1)$ under any symmetry $h$ that commutes with $g$.
These charges form a \emph{discrete invariant} of the symmetry-enriched Ising CFT. We will see an example of this in subsection~\ref{subsec:symfluxHeisenberg}.

\subsubsection{Symmetry fluxes for general universality classes}

In the above two cases, for each $g \in G$, the vector space of symmetry fluxes happened to be one-dimensional. More generally, it might have some dimension $D_g$, such that each basis of symmetry fluxes $\mathcal S^g_{\alpha}$ come with an additional label $\alpha = 1, \dots, D_g$. Hence, the subgroup of elements that commute with $g$ (the stabilizer of $g$) has a higher-dimensional representation on this space of fluxes, $U^h [ \mathcal S^g_\alpha  ] U^{h\dagger} = R^g_{\alpha,\beta}(h) [ \mathcal S^g_\beta ]$, where $[\mathcal S^g]$ denotes the equivalence class defined above. For example, $D_g >1$ generically happens at transitions between distinct SPT phases (or between $G$-CFTs), where the different symmetry fluxes become degenerate.

A higher-dimensional representation can of course still come with discrete labels (in particular, there are cases where all degenerate fluxes have the \emph{same} charges; see Section~\ref{sec:lit}), but extra care has to be taken before one can conclude that these give \emph{invariants} of the $G$-enriched CFT. In particular, in rare cases, some CFTs allow for marginal perturbations which can change $D_g$. For example, the $c=1$ $\mathbb Z_2$-orbifold CFT allows for one-dimensional vector spaces of symmetry fluxes ($D_g=1$), but this CFT can be smoothly tuned to a compact boson CFT where this representation becomes two-dimensional. Such a process is explored in a lattice model in Section~\ref{sec:classification} where we use this to connect apparently distinct $G$-CFTs with $c=1$.

\subsubsection{Implications for phase diagrams}

Identifying \emph{discrete} invariants for symmetry-enriched universality classes has strong implications for the possible structure of phase diagrams. In particular, if two models are described by the same CFT at low energies but by distinct \emph{symmetry-enriched} CFTs, then these two transitions cannot be smoothly connected in a larger phase diagram. This means that any path attempting to connect them must have an intermediate point where the universality class discontinuously changes---either to a distinct CFT (necessarily of higher central charge) or to something that is not a CFT (e.g., a gapless point with dynamical critical exponent $z_\textrm{dyn} \neq 1$). Examples are discussed in subsection~\ref{subsec:phasediagram}.

\begin{figure}
	\includegraphics[scale=.98,trim=3 3 3 2,clip]{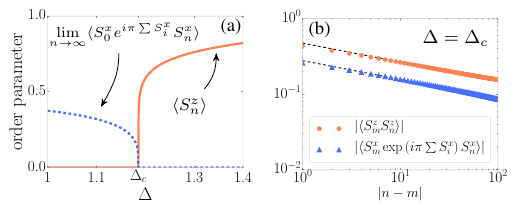}
	\caption{\textbf{Order parameters and symmetry fluxes for the spin-$1$ XXZ chain.} (a) The Haldane string order parameter and Ising order parameter have long-range order in the SPT and symmetry-breaking phase, respectively. (b) At criticality, the long-range order is replaced by algebraic decay. Both operators have the same scaling dimension (i.e., their correlators decay as $\sim 1/|n-m|^{2\Delta}$ with $\Delta = 1/8$; the dashed lines are a guide to the eye) and in the continuum limit correspond to the $\mu$ and $\sigma$ operators of the Ising CFT, respectively. We conclude that $\mu$ has non-trivial charge under other symmetries, which functions as a topological invariant. \label{fig:symflux}}
\end{figure}

\subsection{Topological invariant for the critical spin-1 anisotropic Heisenberg chain \label{subsec:symfluxHeisenberg}}

Having laid out the general structure of symmetry fluxes, we are now in a good position to apply this to a concrete model: the spin-$1$ XXZ chain. As mentioned in the introduction and as shown in Fig.~\ref{fig:xiloc}, as one tunes the easy-axis anisotropy $\Delta$ (not to be confused with a scaling dimension) from the topological Haldane phase to the symmetry-breaking Ising phase, there remains a localized edge mode at the Ising critical point. Here we identify a bulk topological invariant at criticality, establishing that it indeed forms a novel symmetry-enriched Ising CFT. This invariant also explains its edge behavior, as discussed in Section~\ref{sec:edge}.

The gapped Haldane phase is known to be protected by, for example, the $\mathbb Z_2 \times \mathbb Z_2$ group of $\pi$-rotations (represented by $R_\gamma = \prod_n e^{i\pi S^\gamma_n}$ for $\gamma=x,y,z$). This SPT phase has long-range order in $\langle S^\gamma_i \exp ( i \pi \sum_{i<k<j} S^\gamma_k ) S^\gamma_j\rangle$. The symmetry flux of $R_\gamma$ is thus $\prod_{m<n} e^{i\pi S^\gamma_m} S^\gamma_n$. Its long-range order is shown in Fig.~\ref{fig:symflux}(a) for $\gamma=x$, which indeed vanishes at the critical point $\Delta = \Delta_c$. In the Ising phase for $\Delta > \Delta_c$ we have the local order parameter $S^z_n$, characterizing the spontaneous breaking of $R_x$ and $R_y$.

Despite the vanishing of the long-range order of these order parameters, they still play an important role at criticality: they are identified by their scaling dimension $1/8$, as shown in Fig.~\ref{fig:symflux}(b). We thus conclude that---similar to the known gapped case---the symmetry flux of $R_x$ at criticality is still the Haldane string order parameter. As discussed in Section~\ref{subsubsec:ising}, this means that the Ising critical point is non-trivially enriched by $\mathbb Z_2 \times \mathbb Z_2$-symmetry. More concretely, if we denote the symmetry flux of $R_x$ by $\mu$ (using the traditional notation of scaling operators of the Ising CFT)\footnote{In principle we should label it by $\mu_{R_x}$, since the symmetry flux of $R_y$ also has scaling dimension $\Delta =1/8$ and may be denoted by $\mu_{R_y}$.}, we have the non-trivial charge $R_z \mu R_z = - \mu$.

One can also consider the symmetry flux for $R_z$. In fact, since the local fields of the Ising transition are all neutral under $R_z$, this symmetry remains gapped. As a consequence, the topological string order parameter for $R_z$ has long-range order throughout this whole region of the phase diagram. This thus also functions as a topological invariant. This is a general feature of gapless SPT phases which are (partially) protected by gapped degrees of freedom. The case without such an additional gapped sector is thus arguably the most novel: there is no long-range order to appeal to, making the CFT-based approach essential. This is the case, for example, for $H''$ encountered in Section~\ref{sec:overview} and as we will see in Section~\ref{sec:edge}, this difference has an important consequence for the edge modes.

We note that the gapped Haldane phase is also known to be protected by time-reversal symmetry $T_\textrm{spin} = R_y K$ (where $K$ is complex conjugation in the local $z$-basis). Similarly, the Ising criticality at $\Delta = \Delta_c$ is enriched by the $\mathbb Z_2 \times \mathbb Z_2^T$ symmetry generated by $R_x$ and $T_\textrm{spin}$. Indeed, we see that $T_\textrm{spin} \mu T_\textrm{spin} = - \mu$. It would be interesting to define a notion of symmetry flux for anti-unitary symmetries, which in this case would have to be charged under the unitary symmetries $R_\gamma$ as well as under itself!

\begin{figure}
\includegraphics[scale=1]{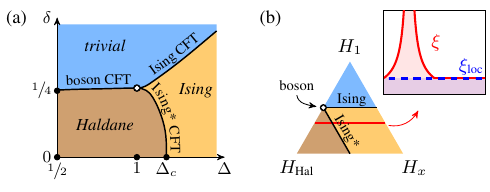}
\caption{\textbf{Transitions between topologically distinct Ising universality classes.} (a) Phase diagram of the bond-alternating $S=1$ XXZ chain. There is a $c=1$ transition between the topologically distinct $c=1/2$ transitions. The tricritical point (hollow marker) is a WZW $SU(2)_1$ CFT, where the trivial and Haldane string order parameters both have scaling dimension $1/8$. (b) Similar phase diagram for an exactly solvable $S=1/2$ model; in this case, the $c=1$ boson CFT is in the free Dirac universality class (inset: analogue of Fig.~\ref{fig:xiloc}(a)). \label{fig:phasediagram}}
\end{figure}

\subsection{Implications for phase diagrams: bond-alternating spin-1 Heisenberg chain \label{subsec:phasediagram}}

Having identified the discrete invariant $R_z \mu R_z = - \mu$ for the spin-$1$ XXZ chain at criticality, we can distinguish it from a trivial Ising criticality where $R_z \mu R_z = \mu$. Whereas the former appears at a phase transition between an Ising phase and a non-trivial SPT phase, the latter appears as one tunes from/to a trivial SPT phase\footnote{Condensing $\mu$ would lead to a symmetric gapped phase, with its symmetry properties determining its phase.}. The discreteness of $R_z \mu R_z = \pm \mu$ means that these two Ising criticalities cannot be smoothly connected, constraining the possible structure of phase diagrams containing such Ising transitions. This can be illustrated by adding, e.g., bond-alternation:
\begin{equation}
H = J \sum_n (1-\delta (-1)^n)\left( S^x_n S^x_{n+1} + S^y_n S^y_{n+1} + \Delta S^z_n S^z_{n+1} \right). \label{eq:XXZv2}
\end{equation}
The two-parameter phase diagram, obtained with iDMRG \cite{White92,Kjall13}, is shown in Fig.~\ref{fig:phasediagram}(a) (see also Refs.~\onlinecite{Yamanaka93,Satoshi18}). For large $\delta$ we can realize a trivial Ising CFT where the symmetry flux of $R_x$ is \emph{not} charged, being given by $\mu \sim e^{i \pi \sum_{m<n} S^x_m}$. We observe that the two topologically-distinct symmetry-enriched Ising CFTs (with central charge $c=1/2$) are separated by a point where the universality class changes (hollow white marker). In this case, this `transition of transitions' is described by the Wess-Zumino-Witten $SU(2)_1$ CFT with central charge $c=1$. Here, the symmetry flux of $R_x$ still has scaling dimension $\Delta_{R_x} = 1/8$, but this space is now \emph{two-dimensional} (i.e., there are two linearly independent symmetry fluxes of $R_x$). Indeed, one symmetry flux is non-trivial, $e^{i \pi \sum_{m<n} S^x_m}S^x_n$, and the other, trivial, $e^{i \pi \sum_{m<n} S^x_m}$. This degeneracy of symmetry fluxes is in fact true along the whole line of $c=1$ CFTs separating the trivial and Haldane SPT phase, which is natural given the emergent duality symmetry at this transition.

Similar phenomenology can already be observed in an exactly-solvable spin-$1/2$ chain which realizes the same three gapped phases, as shown in Fig.~\ref{fig:phasediagram}(b). In this case, the two topologically distinct Ising CFTs are now separated by the $c=1$ free Dirac CFT, where the two symmetry fluxes of $R_x$ again become degenerate (now with scaling dimension $\Delta_{R_x} = 1/4$). The Hamiltonians for these spin-$1/2$ chains are introduced and discussed in detail in Section~\ref{sec:classification} and the phase diagram is calculated in Appendix~\ref{app:freeferm}.

\section{Edge modes at criticality \label{sec:edge}}

In the previous section, we introduced the notion of a symmetry flux and its corresponding charge. In this section, we relate this bulk property to degeneracies in the presence of open boundary conditions. This section is naturally divided in two: in Section~\ref{subsec:halfinf} we focus on a \emph{single} boundary of a half-infinite system, where we explain how a charged symmetry flux can lead to the boundary spontaneously breaking a symmetry; in Section~\ref{subsec:finite}, we study the coupling between such boundary magnetizations and calculate the finite-size splitting of the degeneracy. Throughout this section, we use the Ising CFT as an illustrative example, but the method which we lay out is generally applicable. In fact, at the end of every subsection, we indicate what the necessary analysis would be for a general CFT.

\begin{figure}
\includegraphics[scale=1]{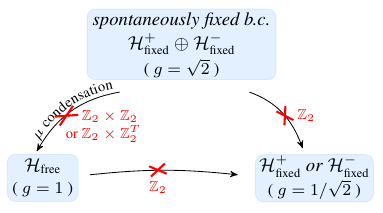}
\caption{\textbf{Topologically protected edge modes in the Ising CFT.} Boundary RG flow for the Ising CFT: usually the free boundary condition is stable (when preserving global $\mathbb Z_2$), but it can be prevented when $\mu$ is charged under additional symmetries. In that case, the spontaneously fixed boundary condition (with a global twofold degeneracy) is stable! \label{fig:edge}}
\end{figure}

\subsection{Half-infinite chain: a single boundary \label{subsec:halfinf}}

Let us consider a \emph{half-infinite} system, allowing us to study a single edge. We furthermore presume that the system is described by the Ising universality class. To study what happens near its boundary, we use the fact that all possible \emph{boundary RG fixed points}---describing the $(0+1)$-dimensional edge of this CFT---are known \cite{Cardy86,Cardy89}. There are three distinct fixed points, as sketched in Fig.~\ref{fig:edge}; these have different behavior with respect to the $\mathbb Z_2$ symmetry intrinsic to the Ising CFT:
\begin{enumerate}
	\item The \emph{free} boundary fixed point: this preserves the $\mathbb Z_2$ symmetry. The symmetry flux of this $\mathbb Z_2$ symmetry, denoted by $\mu(x)$, can be said to have condensed. Indeed\footnote{Note that in the bulk the one-point function of $\mu(x)$ is not well-defined due to it being nonlocal. However, near the boundary $\mu(x)$ becomes local.}, $\langle \mu(x_\textrm{boundary}) \rangle \neq 0$ \cite{Watts01}.
	\item The (explicitly) \emph{fixed}$^\pm$ boundary fixed point: this requires explicitly breaking the $\mathbb Z_2$ Ising symmetry at the level of the Hamiltonian (near its boundary). The local spin operator $\sigma(x)$ points up or down near the edge, $\langle \sigma(x_\textrm{boundary}) \rangle \neq 0$; its sign depends on the explicit breaking of the $\mathbb Z_2$ symmetry. This fixed point is \emph{excluded}  by enforcing $\mathbb Z_2$ symmetry of the Hamiltonian.
	\item The \emph{spontaneously fixed} boundary fixed point: the symmetry is not broken at the level of the Hamiltonian, but the ground state spontaneously magnetizes. More precisely, it is a \emph{two-dimensional} Hilbert space which is the direct sum of the two symmetry-broken orderings.
\end{enumerate}
The latter fixed point\footnote{Note that this is not a Cardy state in the boundary conformal field theory.} is not often mentioned in the literature, but this is for a good reason: it is usually an \emph{unstable}\footnote{Consider, for example, $H = -\sum_{n=1}^\infty Z_n Z_{n+1} - \sum_{n=2}^\infty X_n$ with its spontaneous boundary magnetization $Z_1 = \pm 1$. This is unstable against the perturbation $X_1 \sim \mu(0)$, flowing to a unique ground state.} RG fixed point. A generic perturbation would condense $\mu(x)$ near the boundary, flowing to the free fixed point. From the RG perspective, this is due to the \emph{boundary scaling dimension} of $\mu(x_\textrm{boundary})$ being $1/2$; this is smaller than one, implying it is \emph{relevant} for $0+1$-dimensional RG flows. However, if $\mu(x)$ is charged under some \emph{additional} symmetry, then this RG flow is prohibited and the spontaneous fixed boundary condition---along with its twofold degeneracy---is stabilized! The above is summarized in Fig.~\ref{fig:edge}. This explains the edge modes observed in Section~\ref{sec:overview} and in Fig.~\ref{fig:xiloc}, confirming that these are not fine-tuned features.

One can in principle repeat the above analysis for any CFT. The necessary information to set up the problem is the list of possible conformal boundary conditions, as well as the boundary condition changing operators and their (boundary) scaling dimensions. Supplemented by the symmetry properties of these operators, one can study which boundary RG fixed points are stable and (non)degenerate.

\subsection{Finite-size splitting: coupled boundaries \label{subsec:finite}}

As argued in the previous subsection, if the $\mathbb Z_2$-symmetry flux $\mu(x)$ of the Ising CFT is charged under an additional symmetry, the boundary spontaneously magnetizes. We now investigate the resulting degeneracy---and its finite-size splitting---for a \emph{finite} chain of length $L$ (with $x \in [0,L]$). Since the bulk gap vanishes as $\sim 1/L$, we can only meaningfully speak of degeneracies whose finite-size splitting decays \emph{faster} than this. There are four candidate degenerate ground states, labeled by their boundary magnetizations:
\begin{equation}
|\uparrow_l \uparrow_r\rangle, \; |\uparrow_l \downarrow_r\rangle, \; |\downarrow_l \uparrow_r\rangle, \; |\downarrow_l \downarrow_r\rangle. \label{eq:deg}
\end{equation}
However---presuming the model under consideration has a ferromagnetic sign---the antiferromagnetic states in Eq.~\eqref{eq:deg} are split from the ferromagnetic ones at the scale $\sim 1/L$ \cite{Cardy86,Cardy89}. Intuitively, this is because the edges can sense their (mis)alignment through the critical bulk. We thus only have \emph{two} ground states: $|\uparrow_l \uparrow_r\rangle$ and $|\downarrow_l \downarrow_r\rangle$. In fact, \emph{within the CFT}, these are exactly-degenerate eigenstates. This does not mean that they have no splitting in realistic systems: by definition, a CFT has no length scales, hence the only quantity with units of energy is $1/L$. If we perturb the CFT with RG-irrelevant perturbations---present in any realistic system---then the twofold degeneracy could be split by the smaller energy scales $\sim \xi^\alpha/L^{1+\alpha}$ or $\sim \exp(-L/\xi)$ (where $\xi$ is a constant with units of length). As a consistency check, note that in the RG fixed point limit, $\xi/L \to 0$, confirming that splittings which are faster than $1/L$ are indeed not visible in the CFT.

The purpose of this section is to determine whether the finite-size splitting of the aforementioned twofold degeneracy is algebraic or exponential in system size. This comes down to analyzing the possible perturbations $V$ one can add to the CFT Hamiltonian, $H = H_\textrm{CFT} + V$, mixing $|\uparrow_l \uparrow_r\rangle \leftrightarrow |\downarrow_l \downarrow_r\rangle$. To connect these two states which break the Ising symmetry, we need to perturb with the corresponding symmetry flux. Indeed, $\mu(x_\textrm{boundary})$ is known to be a \emph{boundary-condition-changing} (bcc) operator toggling between the two fixed$^\pm$ boundary conditions \cite{Cardy86,Cardy89}. We have already established that we cannot add $\mu(x)$ due to it being charged under an additional symmetry. However, there is a whole \emph{tower} of bcc operators: the \emph{descendants} of $\mu(x)$, with scaling dimensions $\Delta_n = 1/2 + n$ (where $n = 1,2,3,\cdots$). Note that since $\Delta_n>1$, these are RG-irrelevant for the $(0+1)$-dimensional edge. This means they do not affect the analysis in Fig.~\ref{fig:edge}, but they can indeed contribute to finite-size splitting. Whether it is possible to add such descendants of $\mu(x)$ depends on the the protecting/enriching symmetry. Determining which descendant---if any---is allowed, determines whether the splitting is exponential or algebraic (along with its power), as we explain now.

\subsubsection{Exponential splitting}

If $\mu(x)$ is charged under a symmetry $U$ which is associated to gapped degrees of freedom (d.o.f.), then all its descendants have the same non-trivial charge. To see this, note that the descendants are created by applying the \emph{local} Virasoro generators $L_n$ to $\mu(x)$, and $U$ acts trivially on local gapless d.o.f.. In conclusion, there is no perturbation within the low-energy CFT that can couple $|\uparrow_l \uparrow_r\rangle \leftrightarrow |\downarrow_l \downarrow_r\rangle$. Any effective interaction must hence be mediated through gapped d.o.f., which can lead at most to a finite-size splitting $\sim \exp{(-L/\xi)}$. This applies whenever the protecting symmetry is \emph{unitary}: the only unitary symmetry of the Ising CFT is its $\mathbb Z_2$ symmetry \cite{Ruelle98}, such that any additional unitary symmetry must be associated to gapped d.o.f.. A case in point is the critical spin-$1$ XXZ chain, where $\mu(x)$ is charged under $R_z$ (see Section~\ref{subsec:symfluxHeisenberg}). This explains the exponentially-localized edge mode observed in Fig.~\ref{fig:xiloc}. A similar conclusion was drawn in the work by Scaffidi, Parker and Vasseur \cite{Scaffidi17}, where gapped degrees of freedom stabilized a spontaneous boundary magnetization in a critical Ising chain.

\begin{figure}
\includegraphics[scale=.98,trim=4 4 4 3,clip]{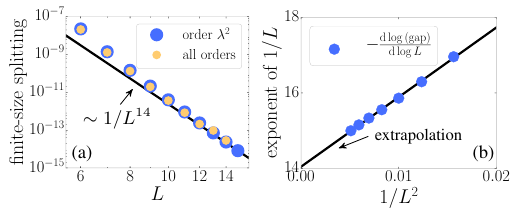}
\caption{\textbf{Finite-size splitting of edge mode in the Ising CFT.} If a unitary symmetry protects the edge mode, finite-size splitting is exponentially small in system size. For an anti-unitary symmetry, however, the field theory analysis in the main text suggests a splitting $\sim 1/L^{14}$. (a) We confirm the latter numerically in a spin chain model where $\mu$ is charged under $\mathbb Z_2^T$. Note that for $L=15$, the gap is around machine precision. (b) To separate out higher-order corrections present in (a), we extrapolate the leading exponent of $1/L$; the data agrees with the predicted splitting $\sim 1/L^{14} \times \left(1  + \alpha/L^{2} \right)$. \label{fig:edgenum}}
\end{figure}

\subsubsection{Algebraic splitting}

In case $\mu(x)$ is charged under an anti-unitary symmetry $T$, then the charge of its descendants can be different. More generally, if $\mathcal O_n$ is the lattice operator for a continuum operator $\varphi(x)$, then its first descendants can be realized by $\partial_x \varphi(x) \sim \mathcal O_{n+1} - \mathcal O_n$ and $\partial_t \varphi(x) \sim i[H,\mathcal O_n]$. Hence, we see that $\partial_x \varphi(x)$ has the \emph{same} charge under $T$ as $\varphi(x)$, whereas it is opposite for $\partial_t \varphi(x)$. For the boundary operator $\mu(0)$, we only have the time-like derivative, i.e., $\mu(0)$ has a unique first descendant with charge $T \partial_t \mu(0) T = + \partial_t \mu(0)$.

We are hence allowed to add a perturbation $H = H_\textrm{CFT} + \lambda \partial_t \mu(0)$. Remarkably, however, this descendant cannot split the twofold ground state degeneracy. Indeed, the perturbation \emph{disappears} after a well-chosen change of basis:
\begin{equation}
H = H_\textrm{CFT} + i \lambda[H,\mu(0)] = \underbrace{e^{-i \lambda \mu(0)} H_\textrm{CFT} e^{i \lambda \mu(0)}}_{\equiv \tilde H_\textrm{CFT}} + O(\lambda^2). \label{eq:rot}
\end{equation}
Note that since $T \mu(0) T = - \mu(0)$, our rotated Hamiltonian $\tilde H_\textrm{CFT}$ is still symmetric: $T \tilde H_\textrm{CFT} T = \tilde H_\textrm{CFT}$. Eq.~\eqref{eq:rot} tells us that the perturbed Hamiltonian $H$ is equivalent to a Hamiltonian with the \emph{same spectrum} as the unperturbed case (plus perturbations given by \emph{higher} descendants at order $O(\lambda^2)$). In particular, the ground state degeneracy is not split by the first descendant.

To summarize the above: the \emph{zeroth} descendant (i.e., $\mu(0)$) was not allowed by symmetry, whereas the symmetry-allowed \emph{first} descendant could be ``rotated'' away using the zeroth descendant. Interestingly, this pattern continues for a while, but not indefinitely. Firstly, it can be shown that any \emph{even} descendant is $T$-odd \cite{Blumenhagen09} and is hence excluded by symmetry. Secondly, whilst all \emph{odd} descendants are allowed by symmetry, many of them can be removed by using the fact that even descendants generate symmetry-preserving rotations. More precisely, if we denote the number of $n^\textrm{th}$ descendants of $\mu(0)$ as $N_n$ (which have scaling dimension $\Delta_n = 1/2+n$), then only $N_{2m+1} - N_{2m}$ descendants\footnote{All $N_{2m}$ descendants generate \emph{independent} rotations since otherwise a linear combination of descendants would be \emph{conserved}---in conflict with the nonzero scaling dimension $\Delta_{2m}$.} at level $2m+1$ cannot be rotated away and can hence actually cause a finite-size splitting! We thus need to determine the smallest \emph{odd} $n$ such that $N_n > N_{n-1}$. Fortunately, the number of descendants of $\mu(0)$ can be read off from the (chiral) partition function, i.e., $\sum_n N_n q^n = q^{-\frac{23}{48}}  \chi_{1,2}(q)$ \cite{DiFrancesco97}, where
\begin{equation}
\begin{split}
q^{-\frac{23}{48}}  \chi_{1,2}(q) = \big(1 & + q+q^2+q^3 \\
 &+ 2q^4 + 2 q^5 + 3 q^6 + 4 q^7 + \cdots \big).
\end{split} \label{eq:partition}
\end{equation}
We see that $N_7-N_6=4-3=1$, leaving us with a single symmetry-allowed seventh descendant of $\mu(0)$, which we denote by $\mu^{(7)}(0)$, that cannot be rotated away. We conclude that the perturbation $V = \lambda(\mu^{(7)}(0) + \mu^{(7)}(L))$ can split the degeneracy. Since we need to flip \emph{both} edges in order to couple $|\uparrow_l \uparrow_r\rangle \leftrightarrow |\downarrow_l \downarrow_r\rangle$, we have to go to second order in $\lambda$ to observe a splitting, i.e., $\lambda^2/L^\beta$. Since this has to have units of energy, we can conclude that $\beta = 1-2[\lambda]$, where the unit of $\lambda$ is $[\lambda] = 1-\Delta_7$. In summary, we find an algebraic splitting with power $\beta = 2 \Delta_7 - 1 = 14$.

This remarkably fast algebraic decay can be confirmed in the spin model $H'' = - \sum_{n=1}^{L-1} Z_{n} Z_{n+1} - \sum_{n=1}^{L-2} Z_{n} X_{n+1} Z_{n+2}$ introduced in Section~\ref{sec:overview}. This fine-tuned model has exact boundary magnetizations $Z_1 = \pm 1 = Z_L$. For this system,
\begin{equation}
\begin{array}{rcl}
\mu(x) &\sim &\cdots X_{n-2} X_{n-1} Y_n Z_{n+1}, \\
\partial_t \mu(x) &\sim &\cdots X_{n-2} X_{n-1} X_n.
\end{array}
\end{equation}
In particular, we see that $\partial_t \mu(0) \sim X_1$. This operator is allowed by symmetry but does not contribute to the splitting, as discussed above. We consider a dressed operator, $X_1 Z_2 Z_3$, which should generically contain all possible descendants of $\mu(0)$ which are $T$-even. Our perturbed Hamiltonian is thus $H = H'' + \lambda(X_1Z_2Z_3 + Z_{L-2}Z_{L-1}X_L)$. In Fig.~\ref{fig:edgenum}(a) we see the finite-size splitting obtained with exact diagonalization for $\lambda = 0.1$. The data is consistent with the CFT prediction $\sim 1/L^{14}$. We can also analytically predict the next-to-leading order contribution, which arises from acting with the seventh descendant of $\mu$ on one end and with the ninth descendant on the other, generating a splitting $\sim 1/L^{\Delta_7 + \Delta_9 -1} = 1/L^{16}$. The gap thus scales as $\sim 1/L^{14} + \alpha/L^{16}$, which means that the effective exponent of the algebraic decay is $L$-dependent: $-\mathrm d \log (\textrm{gap})/\mathrm d \log L \approx 14 + 2 \alpha / L^2$ for large $L$. The numerical data in Fig.~\ref{fig:edgenum}(b) shows perfect agreement with this formula.

The above focused on the illustrative example of the Ising CFT, but the principle is much more general. Once one has studied the boundary RG flow diagram and concluded the presence of edge modes due to charged symmetry fluxes (as in subsection~\ref{subsec:halfinf}), one can study the finite-size splitting by determining the dominant symmetry-allowed bcc operator. In the anti-unitary case, one can use the fact that the contributions of descendants can be rotated away by other descendants---which applies to any CFT. One important difference will be the counting appearing in the tower of states such as in Eq.~\eqref{eq:partition}.

\section{Classifying \texorpdfstring{$\bm G$}{G}-enriched CFTs in 1+1d \label{sec:classification}}

The previous sections have shown how a given bulk universality class (with, in particular, a fixed central charge $c$) can split up into distinct classes when additional symmetries are enforced. The concept of a symmetry flux---and its associated charge---allows us to distinguish such symmetry-enriched CFTs. However, there are \emph{various} invariants one can associate to $G$-symmetric universality classes:
\begin{enumerate}
	\item \textbf{Spontaneously broken symmetries.} At the coarsest level, there is the phenomenon of \emph{symmetry breaking}. The invariant this allows us to define is the subgroup $G_\textrm{eff} \subset G$ of \emph{unbroken symmetries}. In particular, absence of symmetry-breaking means $G_\textrm{eff} = G$.
	\item \textbf{Charges of symmetry fluxes.} As defined and discussed in Section~\ref{sec:symflux}, to any $g \in G_\textrm{eff}$ we can associate a symmetry flux $\mathcal S^g$. We can measure the charge of $\mathcal S^g$ with respect to any other symmetry $h \in G_\textrm{eff}$ which commutes with $g$. The universality class tells us whether this charge is an invariant of the phase (e.g., if the space of $g$-symmetry fluxes is one-dimensional).
	\item \textbf{Gapped symmetries.} Another robust invariant of the phase is the list of the unbroken symmetries which act only on \emph{gapped degrees of freedom}, forming a subgroup $G_\textrm{gap} \subset G_\textrm{eff}$.
	We refer to symmetries in $G_\textrm{gap}$ as being \emph{gapped}.
	(Note that having a larger $G_\textrm{gap}$ is not necessarily correlated with having a smaller central charge; it does, however, usually imply that the critical degrees of freedom are more \emph{unstable}.)
	\item \textbf{Charges of local scaling operators.} Lastly, for any unbroken symmetry $g \in G_\textrm{eff}$ which is \emph{not} gapped (i.e., $g \notin G_\textrm{gap}$), we can study the charges of \emph{local} low-energy CFT observables under $g$. Similar to the charges of symmetry fluxes, these charges can lead to invariants of the phase (e.g., a unique local operator with a particular scaling dimension has an invariant charge). For example, recall the Ising operator $\sigma$ from Section~\ref{sec:overview} that was either real or imaginary.
\end{enumerate}

For gapped phases, the third and fourth invariants are trivial (since $G_\textrm{gap} = G_\textrm{eff}$). Indeed, in most practical cases, the first two provide a complete classification of gapped one-dimensional phases protected by an on-site symmetry group $G$ \cite{Fidkowski11class,Turner11class,Chen10,Schuch11}. A natural question is whether the above invariants are also complete for $G$-enriched CFTs. We show that this is the case for the Ising CFT, which we first illustrate for the symmetry group $\mathbb Z_2 \times \mathbb Z_2$. For Gaussian CFTs, we give a \emph{partial} answer, showing that the $c=1$ transitions arising between gapped $\mathbb Z_2 \times \mathbb Z_2$-symmetric phases can all be smoothly connected. We also use the latter as an instructive example for the more subtle points encountered in Section~\ref{sec:symflux}, showing that $G$-CFTs with \emph{apparently} distinct charges for their symmetry fluxes can be part of the \emph{same} $G$-CFT.
In this section, we limit ourselves to unitary symmetries.

Throughout this section, we discuss the example $G=\mathbb Z_2 \times \mathbb Z_2$ in the context of a spin-$1/2$ chain, with the symmetries realized by the group of $\pi$-rotations, $R_\gamma = \prod_n R_{\gamma}^{(n)}$, where
\begin{equation*}
R_{x}^{(n)} = X_{2n-1}X_{2n}, \; R_{y}^{(n)} = Y_{2n-1}Y_{2n}, \; R_{z}^{(n)} = Z_{2n-1}Z_{2n}.
\end{equation*}
Note that the way we write these symmetries betrays that we have fixed a \emph{unit cell}, such that the $R_\gamma^{(n)}$ symmetries indeed \emph{commute}. This ensures that this defines a \emph{linear}, \emph{on-site} representation of $\mathbb Z_2 \times \mathbb Z_2$.

In Section~\ref{subsec:gap}, we recall the six gapped phases that can occur for this symmetry group, along with solvable models which are used in the following two subsections. In Section~\ref{subsec:ising}, we classify the symmetry-enriched Ising CFTs. In Section~\ref{subsec:gaussian}, we discuss the case of the Gaussian CFT.

\subsection{Gapped bulk \label{subsec:gap}}

To keep this work self-contained, we briefly review the gapped phases with this symmetry group. According to the classification for spin chains with $\mathbb Z_2 \times \mathbb Z_2$ symmetries, there are exactly two gapped phases where the ground state preserves the full symmetry group \cite{Fidkowski11class,Turner11class,Chen10,Schuch11}. These are the trivial phase and the topological Haldane phase, which can be realized in a spin-$1/2$ chain, respectively, as follows:
\begin{align*}
H_1 &= \sum_n \left( X_{2n-1} X_{2n} + Y_{2n-1} Y_{2n} \right), \\
H_\textrm{Hal} &= \sum_n \left( X_{2n} X_{2n+1} + Y_{2n} Y_{2n+1} \right).
\end{align*}
For both Hamiltonians, the ground state is a product of singlets, but for $H_1$ each singlet is \emph{within} a unit cell, whereas for the latter it is \emph{across} unit cells. The latter is reminiscent of the ground state of the well-known spin-$1$ Affleck-Kennedy-Lieb-Tasaki (AKLT) model; indeed, by introducing a term that penalizes spin-$0$ states in each unit cell, $H_\textrm{Hal}$ can be adiabatically connected to the AKLT model \cite{Affleck88,Hida92} (which in turn can be connected to the spin-$1$ Heisenberg chain in Eq.~\eqref{eq:XXZ}). Relatedly, while both models clearly have a unique ground state for periodic boundary conditions, $H_\textrm{Hal}$ has a zero-energy spin-$1/2$ degree of freedom at each open boundary. (Note that to keep the \emph{on-site} representation of $\mathbb Z_2 \times \mathbb Z_2$ well-defined, we can only cut the chain \emph{between} unit cells.) These two symmetric phases can be distinguished by their symmetry fluxes: e.g., for $H_1$ the symmetry flux of $R_x$ is $\mathcal S^x = \cdots R_x^{(n-2)} R_x^{(n-1)} R_x^{(n)}$, whereas for $H_\textrm{Hal}$ it is $\mathcal S^x =\cdots R_x^{(n-2)} R_x^{(n-1)} X_{2n-1}$. These clearly have different charges under $R_y$ and $R_z$.

In addition to these two symmetry-\emph{preserving} phases, there are four symmetry-\emph{breaking} phases (which can be labeled by $G_\textrm{eff}$). Three of these preserve a $\mathbb Z_2$ subgroup generated by one of the $R_{\gamma=x,y,z}$:
\begin{equation*}
H_x = \sum_n X_n X_{n+1}, \; H_y = \sum_n Y_n Y_{n+1}, \; H_z = \sum_n Z_n Z_{n+1}.
\end{equation*}
We label these three Ising phases as $I_x$, $I_y$ and $I_z$, respectively. The fourth breaks the complete $\mathbb Z_2 \times \mathbb Z_2$ symmetry group (i.e., $G_\textrm{eff} = \{ 1 \}$) and is realized by
\begin{equation}
H_0 = -\sum_n \left( X_{2n-1} X_{2n+1} + Y_{2n} Y_{2n+2} \right). \label{eq:zero}
\end{equation}
These six Hamiltonians satisfy a very useful duality property: there is a non-local change of variables which effectively interchanges $H_x \leftrightarrow H_1$, $H_y \leftrightarrow H_\textrm{Hal}$ and $H_z \leftrightarrow H_0$. We give an explicit lattice construction in Appendix~\ref{app:duality}. Physically, this transformation can be interpreted as gauging $R_z$ (whilst keeping $R_x$ fixed).

Lastly, we note that phases of matter can be \emph{stacked}. Two $G$-symmetric models can be combined into a two-leg ladder; the new on-site symmetry is simply the tensor product of the two individual on-site symmetries. One can then study what phase the stacked model belongs to. E.g., stacking the trivial phase onto any other phase leaves the latter invariant; it acts as the identity element, explaining our notation $H_1$. Oppositely, stacking any phase onto the phase that breaks all symmetries, remains in the latter phase; it hence acts as the zero element\footnote{Together with the relations $\textrm{Hal}*\textrm{Hal} = 1$, $I_\gamma * \textrm{Hal} = I_\gamma$ and $I_\gamma * I_\gamma = I_\gamma$, the abelian semigroup of $\mathbb Z_2 \times \mathbb Z_2$-phases is completely specified.}, explaining the label $H_0$.

\subsection{Ising criticality (\texorpdfstring{$c=1/2$}{c=1/2}) \label{subsec:ising}}

We classify $\mathbb Z_2 \times \mathbb Z_2$-enriched Ising CFTs, with a straightforward generalization to general symmetry groups $G$. This will show that the four invariants mentioned above form a \emph{complete} set of invariants for this universality class. Our principal focus is on symmetry-enriched Ising CFTs which naturally occur as phase transitions between the aforementioned gapped phases. Practically, this means that such Ising CFTs have a \emph{single} symmetry-allowed bulk perturbation which can open up a gap (a \emph{relevant} operator). I.e., only one parameter needs to be tuned to achieve criticality; such CFTs are said to have \emph{codimension one}. Ising CFTs with higher codimension can occur but physically correspond to accidental criticalities in a phase diagram---we will discuss them at the end of this subsection. Ising CFTs of codimension one are characterized by having an \emph{unbroken} symmetry $\mathbb Z_2 \subset G_\textrm{eff}$ which anticommutes with the local Ising scaling operator $\sigma$ (i.e., this $\mathbb Z_2 \not\subset G_\textrm{gap}$). Indeed, this forbids us from adding $\sigma$ as a perturbation, such that there is only one relevant symmetric operator which can open up a gap (commonly denoted by $\varepsilon$).

We now demonstrate how the four labels we proposed allow us to derive that there are \emph{at least nine} codimension one Ising CFTs---afterwards we confirm that this list is complete. Firstly, we have to determine the group of unbroken symmetries. If there \emph{is} symmetry-breaking at the critical point, we have $G_\textrm{eff} \cong \mathbb Z_2$ (the symmetry group cannot be completely broken if we consider a codimension one CFT). There are \emph{three} choices for the unbroken symmetry $R_\gamma$ ($\gamma = x,y,z$). Note that the remaining invariants are already determined: $\sigma$ must be odd under this remaining symmetry, and a single bosonic $\mathbb Z_2$ on-site symmetry flux cannot be charged under itself, i.e., $\chi_g(g) =1$. We can also determine which gapped phases this CFT can be perturbed into. If we condense $\sigma$ (corresponding to perturbing with, say, $-\varepsilon$), we arrive at the phase breaking \emph{all} symmetries (labeled by $0$), whereas if we condense $\mu$ (perturbing with $+\varepsilon$) then we flow to a gapped phase with a single $\mathbb Z_2$ symmetry $R_\gamma$, which can only be the Ising phase $I_\gamma$. We use the shorthand notation [0,$I_\gamma$] to denote this transition, which is realized by, for example, the lattice Hamiltonian $H_0 + H_\gamma$.

If there is \emph{no} symmetry-breaking, then $G_\textrm{eff} = \mathbb Z_2 \times \mathbb Z_2$. We can then continue to determine its gapped subgroup $G_\textrm{gap}$. This cannot be empty, since it is known that the Ising CFT (without gapped degrees of freedom) does not have a $\mathbb Z_2 \times \mathbb Z_2$ symmetry at low energies \cite{Ruelle98}. Moreover, $G_\textrm{gap}$ cannot be the full group if we are to have a codimension one Ising CFT. We thus conclude that $G_\textrm{gap} \cong \mathbb Z_2$. There are \emph{three} choices for this gapped symmetry $R_\gamma$ ($\gamma=x,y,z$). Note that this implies that $\sigma$ is odd under the non-gapped symmetries. Physically, one side of the Ising transition---where we condense $\sigma$---is thus the Ising phase $I_\gamma$. The remaining invariant of the $G$-CFT is the charge of the symmetry fluxes. These symmetry fluxes determine the nearby gapped \emph{symmetric} phase (where $\mu$ is condensed), hence we know that there are only \emph{two} distinct choices, labeled by whether the symmetry flux of $R_\gamma$ is even or odd under one of the other symmetries. We thus arrive at the transitions [1,$I_\gamma$] and [Hal,$I_\gamma$], respectively; these are realized by $H_1 + H_\gamma$ and $H_\textrm{Hal} + H_\gamma$.

\begin{figure}
\includegraphics[scale=.99,trim=3 3 3 3,clip]{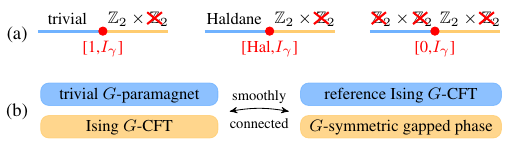}
\caption{\textbf{Classifying $\bm{\mathbb Z_2 \times \mathbb Z_2}$-enriched Ising CFTs.} (a) There are nine distinct $\mathbb Z_2 \times \mathbb Z_2$-(Ising CFTs) of codimension one, coming in groups of three: one nearby gapped phase is in the trivial/non-trivial SPT phase or the phase that breaks the \emph{full} $\mathbb Z_2 \times \mathbb Z_2$; the other side is an Ising phase preserving a $\mathbb Z_2$ symmetry $R_\gamma$ ($\gamma=x,y,z$). All nine can be distinguished by symmetry properties of local operators and/or symmetry fluxes. (b) To see that these nine exhaust all options, we prove that all $G$-enriched Ising CFTs (Ising $G$-CFTs, in short) can be obtained by stacking gapped phases on top of a reference Ising $G$-CFT; the latter is chosen to be a transition between the trivial gapped phase and an Ising phase. \label{fig:classificationA}}
\end{figure}

These nine distinct symmetry-enriched CFTs are summarized in Fig.~\ref{fig:classificationA}(a). Three of these, namely [Hal,$I_\gamma$] ($\gamma=x,y,z$), are topologically non-trivial with exponentially localized edge modes. We identify the critical spin-$1$ XXZ chain in Eq.~\eqref{eq:XXZ} as being in the class [Hal,$I_z$]. A priori, it could have been that there are additional (codimension one) Ising CFTs which cannot be smoothly connected to one of these nine, meaning that they would fall outside our proposed set of invariants. However, one can prove (see Appendix~\ref{app:baitandswitch}) that once one fixes the unbroken symmetry group $G_\textrm{eff}$ and a list of charges for $\sigma$, then any $G$-enriched Ising CFT can be obtained by stacking with (and coupling to) a $G$-symmetric gapped phase; see Fig.~\ref{fig:classificationA}(b). Note that the previous list is already closed under this action---e.g., stacking the Haldane phase on top of [1,$I_\gamma$] gives us [Hal,$I_\gamma$]---and is hence complete.

To generalize the above discussion to a general symmetry group $G$, note that the first step involved determining the unbroken symmetry group $G_\textrm{eff}$. The possible choice of gapped symmetries, $G_\textrm{gap}$, was then constrained by ensuring that our Ising CFT would have codimension one. The only remaining non-trivial labels are the charges of symmetry fluxes, which can be related to the topological properties of the nearby gapped phases. Altogether, this straightforwardly leads to a general classification (which is complete due to the \emph{bait-and-switch lemma}; see Fig.~\ref{fig:classificationA}(b) and Appendix~\ref{app:baitandswitch}):
\begin{shaded}
	\emph{\textbf{Classification of symmetry-enriched Ising CFTs of codimension one for a unitary group $\bm G$:} these CFTs are labeled by a choice of nested subgroups $G_\textrm{gap} \subset G_\textrm{eff} \subset G$ such that $G_\textrm{eff}/G_\textrm{gap} \cong \mathbb Z_2$, and by an element $\omega \in H^2(G_\textrm{eff},U(1))$, i.e., a class of projective representations of $G_\textrm{eff}$.}
\end{shaded}
Note that the quotient of unbroken symmetries and gapped symmetries, $G_\textrm{eff}/G_\textrm{gap}$, can be identified with the $\mathbb Z_2$ symmetry that is intrinsic to the Ising CFT, which ensures that the CFT has codimension one (note that $\sigma$ is odd under any $g\in G_\textrm{eff}-G_\textrm{gap}$). The projective class specifies the symmetry fluxes of $G_\textrm{eff}$ such that if we condense $\mu$, we enter the gapped phase labeled by this projective class. If we instead condense $\sigma$, we break the symmetry group down to $G_\textrm{gap}$, with the resulting gapped phase identified by $\omega|_{G_\textrm{gap}} \in H^2(G_\textrm{gap},U(1))$. Let us illustrate the above general classification for $G = \mathbb Z_2 \times \mathbb Z_2$. One choice of nested subgroups is $G_\textrm{gap} = \{1\} \subset G_\textrm{eff} = \mathbb Z_2 \subset G = \mathbb Z_2 \times \mathbb Z_2$, with \emph{three} choices for the unbroken symmetry $G_\textrm{eff}$ and \emph{no} choice for an element in $H^2(\mathbb Z_2,U(1)) \cong \{ 1\}$. The other choice is $G_\textrm{gap} = \mathbb Z_2 \subset G_\textrm{eff} = G$, with \emph{three} choices for the gapped symmetry $G_\textrm{gap}$ and \emph{two} choices of $\omega \in H^2(\mathbb Z_2 \times \mathbb Z_2, U(1)) \cong \mathbb Z_2$. We thus recover all $3+3\times2 = 9$ distinct $\mathbb Z_2 \times \mathbb Z_2$-enriched Ising CFTs of codimension one. 

Let us briefly comment on $G$-enriched Ising CFTs of codimension two. These occur if $G_\textrm{gap} = G_\textrm{eff}$, since then both $\sigma$ and $\varepsilon$ can open up a gap. This is the maximal codimension. Since such an Ising CFT requires the fine-tuning of two parameters, it generically appears in two-dimensional phase diagrams as an isolated point (with a first-order line emanating from it). The symmetry group acts only on gapped degrees of freedom, and the latter can realize any of its phases labeled by a choice of $G_\textrm{eff} \subset G$ and $\omega \in H^2(G_\textrm{eff},U(1))$. For $\mathbb Z_2 \times \mathbb Z_2$, there are thus six such symmetry-enriched Ising CFTs of codimension two. More generally, for any CFT one can consider the symmetry-enrichment where $G = G_\textrm{gap}$. However, this is not a very interesting example: it recovers the gapped classification and the CFT is a mere spectator. In particular, one can apply the usual mechanism of symmetry fractionalization, implying that the degeneracies with open boundary conditions (and their finite-size splitting) exactly coincide with that of the purely gapped case.

\subsection{Gaussian criticality (\texorpdfstring{$c=1$}{c=1}) \label{subsec:gaussian}}

There are six gapped phases for $\mathbb Z_2 \times \mathbb Z_2$-symmetric Hamiltonians and thus at least ${6 \choose 2} = 15$ direct transitions to consider. Six of these are Ising transitions of codimension one, such that they are indeed neighboring only two gapped phases; these were considered in Section~\ref{subsec:ising} and were all found to be distinct. However, we will see that the remaining direct transitions are Gaussian CFTs (where the central charge $c=1$), but there is no codimension one Gaussian CFT with $\mathbb Z_2 \times \mathbb Z_2$ symmetry. As we will see, the smallest codimension is two, and these phases are proximate to three or four distinct gapped phases. Hence, there are more than the naive $15-6=9$ remaining direct transitions. Nevertheless, in this section we will see that all of these can be smoothly connected. In particular, this means that the set of gapped phases which are proximate to a given (symmetry-enriched) Gaussian CFT are not an invariant: they can change when tuning through the moduli space of CFTs. Relatedly, charge assignments of dominant (local and nonlocal) scaling operators can change when tuning through special points of this moduli space. Note that this phenomenon of level crossing in the universal spectrum is very particular to CFTs with symmetry-allowed marginal deformations.

\begin{figure}
\includegraphics[scale=1,trim=3 4 3 3,clip]{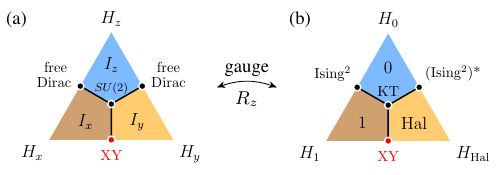}
\caption{\textbf{Classifying $\bm{\mathbb Z_2 \times \mathbb Z_2}$-enriched Gaussian CFTs.} All $\mathbb Z_2 \times \mathbb Z_2$-CFTs with $c=1$ can be connected (for $G_\textrm{gap} = \{1\}$ and $G_\textrm{eff} = G$). The $U(1)$-symmetric spin-$1/2$ XY chain appears in \emph{both} panels (red dot). (a) The black lines are $c=1$ CFTs whose (dominant) \emph{local} operators have different symmetry properties; nevertheless, these operators become degenerate at the $SU(2)$-symmetric point, smoothly connecting these $c=1$ CFTs. (b) Obtained from the former by a (non-local) unitary transformation which can be interpreted as gauging $R_z$ symmetry. Two of the three $c=1$ lines are described by orbifold CFTs; the dominant symmetry flux is symmetry-even (Ising$^2$) or symmetry-odd ($(\textrm{Ising}^2)^*$), becoming degenerate at the KT point. \label{fig:classificationB}}
\end{figure}

Proving that all these symmetry-enriched Gaussian CFTs can be connected is somewhat technical---requiring a certain familiarity with these CFTs---and is explained in Appendix~\ref{app:core}. In the main text we focus on illustrative examples which contain all the key ingredients for the general case: for each of the six aforementioned pairs of gapped phases we consider a Gaussian CFT that separates them. We note that different types of direct transitions can be realized as direct interpolations between the fixed point Hamiltonians introduced in subsection~\ref{subsec:gap}, as shown in Fig.~\ref{fig:classificationB}. We first discuss these transitions from the lattice perspective, which has the benefit of being concrete and constructive. Afterwards, we rephrase it in the field-theoretic language, which emphasizes the generality of these examples and makes a direct link with established CFT methods.

\subsubsection{A lattice perspective}

One generically needs to tune two parameters to find the direct transitions in Fig.~\ref{fig:classificationB}---otherwise they split up into two separate Ising transitions. Relatedly, each of these transitions is proximate to \emph{two other gapped phases}, which are not shown. The clearest instance of this is the $XY$ model $H_{XY}=\sum_n (X_n X_{n+1} + Y_n Y_{n+1})$ which appears \emph{twice} in Fig.~\ref{fig:classificationB}, once in each panel (red dot); indeed, $H_{XY} = H_x+ H_y = H_1 + H_\textrm{Hal}$. If one breaks its $U(1)$ symmetry, one flows to one of the $I_x$ or $I_y$ phases (left panel). If one instead dimerizes the chain, one flows to either the trivial or topological Haldane phase (right panel); see also the discussion in Section~\ref{subsec:gap}. To ensure that the CFT has codimension two, we need the full $\mathbb Z_2 \times \mathbb Z_2$ symmetry to act non-trivially on the CFT; see Section~\ref{subsubsec:gaussian}. In other words, $G_\textrm{eff} = \mathbb Z_2 \times \mathbb Z_2$ and $G_\textrm{gap} = \{1\}$. The remaining invariants which we proposed are thus the charges of local operators and of symmetry fluxes.

At first glance, such charges seem to indicate distinct Gaussian $\mathbb Z_2 \times \mathbb Z_2$-CFTs. For instance, let us compare the [$I_x$,$I_y$] transition to the [$I_x$,$I_z$] transition.
The former has a $U(1)$ symmetry in the $xy$-plane, and indeed, $X_n$ and $Y_n$ have the same scaling dimension $\Delta_\theta=1/4$ for $H_{XY}$, whereas $Z_n$ has $\Delta_\varphi = 1$. The roles are reversed in the latter, where $Y_n$ is now the one with larger scaling dimension. Does this not give us a discrete invariant? I.e., for $H_{XY}$ we can say that \emph{all local operators of lowest scaling dimension are odd under $R_z$}. This \emph{discrete} statement is surely \emph{true}, but not \emph{invariant}: the Gaussian CFT has a marginal tuning parameter that smoothly connects a whole line of universality classes \cite{Ginsparg88,DiFrancesco97}. Indeed, this is well-known for the spin-$1/2$ XXZ chain, $H = \sum_n (X_n X_{n+1} + Y_n Y_{n+1} + \tilde \Delta Z_n Z_{n+1})$, which is described by a range of Gaussian CFTs for $-1 < \tilde \Delta \leq 1$. As one tunes $\tilde \Delta$ from zero to one, the scaling dimensions of $X_n$, $Y_n$ and $Z_n$ \emph{smoothly} evolve toward the same value (this being $1/2$); at $\tilde \Delta=1$, the three operators are related by the $SU(2)$ symmetry of the model! All three [$I_\gamma$,$I_{\gamma'}$] transitions can clearly be connected through this point, as shown in Fig.~\ref{fig:classificationB}. The take-away message is that the space of local operators with smallest scaling dimension can change: this two-dimensional space becomes part of a larger-dimensional vector space at the $SU(2)$-symmetric point. The above attempt at defining a \emph{discrete invariant charge} was thus not well-founded.

The discussion above demonstrates a subtlety in associating invariant charges to local observables. Something similar happens for the charges of symmetry fluxes. To find a Gaussian CFT whose symmetry fluxes have non-trivial charge, we need to consider transitions which are \emph{not} invariant under stacking with the gapped Haldane phase: since the latter comes with charged symmetry fluxes, any transition that is not affected by this could not have had well-defined charges for its symmetry fluxes to begin with. To make clear what we mean, consider [1,Hal], as realized by the XY-chain. At this transition, the two distinct symmetry fluxes of $R_\gamma$ for $H_1$ and $H_\textrm{Hal}$---as described in subsection~\ref{subsec:gap}---become scaling operators with the \emph{same} scaling dimension $\Delta = 1/4$. We can thus not associate a unique charge to the symmetry flux of $R_\gamma$. This is related to the fact that [1,Hal]$*$Hal = [1,Hal], since it would merely interchange both sides of the transition. Similarly, the [$I_\gamma$,$I_{\gamma'}$] transitions considered above would also \emph{absorb} such an SPT phase (since $I_\gamma *\textrm{Hal} = I_\gamma$).

We can, however, find $c=1$ transitions which are (seemingly) not invariant under such a stacking: consider the transition between the trivial phase and the phase that completely breaks $\mathbb Z_2 \times \mathbb Z_2$, i.e., [1,0], realized by $H_1 + H_0$. Conceptually, one can think of this as two separate Ising transitions occuring at the same point. Indeed, the field theory describing this critical point is not a usual \emph{Gaussian CFT}, but rather the direct product of \emph{two Ising CFTs}, usually denoted by (Ising)$^2$. Accordingly, each $R_\gamma$ has a \emph{unique} symmetry flux and hence a well-defined charge! For [1,0], these fluxes commute with all symmetries, whereas for [Hal,0], they anticommute. Indeed, stacking with the Haldane phase interchanges these two transitions. However, despite having distinct discrete charge in these two cases, these models can be smoothly connected by tuning to $H_1 + H_\textrm{Hal} + H_0$, as shown in Fig.~\ref{fig:classificationB}(b). The catch is that at this special point, the space of symmetry fluxes becomes two-dimensional, and the aforementioned charge is no longer well-defined---this arises naturally from a careful study of the Gaussian universality class (see subsection~\ref{subsubsec:gaussian}). Note that the novel phase diagram in Fig.~\ref{fig:classificationB}(b) can be obtained from the known phase diagram in Fig.~\ref{fig:classificationB}(a) by applying the duality transformation mentioned in subsection~\ref{subsec:gap}.

In conclusion, all $\mathbb Z_2\times \mathbb Z_2$-enriched CFTs with central charge $c=1$ can be smoothly connected if they are of codimension two (see Appendix~\ref{app:core} for further details). For higher codimension, there are distinct symmetry-enriched CFTs. Examples have already appeared in the literature \cite{Kestner11,Parker18}, and we will touch upon this in Section~\ref{sec:lit}.

\subsubsection{A field-theoretic perspective \label{subsubsec:gaussian}}

We now give a field-theoretic interpretation of Fig.~\ref{fig:classificationB}, which is complementary to the previous discussion.

Let us recall the compact boson CFT (also known as the one-component Luttinger liquid), which contains a phase field $\theta(x)$ and its conjugate field $\partial_x \varphi(x)$ that generates shifts in $\theta(x)$, i.e., $[\partial_x \varphi(x),\theta(y)] = 2 \pi i \delta(x-y)$ \cite{Ginsparg88,DiFrancesco97}. Both fields are $2\pi$-periodic and the Hamiltonian is given by
\begin{equation}
H = \frac{1}{2\pi} \int \left( \frac{1}{4K} (\partial_x \varphi)^2 +  K (\partial_x \theta)^2 \right) \mathrm dx. \label{eq:boson}
\end{equation}
Here, $K$ is the stiffness or Luttinger liquid parameter (sometimes one instead speaks of the compactification radius\footnote{This name arises due to the rescaled field $\tilde \theta \equiv \sqrt{K} \theta$ being $2\pi r_c$-periodic.} $r_c = \sqrt{K}$). There is thus a one-parameter family of compact boson CFTs with a duality $K \leftrightarrow \frac{1}{4K}$ which interchanges the two fields.
Particularly important are the primary vertex operators $e^{\pm i(m \varphi + n \theta)}$ which have scaling dimension $\Delta_{m,n} = m^2 K + n^2/(4K)$ (and conformal spin $s_{m,n}=mn$). These vertex operators are local if and only if $n,m \in \mathbb Z$.

To see how the $\mathbb Z_2 \times \mathbb Z_2$ symmetry acts in this field theory, it is useful to use a lattice-continuum correspondence. Let us consider the XY-chain $H_{XY} = -\sum_n(X_n X_{n+1} + Y_n Y_{n+1})$: this can be mapped to free fermions, allowing for a straight-forward continuum limit of the lattice model, which ends up with the compact boson CFT where $K=r_c= 1$, also called the \emph{free Dirac CFT}. This route gives a direct relationship between lattice operators and operators of the CFT \cite{Affleck88b,Sachdev01,Giamarchi04}. For instance, the generator of the on-site lattice $U(1)$ symmetry, $Z_n$, maps to the generator in the continuum, $\partial_x \varphi$. Similarly useful correspondences are $X_n \sim \cos \theta$ and $Y_n \sim \sin \theta$. From such relations---and from knowing how the symmetries act on the lattice operators---we can infer that $R_x$, $R_y$ and $R_z$ act in the continuum as\footnote{In Appendix~\ref{app:core} we show that up to relabeling the CFT fields, this is the only $\mathbb Z_2\times \mathbb Z_2$ action on the $\varphi,\theta$ fields of the compact boson CFT.}
\begin{equation}
R_x: \left\{
\begin{array}{l}
\varphi \to -\varphi \\
\theta \to -\theta,
\end{array} \right.
\;
R_y: \left\{
\begin{array}{l}
\varphi \to -\varphi \\
\theta \to \pi-\theta,
\end{array} \right.
\;
R_z: \left\{
\begin{array}{l}
\varphi \to \varphi \\
\theta \to \theta+\pi.
\end{array} \right. \label{eq:sym1}
\end{equation}
(Note that $R_z = R_xR_y$, as required.) Enforcing these symmetries constrains the possible perturbations of the CFT. In fact, only two relevant operators commute with this symmetry, namely the vertex operators\footnote{On the lattice, these correspond to $X_n X_{n+1} - Y_n Y_{n+1}$ and $(-1)^n (X_n X_{n+1} + Y_n Y_{n+1})$, respectively.} $\cos(2\theta)$ and $\cos(\varphi)$ with scaling dimension $\Delta =1<2$ at $K=1$. Hence, this $\mathbb Z_2 \times \mathbb Z_2$-enriched $c=1$ CFT has \emph{codimension two}. For example, $\sin(2\theta)$---which is also relevant---is odd under $R_x$ and is thus forbidden\footnote{The lattice-continuum correpondence is $X_n Y_{n+1} + Y_n X_{n+1} \sim \sin(2\theta)$.}. We note that $\pm \cos(2\theta)$ drives to $I_x$ and $I_y$, whereas $\pm\cos(\varphi)$ drives to the trivial and Haldane phase; this is consistent with the red dot in Fig.~\ref{fig:classificationB}(a) and (b).

The above action of $\mathbb Z_2 \times \mathbb Z_2$ was established at $K=1$, but we see that it must continue to hold for any value of $K$ (since $R_\gamma^2 =1$). Nevertheless, we obtain other actions when considering, e.g., a transition [$I_x$,$I_z$] instead of [$I_x$,$I_y$]. Indeed, for $H = -\sum_n(X_n X_{n+1} + Z_n Z_{n+1})$ we would \emph{also} arrive at the Dirac CFT in the continuum limit, where we now denote the fields by $\tilde \varphi$ and $\tilde \theta$ to avoid confusion. The lattice-continuum correspondence, including $Y_n \leftrightarrow \partial_x \tilde \varphi$ and $Z_n \leftrightarrow \sin(\tilde \theta)$, implies
\begin{equation}
R_x: \left\{
\begin{array}{l}
\tilde \varphi \to -\tilde \varphi \\
\tilde \theta \to -\tilde \theta,
\end{array} \right.
\;
R_y: \left\{
\begin{array}{l}
\tilde \varphi \to \tilde \varphi \\
\tilde \theta \to \tilde \theta+\pi,
\end{array} \right.
\;
R_z: \left\{
\begin{array}{l}
\tilde \varphi \to -\tilde \varphi \\
\tilde \theta \to \pi-\tilde \theta.
\end{array} \right. \label{eq:sym2}
\end{equation}
The \emph{same} lattice symmetries $R_{\gamma=x,y,z}$ thus act \emph{differently} in the low-energy compact boson CFTs for these distinct transitions, as evidenced by Eqs.~\eqref{eq:sym1} and \eqref{eq:sym2}. In particular, this tells us that the local vertex operators of these CFTs (identified by their scaling dimensions) carry seemingly distinct charges, as already observed in the previous subsection.

Nevertheless, these two compact boson CFTs are merely two extremes of a single unified CFT. To see this, we use that tuning $K$ preserves criticality. This corresponds to the black lines in Fig.~\ref{fig:classificationB}(a) emerging from the phase diagram's edges. As $K\to 1/2$, we reach the center of the phase diagram. To see that the CFTs in terms of $\varphi,\theta$ and $\tilde \varphi, \tilde \theta$ truly \emph{coincide} at this point, we use the fact that the self-dual point $K=1/2$ is known to have an emergent $SU(2) \times SU(2)$ symmetry. As we show in Appendix~\ref{app:SU2}, this allows us to perform a change of variables that leaves the action invariant:
\begin{equation}
\partial_x \tilde \theta = -2\cos \theta \sin \varphi, \qquad \partial_x \tilde \varphi = - 2 \sin \theta \cos \varphi. \label{eq:identification}
\end{equation}
Note that under this correspondence, Eqs.~\eqref{eq:sym1} and \eqref{eq:sym2} coincide! The identification in Eq.~\eqref{eq:identification} tells us how these seemingly distinct compact boson CFTs are glued together. This can also be understood from the fact that at $K=1/2$, the CFT has \emph{nine} marginal perturbations \cite{Ginsparg88}. Only \emph{three} of these are compatible with $\mathbb Z_2 \times \mathbb Z_2$ symmetry, corresponding to the three gapless lines emanating from this point in Fig.~\ref{fig:classificationB}(a).

Let us now consider the other $c=1$ transitions, shown in Fig.~\ref{fig:classificationB}(b). As mentioned in the previous subsection, this phase diagram is obtained from Fig.~\ref{fig:classificationB}(a) by a non-local change of variables which maps the XY-chain to itself (see Appendix~\ref{app:duality}). In the field theory, this can be interpreted as gauging $R_z$, the effect of which depends on how $R_z$ acts on the fields. If $R_z$ is a subset of the $U(1)$ rotation $\theta \to \theta + \alpha$---as happens for [$I_x$,$I_y$] (see Eq.~\eqref{eq:sym1})---then gauging $R_z$ amounts\footnote{This can be seen as a concatenation of halving the compactification radius and a duality transform: $r_c \to r_c/2 \to \frac{1}{2(r_c/2) }= 1/r_c$.} to $r_c \to 1/r_c$, or, equivalently, $K \to 1/K$ \cite{Ginsparg88}. This indeed keeps the free Dirac point with $K=1$ invariant (the red dots in Fig.~\ref{fig:classificationB}). The self-dual point $K=1/2$---in the middle of the phase diagram---maps to $K=2$, which is called the Kosterlitz-Thouless (KT) point since at this point there are \emph{no} relevant $U(1)$-symmetric operators. However, if $R_z$ negates the fields---as happens for [$I_x$,$I_z$] (see Eq.~\eqref{eq:sym2})---then it is known that gauging this is \emph{no longer the usual compact boson CFT}. Instead, we obtain a so-called \emph{orbifold CFT} (this can also be seen as the boson CFT compactified on $S^1/\mathbb Z_2$) \cite{Ginsparg88}. In particular, the free Dirac CFT in Fig.~\ref{fig:classificationB}(a) maps to a stack of two Ising CFTs in Fig.~\ref{fig:classificationB}(b), usually denoted by (Ising)$^2$, which is a $c=1$ orbifold CFT with orbifold radius $r_\textrm{orbifold} = 1$ \cite{Ginsparg88}. This remains critical as one tunes $r_\textrm{orbifold} \to 1/\sqrt{2}$, reaching the center of the phase diagram. Indeed, it is known that at the particular value $r_\textrm{orbifold} = 1/\sqrt{2}$, the $c=1$ orbifold CFT \emph{coincides} with the aforementioned KT point of the compact boson CFT \cite{Ginsparg88}. The interpolation between the three models $H_1$, $H_\textrm{Hal}$ and $H_0$ thus provides a concrete lattice realization---incidentally solvable by Bethe ansatz---of the deep fact that the $c=1$ orbifold CFT and compact boson CFT can be connected by a marginal perturbation, which in this case also preserves $\mathbb Z_2 \times \mathbb Z_2$.

Lastly, we comment on symmetry fluxes from the perspective of field theory. For the compact boson CFT, the space of symmetry fluxes for $R_\gamma$ is two-dimensional. This is easiest to see for $R_z$ in the case where it acts as $\theta \to \theta + \pi$ (see Eq.~\eqref{eq:sym1}). The two independent fluxes are then $\cos(\varphi/2)$ and $\sin(\varphi/2)$ with dimension $\Delta = K/4$ (on the lattice, these correspond to $\cdots Z_{n-2} Z_{n-1} Z_n$ with $n$ even and odd, respectively). The former (latter) is even (odd) under $R_x$ and $R_y$. We thus cannot associate a unique charge to the symmetry flux of $R_z$. The situation is seemingly different for the orbifold CFT, which we illustrate for the (Ising)$^2$ CFT as realized by $H = H_1 + H_0$ (appearing in Fig.~\ref{fig:classificationB}(b)). Recalling the definition of $H_0$ in Eq.~\eqref{eq:zero}, we see that if $\sigma_i$ denotes the spin operators of the two copies of the Ising CFT, then $\sigma_1$ is odd under $R_y$ and $R_z$ (but even under $R_x$), and similarly for $\sigma_2$ with $R_x$ and $R_y$ interchanged. From this, we can infer that the symmetry flux for $R_z$ is $\mu_1 \mu_2$ (with dimension $\Delta =1/4$). Most importantly: this is \emph{unique}. Indeed, its dimension is well-separated\footnote{Similarly, the unique symmetry flux of $R_x$ is $\mu_2$ with $\Delta= 1/8$; its subdominant flux is $\sigma_1 \mu_2$ with $\Delta=1/4$.} from that of $(\mu_1 \mu_2) \times (\sigma_1 \sigma_2) \sim \psi_1 \psi_2$ for which $\Delta = 1$. As discussed in Section~\ref{sec:symflux}, we can associate charges to unique symmetry fluxes. For the transition [1,0], it is easy to see that the symmetry flux of $R_z$ is \emph{even} under $R_x$ and $R_y$, whereas for [Hal,0] it is \emph{odd}. Hence, the two orbifold lines in Fig.~\ref{fig:classificationB}(b) seem topologically distinct! The catch, however, is that both lines connect to the compact boson line, as discussed above. At that point, the scaling dimensions of $\mu_1 \mu_2$ and $\psi_1 \psi_2$ meet, coinciding with the aforementioned $\cos(\varphi/2)$ and $\sin(\varphi/2)$.
Hence, to realize topologically non-trivial symmetry-enriched $c=1$ CFTs, one either needs to go to higher codimension, or larger symmetry groups; examples for both are discussed in Section~\ref{sec:lit}.

\section{Other examples in 1+1d \label{sec:other}}

Although the framework we have introduced is generally applicable, we have thus far shown examples only for symmetry groups $\mathbb Z_2 \times \mathbb Z_2$ and $\mathbb Z_2 \times \mathbb Z_2^T$. The purpose of this section is to add two more illustrative examples to this list: a bosonic $\mathbb Z_3 \times \mathbb Z_3$ and a fermionic $\mathbb Z_2^f \times \mathbb Z_2^T$. Various other symmetry groups will appear in the next section (Section~\ref{sec:lit} on previously studied examples in the literature).

\subsection{$\mathbb Z_3 \times \mathbb Z_3$ example \label{sec:Z3}}

We consider a one-dimensional chain where each site consists of a qutrit. For each qutrit, let $X$ and $Z$ denote the analogues of the Pauli matrices (also known as the clock matrices):
\begin{equation}
X = \left( \begin{array}{ccc}
0 & 1 & 0 \\
0 & 0 & 1 \\
1 & 0 & 0
\end{array} \right) \quad \textrm{and} \quad
Z = \left( \begin{array}{ccc}
1 & 0 & 0 \\
0 & \omega & 0 \\
0 & 0 & \omega^2
\end{array} \right)
\end{equation}
where $\omega = e^{2\pi i/3}$. Note that $XZ = \omega ZX$. Subdividing the chain into two sublattices (labeled $A$ and $B$), we define the Hamiltonian
\begin{equation}
H = \eta_\textrm{triv} H_\textrm{triv} + \eta_\textrm{SSB} H_\textrm{SSB} + \eta_\textrm{SPT} H_\textrm{SPT} \label{eq:Z3Z3}
\end{equation}
where
\begin{align}
H_\textrm{triv} &= - \sum_n \left( X_{A,n} +X_{B,n} \right) + h.c.\\
H_\textrm{SSB} &= - \sum_n Z_{B,n}^\dagger Z_{B,n+1} + h.c.  \\
H_\textrm{SPT} &=- \sum_n \Big( Z_{B,n-1}^\dagger X_{A,n} Z_{B,n}\\
&\qquad \qquad + Z_{A,n} X_{B,n} Z_{A,n+1}^\dagger \Big) + h.c. \; .  \notag 
\end{align}
For all parameters, this Hamiltonian has a $\mathbb Z_3 \times \mathbb Z_3$ symmetry group generated by $P_A = \prod_n X_{A,n}$ and $P_B = \prod X_{B,n}$. The fixed point Hamiltonian $H_\textrm{triv}$ has a product state ground state whereas $H_\textrm{SSB}$ spontaneously breaks $P_B$ such that the remaining symmetry group is $\mathbb Z_3$ generated by $P_A$.
More interestingly, $H_\textrm{SPT}$ is a generalization of the cluster model encountered in Section~\ref{sec:overview} and is one of the two non-trivial SPT phases protected by $\mathbb Z_3\times \mathbb Z_3$ \cite{Geraedts14,Santos15} (the other being given by the site-centered inversion of $H_\textrm{SPT}$).

\begin{figure}
	\includegraphics[scale=1]{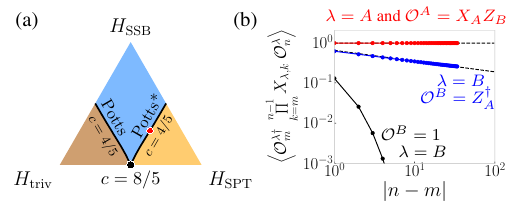}
	\caption{\textbf{$\mathbb Z_3 \times \mathbb Z_3$-enriched Potts universality with topological edge modes.} (a) Phase diagram of Eq.~\eqref{eq:Z3Z3} with $\mathbb Z_3 \times \mathbb Z_3$ symmetry generated by $\prod_n X_{A,n}$ and $\prod_n X_{B,n}$; obtained with DMRG. The blue phase spontaneously breaks the latter. The two black lines (despite appearances these are not straight) are two topologically-distinct Potts transitions, separated by a CFT with central charge $c=8/5$. (b) String correlation functions for the red point in (a), given by $(\eta_\textrm{triv},\eta_\textrm{SSB},\eta_\textrm{SPT}) \approx (0.8,1,1.775)$. The symmetry flux for $\prod_n X_{A,n}$ has long-range order, whereas the one for $\prod_n X_{B,n}$ has algebraic decay; their endpoint operators are non-trivially charged under the global symmetries, protecting a threefold degeneracy with open boundaries (see main text). The dashed lines are the analytic predictions; the algebraic decay has dimension $\Delta= 2/15$. Without the charged endpoint operator, the decay is exponential (the black line was multiplied $\times 10$ for visibility). \label{fig:Z3Z3}}
\end{figure}

We only consider non-negative tuning parameters in Eq.~\eqref{eq:Z3Z3}. There are two limiting cases of this model which have been studied before. Firstly, if $\eta_\textrm{SPT}=0$, then the $A$-sites remain in a decoupled product state and the $B$-sites are described by the three-state Potts chain. In particular, the transition between the trivial and SSB phase ($\eta_\textrm{triv} = \eta_\textrm{SSB}$) is known to be described by the three-state Potts CFT with central charge $c=4/5$ \cite{Zamolodchikov85,Gehlen86,DiFrancesco97}. Secondly, if $\eta_\textrm{SSB}=0$, then we have a direct SPT transition which is known to be described by an orbifold CFT with central charge $c=8/5$ \cite{Tsui17}. This gives us two edges of the phase diagram in Fig.~\ref{fig:Z3Z3}(a); we obtained the rest of the phase diagram using DMRG. We observe that the Potts criticality is stable over a whole line; this is consistent with the Potts CFT only having a single relevant perturbation in the presence of $\mathbb Z_3$ symmetry (in this case being $P_B$, which is spontaneously broken on one side of the transition). Note that with periodic boundary conditions, there is a duality\footnote{Under the anti-unitary $\mathbb Z_2^T$ mapping $\tilde X_{A,n} := Z_{B,n-1}^\dagger X_{A,n} Z_{B,n} $, $\tilde X_{B,n}=Z_{A,n} X_{B,n} Z_{A,n+1}^\dagger$, $\tilde Z_{A,n} := Z_{A,n}^\dagger$ and $\tilde Z_{B,n}:= Z_{B,n}^\dagger$, we have $H_\textrm{triv} \leftrightarrow H_\textrm{SPT}$ with $H_\textrm{SSB}$ being invariant.} that mirrors the phase diagram ($\eta_\textrm{triv} \leftrightarrow \eta_\textrm{SPT}$); hence, the critical line between the SPT phase and the SSB phase is also three-state Potts criticality. Nevertheless, we will see that the these two Potts criticality are distinct symmetry-enriched versions, the latter being topologically non-trivial.

Let us focus on the symmetry flux for $P_B$. In the trivial gapped phase, its endpoint operator is trivial, whereas in the gapped SPT phase, we have\footnote{As in the $\mathbb Z_2 \times \mathbb Z_2$ case in Section~\ref{sec:overview}, this follows from noting that the limiting Hamiltonian $H_\textrm{SPT}$ is a stabilizer code such that $Z_{A,n} X_{B,n} Z_{A,n+1}^\dagger=+1$ in the ground state.} long-range order in the string
\begin{equation}
\mathcal S^B_n = \cdots X_{B,n-2} X_{B,n-1} \mathcal O^B_n \; \textrm{with }\mathcal O^B_n = Z_{A,n}^\dagger. \label{eq:Z3flux}
\end{equation}
Since $P_A \mathcal O^B_n P_A^\dagger= \omega \mathcal O^B_n$, long-range order in this charged symmetry flux indeed implies a non-trivial SPT phase. We expect that these symmetry fluxes will still distinguish the two Potts transitions from the trivial or SPT phase to the symmetry-breaking phase.
In particular, in Fig.~\ref{fig:Z3Z3}(b) we show that at the critical point neighboring the SPT phase, the charged symmetry flux in Eq.~\eqref{eq:Z3flux} has the slowest possible decay: $\Delta = 2/15$ is the smallest scaling dimension of the Potts CFT \cite{DiFrancesco97}. If we leave out the charged endpoint operator, the correlation function decays exponentially. This is in contrast to the vanilla-flavored Potts criticality between the trivial and symmetry-breaking phases, where it is the uncharged string that has the slowest possible decay ($\Delta =2/15$). Similar to the Ising CFT, the Potts CFT has a unique nonlocal scaling operator with this dimension \cite{DiFrancesco97}.

In conclusion, we have found two topologically-distinct Potts transitions which are enriched by $\mathbb Z_3 \times \mathbb Z_3$ symmetry; these are distinguished by the symmetry properties of certain nonlocal scaling operators. One consequence is that despite both transitions being a Potts universality with central charge $c=4/5$, they have to be separated by a different universality class. For instance, in Fig.~\ref{fig:Z3Z3}(a) the two phase transitions meet at a multi-critical point with central charge $c=8/5$. Another consequence of this topological invariant is that with open boundaries, the transition between the SPT and symmetry-breaking phases host exponentially-localized edge modes, as explained in Section~\ref{sec:edge}.

\subsection{A fermionic example \label{sec:ferm}}

Thus far, we have discussed examples of bosonic $G$-CFTs. However, the concept also applies to fermionic systems. We illustrate this by reinterpreting the non-interacting critical chains of Ref.~\onlinecite{Verresen18} as Majorana CFTs enriched by fermionic parity symmetry $\mathbb Z_2^f$ and spinless time-reversal symmetry $\mathbb Z_2^T$; this demonstrates that it is stable to interactions. The non-trivial case hosts zero-energy Majorana edge modes. (The models we will discuss are related to the spin models encountered in Section~\ref{sec:overview} by a Jordan-Wigner transformation; however, since this non-local mapping changes the \emph{physical interpretation}, we will keep this section self-contained within the fermionic language.)

\begin{figure}
	\includegraphics[scale=1]{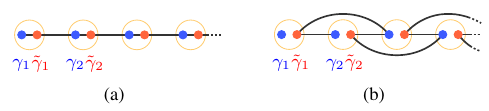}
	\caption{\textbf{Examples of $\bm{\mathbb Z_2^f\times \mathbb Z_2^T}$-enriched Majorana CFTs.} Both Majorana chains are described by the Majorana CFT ($c=1/2$) but fall into distinct symmetry-enriched classes. (a) For the standard translation-invariant Majorana chain, the $P$-even symmetry flux of $P$ is \emph{even} under spinless time-reversal symmetry $T$. (b) For this chain, the $P$-even symmetry flux of $P$ is \emph{odd} under $T$; this protects a Majorana edge mode with finite-size splitting $\sim 1/L^{14}$ (if the system is non-interacting, the finite-size splitting is exponentially small). \label{fig:ferm}}
\end{figure}

We will consider spinless fermions, $\{c_n^\dagger,c_m\} = \delta_{nm}$, where spinless time-reversal symmetry $T$ acts in the occupation basis, i.e., $Tc_n T = c_n$ and $T c_n^\dagger T = c_n^\dagger$. It is useful to introduce the basis of Majorana modes, $\gamma_n = c_n^\dagger + c_n$ and $\tilde \gamma_n = i( c_n^\dagger - c_n )$.
We see that $T \gamma_n T = \gamma_n$ and $T \tilde \gamma_n T = - \tilde \gamma_n$. Using these modes, we now define two critical Majorana chains which are symmetric with respect to $T$ and fermionic parity symmetry $P = \prod P_n$ (where $P_n = i \tilde \gamma_n \gamma_n = (1-2 c^\dagger_n c_n$)):
\begin{equation}
\begin{array}{ccl}
H &=& i \sum_n \tilde \gamma_n (\gamma_{n} + \gamma_{n+1}), \\
H'' &=& i \sum_n \tilde \gamma_n (\gamma_{n+1} + \gamma_{n+2}).  
\end{array} \label{eq:ferm}
\end{equation}
These are sketched in Fig.~\ref{fig:ferm} (for a half-infinite chain). In the bulk, both systems are described by the Majorana CFT with central charge $c=1/2$ \cite{Verresen18}. Nevertheless, they show distinct behavior near their boundaries: $H''$ has a dangling Majorana edge mode, as shown in Fig.~\ref{fig:ferm}(b). In Ref.~\onlinecite{Verresen18}, which only considered \emph{non-interacting} systems, it was proven that as long as we preserve $P$ and $T$ symmetry and keep the bulk in the universality class of the Majorana CFT, then this edge mode remains \emph{exponentially localized}---despite the absence of well-defined gapped degrees of freedom!

We now illustrate how these two models are distinguished by the charges of the symmetry fluxes of $P$. Interestingly, for fermionic systems it is possible for a symmetry flux to be charged \emph{under itself}: the symmetry flux of fermionic parity can itself be fermionic. Indeed, both $H$ and $H''$ have one symmetry flux of $P$ which is \emph{odd} under $P$, and one which is \emph{even}. The former is the same for both models, being given by $\cdots P_{n-2} P_{n-1} \gamma_n $ (with scaling dimension $\Delta=1/8$) \cite{Jones19}. Physically, this corresponds to both phases being proximate to the topological Kitaev phase where this symmetry flux has long-range order. The symmetry flux of $P$ which is \emph{even} under $P$ (with the same scaling dimension), is given by \cite{Jones19}
\begin{equation}
\begin{array}{ccll}
\mathcal S^P_n &=& \cdots P_{n-2} P_{n-1} P_n & \textrm{(for $H$),} \\
\mathcal S^P_n &=& \cdots P_{n-2} P_{n-1} (i\gamma_n \gamma_{n+1}) &\textrm{(for $H''$).}
\end{array}
\end{equation}
(The prefactors of the endpoint operators were chosen such that the symmetry fluxes are hermitian; see Section~\ref{sec:symflux}.) We observe that the former is \emph{even} under $T$, whereas the latter is \emph{odd} under $T$. This defines a discrete bulk invariant. Analogous to the discussion in Section~\ref{sec:edge}, the latter protects a Majorana edge mode at each boundary, giving a global two-fold degeneracy with a finite-size splitting $\sim 1/L^{14}$. Indeed, under a Jordan-Wigner transformation---which is exact for \emph{open} boundary conditions---the models in Eq.~\eqref{eq:ferm} are mapped onto the spin chains in Eq.~\eqref{eq:example} discussed in Section~\ref{sec:overview}. It would be interesting to work out the analogous case for phase transitions between distinct SPT phases in parafermionic chains \cite{Motruk13,Meidan17}.

\section{Application to previous works \label{sec:lit}}

In this penultimate section, we demonstrate how previous works on critical systems with edge modes can be placed into the framework of symmetry-enriched CFTs proposed in this work. Considering the extent of the literature on topologically non-trivial gapless phases 
\cite{Kestner11,Cheng11,Fidkowski11longrange,Sau11,Ruhman12,Grover12,Kraus13,Ortiz14,Keselman15,Ruhman15,Kainaris15,Iemini15,Lang15,Ortiz16,Montorsi17,Wang17,Ruhman17,Scaffidi17,Guther17,Kainaris17,Jiang18,Zhang18,Verresen18,Parker18,Keselman18,Chen18}, a \emph{complete} demonstration would constitute a work on its own. Here, we limit ourselves to a few (chronologically-ordered) case studies. We point out how these systems can be interpreted as non-trivial symmetry-enriched CFTs, for which we identify bulk invariants.

For clarity, let us mention that in 2011 there were several contemporaneous works discussing algebraically-ordered superconductors with topological edge modes \cite{Fidkowski11longrange,Cheng11,Sau11}. However, these are \emph{not} examples of edge modes protected by a symmetry-enriched CFT. This can be seen from the fact that in these cases, the edge modes have algebraic finite-size splitting $\sim 1/L^\beta$ where the power $\beta$ is proportional to the Luttinger liquid parameter $K$. This means that the edge modes can be destroyed by \emph{smoothly} tuning $K$ to a small enough value. Hence, there cannot be a \emph{discrete bulk invariant} associated to the $c=1$ CFT. This is consistent with symmetries not being important to stabilize the edge modes in Refs.~\onlinecite{Fidkowski11longrange,Cheng11,Sau11} (aside from the $U(1)$ symmetry stabilizing criticality).

\subsection{``\emph{Prediction of a gapless topological Haldane liquid phase in a one-dimensional cold polar molecular lattice}''}

In Ref.~\onlinecite{Kestner11}, Kestner, Wang, Sau and Das Sarma introduced a gapless analogue of the spin-$1$ Haldane phase. Due to the context of the work (i.e., dipolar gases), long-range interactions were considered. Nevertheless, a nearest-neighbor model is straight-forward:
\begin{equation}
H = \sum_n \left[ (S^x_n S^x_{n+1} + S^y_n S^y_{n+1})(S^z_n + S^z_{n+1})^2 + \Delta S^z_n S^z_{n+1} \right]. \label{eq:XXZS2}
\end{equation}
The Hamiltonian looks similar to the spin-$1$ XXZ chain, with an extra factor in the spin-hopping term such that $\sum_n (S^z_n)^2$ is conserved---this makes all the difference. For $0 < \Delta <1$, the system is critical with central charge $c=1$, and using the methods of Ref.~\onlinecite{Kestner11}, it can be shown that the system is topologically non-trivial: there is still long-range order in the Haldane string order parameter $\cdots R^z_{n-2} R^z_{n-1} S^z_n$ and the ground state is twofold degenerate with open boundary conditions.

We note that this model is an example of a $c=1$ CFT enriched by the $\mathbb Z_2 \times \mathbb Z_2$ group of $\pi$-rotations $R^x$, $R^y$ and $R^z$ (in fact the latter symmetry is enhanced to a full $U(1)$ generated by $\sum_n \left(S^z_n\right)^2$). Indeed, the fact that $\cdots R^z_{n-2} R^z_{n-1} S^z_n$ has long-range order, tells us, firstly, that the symmetry flux of $R^z$ is \emph{unique} and, secondly, that it has a non-trivial charge with respect to $R^x$ and $R^y$. This thus gives us a \emph{discrete topological invariant} for the critical bulk. In Section~\ref{subsec:gaussian} we saw that there are no topologically non-trivial $\mathbb Z_2 \times \mathbb Z_2$-enriched CFTs of minimal codimension. Indeed, Eq.~\eqref{eq:XXZS2} is of higher codimension and is not naturally interpreted as a critical point between phases with $\mathbb Z_2 \times \mathbb Z_2$ symmetry.

However, it \emph{is} a $\mathbb Z_2 \times \mathbb Z_2 \times \mathbb Z_2^T$-enriched CFT of minimal codimension! More precisely, it occurs as a direct transition between two gapped phases which are \emph{topologically distinct} with respect to $T_\textrm{spin} = R^y K$ but which are \emph{topologically identical} and \emph{non-trivial} with respect to $\mathbb Z_2 \times \mathbb Z_2$. To see this---and to shed light on the unusual conservation law---it is useful to consider the unitary $U \equiv \exp \left( i \frac{\pi}{2} \sum_n (S^z_n)^2 \right)$. It can be shown that this commutes with $R^{\gamma=x,y,z}$ but not with $T_\textrm{spin}$, i.e., $U T_\textrm{spin} U^\dagger = R^x K \neq T_\textrm{spin}$. Hence, if $H_\textrm{XXZ}$ is the spin-$1$ XXZ chain---which is non-trivial with respect to $\pi$-rotations \emph{and} time-reversal---then $U H_\textrm{XXZ} U^\dagger$ is still non-trivial with respect to the former, but \emph{not} the latter. Remarkably, the model in Eq.~\eqref{eq:XXZS2} is exactly the halfway interpolation between these two, $H = \frac{1}{2} \left( H_\textrm{XXZ} + U H_\textrm{XXZ} U^\dagger \right)$. Indeed, since $U$ toggles between the two gapped phases, the transition must occur at the $\sum (S^z_n)^2$-symmetric point. Lastly, note that both gapped phases have the same symmetry flux for $R^z$ (this can hence remain long-range ordered at criticality) but have distinct symmetry fluxes for $R^x$, i.e., $\cdots R^x_{n-2} R^x_{n-1} S^x_n$ and $\cdots R^x_{n-2} R^x_{n-1} \left( S^y_n S^z_n -  S^z_n S^y_n\right)$. At the critical point, these symmetry fluxes for $R^x$ become algebraically ordered with the same scaling dimension. They are thus degenerate, but both have the same non-trivial charges with respect to $R^y$ and $R^z$, consistent with this being a non-trivially-enriched CFT.

\subsection{``\emph{Quantum Criticality in Topological Insulators and Superconductors: Emergence of Strongly Coupled Majoranas and Supersymmetry}''}

One of the phase transitions considered by Grover and Vishwanath in Ref.~\onlinecite{Grover12} is between a topological superconductor in class DIII (protected by an anti-unitary symmetry satisfying $T^2 = P$) and an Ising phase which spontaneously breaks $T$. It was demonstrated that the edge flows to an unusual fixed point, exhibiting itself in unusual fermionic correlations near the boundary (as a function of time).

We point out that this transition is an Ising CFT enriched by $T$ and fermionic parity symmetry $P$. The clearest way of seeing this is that the symmetry flux of $P$ (which has long-range order for this bosonic transition) can be shown to be \emph{odd} under $T$. In an approximate converse, the $\mu$ operator of the Ising CFT is \emph{odd} under fermionic parity symmetry: $P\mu P = - \mu$; see Appendix~\ref{app:grover} for details. As discussed in Fig.~\ref{fig:edge}, this stabilizes the spontaneously-fixed boundary condition. According to the analysis in Section~\ref{sec:edge}, we thus expect that this class of transition generically has a global twofold degeneracy for open boundaries whose finite-size splitting is exponentially small in system size.

\subsection{``\emph{Gapless symmetry-protected topological phase of fermions in one dimension}''}

In Ref.~\onlinecite{Keselman15}, Keselman and Berg show that spinful fermions with attractive triplet-pairing stabilize a $c=1$ CFT with exponentially-localized edge modes protected by fermionic parity symmetry $P$ (enhanced to a full $U(1)$ in their particular model) and time-reversal symmetry $T$ (obeying $T^2 = P$). In addition to a field-theoretic analysis, this was numerically demonstrated in a simple lattice model:
\begin{equation}
H = - \sum_{n,\sigma} \left( c^\dagger_{n,\sigma} c^{\vphantom \dagger}_{n+1,\sigma} + h.c. \right) + U \sum_n \Delta_n^\dagger \Delta_n^{\vphantom \dagger}, \label{eq:Berg}
\end{equation}
where $\sigma \in \{ \uparrow,\downarrow \}$ and $\Delta_n = c_{n,\uparrow} c_{n+1,\downarrow} + c_{n,\downarrow} c_{n+1,\uparrow}$ (on the lattice, the symmetries are given by $P = \prod_j P_j$ where $P_j = e^{i \pi ( n_{j,\uparrow} + n_{j,\downarrow })}$ and $T = \prod_j  e^{i \pi S^y_j} K $ with $S^y_j = \frac{i}{2} \big( c_{j,\downarrow}^\dagger c_{j,\uparrow}^{\vphantom \dagger}  - c_{j,\uparrow}^\dagger c_{j,\downarrow}^{\vphantom \dagger} \big) $). For $U<0$, the spin sector is gapped, whereas charges remain gapless. The stability of the edge mode was argued to be a consequence of the gap to single-fermion excitations (which moreover implies that the phenomenon persists upon adding spin-orbit coupling). Moreover, in Ref.~\onlinecite{Keselman15} it was shown that fermionic parity symmetry anticommutes with $T$ \emph{near the edges}, implying edge modes.

We remark that this $c=1$ CFT is non-trivially enriched by $P$ and $T$. We find that for $U<0$, the symmetry flux of $P$ has long-range order and is given by $\mathcal S^P_j = \cdots P_{j-2} P_{j-1} ( n_{j,\uparrow} - n_{j,\downarrow} )$, which obeys $T \mathcal S^P T^{-1} = - \mathcal S^P$. See Appendix~\ref{app:Berg} for numerical details. This generalizes the observation of Ref.~\onlinecite{Keselman15} that $P$ and $T$ anticommute near the \emph{boundary} to a topological invariant \emph{in the bulk}. This moreover confirms that the phase is protected by virtue of $P$ being a gapped symmetry. The fact that the non-trivial symmetry flux is for a \emph{gapped} symmetry is naturally related to the exponentially small finite-size splitting observed in Ref.~\onlinecite{Keselman15}.

\subsection{``\emph{Gapless symmetry-protected topological order}''}

In Refs.~\onlinecite{Scaffidi17,Parker18}, Scaffidi, Parker and Vasseur extended the well-known mechanism of creating gapped SPT phases by decorating domain walls \cite{Chen14} to critical systems. The idea is illustrated straightforwardly with an example (their mechanism also applies to higher dimensions, but here we limit ourselves to one dimension). Let $U$ be the unitary which maps the gapped $\mathbb Z_2 \times \mathbb Z_2$-paramagnet to the $\mathbb Z_2 \times \mathbb Z_2$-SPT phase. As a starting point, take a Hamiltonian $H$ which consists of trivial gapped degrees of freedom decoupled from a critical bulk; moreover, presume that \emph{one} of the $\mathbb Z_2 \subset \mathbb Z_2 \times \mathbb Z_2$ acts only on the former gapped chain. The claim is then that $U H U^\dagger$---which is clearly still critical---is topologically non-trivial. This was diagnosed in terms of edge properties. In particular, in the above case, the ground state will be twofold degenerate for open boundary conditions (with exponentially small finite-size splitting). Through a combination of perturbative arguments and numerical calculations, this topological phenomenon was shown to be stable.

The above procedure---which can be applied for any symmetry group $G$---indeed creates non-trivial $G$-enriched CFTs. This can be seen from the perspective of symmetry fluxes: by construction, part of the symmetry group will be \emph{gapped}, which ensures that the associated symmetry fluxes are \emph{unique}. This means that the non-trivial charge endowed by the SPT-entangler $U$ will be a well-defined topological invariant, even at criticality, as discussed in Section~\ref{sec:symflux}. Using the boundary RG analysis of Section~\ref{sec:edge}, this non-trivial symmetry flux can be used to argue the presence of edge modes. The notion of \emph{charged symmetry flux} thus gives a \emph{topological bulk invariant} for the cases studied in Refs.~\cite{Scaffidi17,Parker18}. We note that the presence of gapped degrees of freedom are not essential to make domain wall decoration work at criticality (which goes beyond the scope of Ref.~\cite{Scaffidi17}): the models $H$ and $H''$ in Section~\ref{sec:overview} are related by the SPT-entangler $U = \prod_n (CZ)_{n,n+1}$; in this case, we found the finite-size splitting to be algebraic.

The one-dimensional system which Ref.~\onlinecite{Scaffidi17} focus on is the cluster SPT model with an Ising coupling which preserves $\mathbb Z_2 \times \mathbb Z_2$ symmetry. We identify its bulk topological invariant in Appendix~\ref{app:Scaffidi}. We note that Ref.~\onlinecite{Scaffidi17} also presented a $2+1d$ example, which we will revisit in Section~\ref{sec:higher} where we generalize our topological invariants to higher dimensions.

\subsection{``\emph{Topology and edge modes in quantum critical chains}''}

In Ref.~\onlinecite{Verresen18}, a subset of the current authors showed that exponentially localized edge modes can exist in critical systems without gapped degrees of freedom. This work was restricted to non-interacting fermions, where arbitrary models with fermionic parity symmetry $P$ and spinless time-reversal symmetry $T$ ($T^2=1$)---realizing the so-called BDI class---could be described in terms of a complex function $f(z)$. The number of roots of this function on the unit circle tells us the central charge of the bulk CFT. The number (and norm) of roots stictly within the unit disk tell us about the number of topological edge modes (and their individual localization lengths). In particular, this shows that in this non-interacting set-up one can have edge modes for an arbitrary (half-integer) central charge $c$.

These critical systems can indeed be seen as CFTs enriched by $P$ and $T$ symmetry. Since there are no additional gapped degrees of freedom, all symmetry fluxes are algebraic. Nevertheless, these can still have non-trivial charges. In particular, as shown in Section~\ref{sec:symflux}, the Ising CFT ($c=1/2$) has unique symmetry fluxes. And indeed, as discussed in Section~\ref{sec:ferm}, this allows to extend some of the non-trivial CFTs considered in Ref.~\onlinecite{Verresen18} to the \emph{interacting case}, where the finite-size splitting becomes algebraic $\sim 1/L^{14}$ (the algebraic splitting does not show up in the non-interacting limit since the descendant responsible for the splitting is intrinsically interacting). For more general CFTs ($c\geq 1$), a case-by-case study is necessary: some remain non-trivial with interactions, others do not, which can be determined by examining the symmetry action on the higher-dimensional space of symmetry fluxes. We leave an exhaustive discussion of these more general situations to future work.

\section{On twisted sectors and higher dimensions \label{sec:higher}}

Thus far, we have discussed topological invariants and edge modes in one-dimensional critical systems in great detail. The purpose of this last section is to give some higher-dimensional generalizations. A complete characterization of gapless topological phases is a challenging endeavor, since---as we will argue---it would require the knowledge of topological defects in conformal field theories, which in higher dimensions is itself a nascent field with a great promise for rich physics \cite{McAvity93,McAvity95,Liendo13,Billo13,Gaiotto14,Gliozzi15,Billo16,Fukuda18,Lauria19,Metlitski20,Dey20,Antunes21,Lauria21,Herzog21}. We first give an equivalent reformulation of our one-dimensional invariant (restricted to unitary symmetry groups) in Section~\ref{subsec:reformulation}, which is then extended to higher dimensions in Section~\ref{subsec:higher}.

\subsection{Reformulation in terms of twisted sectors \label{subsec:reformulation}}

Let us first explain why a reformulation is natural when trying to generalize to higher dimensions. In Section~\ref{sec:symflux}, topological invariants at criticality were defined in terms of the symmetry charges of (slowest-decaying) string operators. Since these strings consist of a symmetry $\prod_n U_n$, one possible generalization to 2D would be to consider membrane operators corresponding to a symmetry $\prod_{m,n} U_{m,n}$. One could thus attempt to define a `slowest-decaying' membrane operator and study its symmetry properties. However, this is a less natural object to define and study than in 1D since the boundary of a $L \times L$ membrane grows like $L$. This means that generically the expectation value of the membrane operator will decay \emph{exponentially} in both the gapped\footnote{By symmetry fractionalization, gapped phases admit a fine-tuned choice of unitary string operator which would give the membrane operator long-range order---indeed it gives one route to the classification of 2D gapped SPT phases \cite{Chen13}---but this object is exceedingly difficult to access in practice (away from fixed-point limits).} and gapless case \cite{Zhao21,Wu21}, making the notion of `slowest-decaying' less practical and straightforward compared to 1D.

Fortunately, we can avoid this subtlety by instead working with static/external symmetry defects, which are more natural objects in CFTs (in arbitrary dimensions). In 1D, this will turn out to be equivalent to what we have already discussed. To define a symmetry defect (or equivalently, symmetry twist) in 1D, consider a local Hamiltonian $H = \sum_n h_n$ and an on-site unitary symmetry $U^g = \prod_n U_n^g$ (i.e., $U^g h_n U^{g\dagger} = h_n$). A system with a $g$-defect or -twist at site $n_0$ in an infinite chain is defined as the Hamiltonian $H_{\textrm{twist}} := V_{n_0} H V_{n_0}^\dagger$ with $V_{n_0} = \prod_{n \leq n_0} U^g_n$. Crucially, since $H$ is local and symmetric, $H_\textrm{twist}$ differs from $H$ \emph{only} near the defect $n \approx n_0$. (See also the discussion in Ref.~\cite{Zaletel14}.) This thus also defines a $g$-defect for a finite system on a ring (as long as the system size is larger than the support/range of the terms in the Hamiltonian). One can refer to $H_\textrm{twist}$ as the \emph{twisted sector}, since a periodic system with a $g$-defect can be interpreted as having a boundary condition which has been twisted by $g$.

\begin{figure}
	\includegraphics[scale=1]{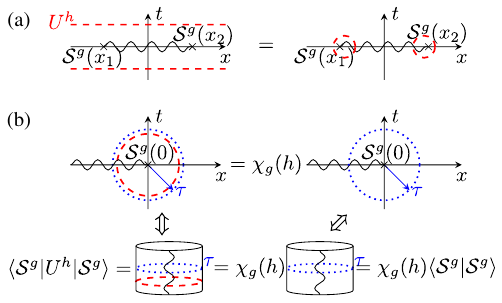}
	\caption{\textbf{Operator-state correspondence for symmetry fluxes.} (a) The black wiggly line represents a two-point function of a symmetry flux operator $\mathcal S^g$ as defined in Section~\ref{sec:symflux}. The red dashed lines represent a commuting symmetry $U^h$; the figure shows how $U^h \mathcal S^g U^{h \dagger}$ can be represented in spacetime as $\mathcal S^g$ with a $U^h$ loop wrapped around its endpoint. (b) The first equality is simply the definition of the charge $\chi_g(h)$ of the symmetry flux, as defined in Section~\ref{sec:symflux}. For operator-state correspondence \cite{DiFrancesco97}, we perform radial quantization (i.e., the new time $\tau$ runs radially; the blue dotted line is a spatial slice); by conformally mapping the plane to the cylinder, the operator $\mathcal S^g$ acts at $\tau = -\infty$, such that on the dashed line we have a state denoted by $|\mathcal S^g \rangle$, which is defined for $g$-twisted boundary conditions. Due to the symmetry flux being the \emph{slowest} decaying string operator, $|\mathcal S^g\rangle$ is the \emph{ground} state of the twisted sector. We see that $U^h|\mathcal S^g \rangle = \chi_g(h) |\mathcal S^g\rangle$, i.e., $\chi_g(h)$ can be obtained by considering a lattice model with $g$-twisted boundary conditions and measuring the $h$-charge of its ground state. \label{fig:correspondence}}
\end{figure}

A key claim is that the topological invariant which we defined in Section~\ref{sec:symflux}---i.e., the charge\footnote{Reminder: we take $g,h \in G$ to be two commuting symmetries.} under $U^h$ of the symmetry flux $\mathcal S^g$ associated to $U^g$---is identical to the charge under $U^h$ of the ground state of the $g$-twisted sector $H_\textrm{twist}$. Indeed, this is a direct consequence of the operator-state correspondence. To see this, note that this correspondence---which involves conformally mapping the (punctured) plane to the spacetime cylinder \cite{Cardy86operatorcontent,DiFrancesco97}---relates the scaling dimensions of local operators to the finite-size energy spectrum on a circle \cite{DiFrancesco97}. Similarly, as illustrated in Fig.~\ref{fig:correspondence}, scaling dimensions of nonlocal string operators whose string consists of a symmetry operator are in 1-to-1 correspondence with the finite-size energy spectrum on a circle with a twisted boundary condition, i.e., the spectrum of $H_\textrm{twist}$. In particular, the leading scaling dimension (i.e., the `slowest-decaying' string operator) then naturally corresponds to the ground state of $H_\textrm{twist}$. The claimed identification of their charges thus follows. Note that the symmetry twist on a ring can equivalently be interpreted as threading a flux through the ring; this explains our nomenclature of \emph{symmetry flux} in Section~\ref{sec:symflux}. Indeed, in the case of gapped SPT phases it is well-known that SPT invariants are encoded in the charges attached to external fluxes \cite{Levin12,Hung13,Chen13,Barkeshli13,Wen14,Cheng14,Zaletel14,Teo14,Else14,Kapustin14,Kapustin14b,Wang15,Tarantino16,Tiwari18}.

Let us illustrate this equivalence for the cluster model with a single-sublattice Ising coupling \cite{Scaffidi17}:
\begin{align}
H = &- \sum_n \left( Z_{A,n} X_{B,n} Z_{A,n+1} + Z_{B,n-1} X_{A,n} Z_{B,n} \right) \notag \\
&- J \sum_n Z_{A,n} Z_{A,n+1} . \label{eq:Scaffidi} 
\end{align}
This has a $\mathbb Z_2 \times \mathbb Z_2$ symmetry generated by $P_A = \prod_n X_{A,n}$ and $P_B = \prod_n X_{B,n}$. Ref.~\cite{Scaffidi17} showed that the edges mode of the cluster SPT phase ($|J|<1$) persist at the Ising critical point at $|J|=1$ (beyond which there is a phase that spontaneously breaks $P_A$), although no bulk invariant was identified. Using the methods introduced in Section~\ref{sec:symflux}, one can straightforwardly derive a bulk topological invariant for these two cases (i.e., $|J| \leq 1$): we find that the symmetry flux of $P_A$ is \emph{odd} under $P_B$ (and vice versa); the derivation can be found in Appendix~\ref{app:Scaffidi}.

We now demonstrate how to obtain the same non-trivial invariant using symmetry twists. Twisting Eq.~\eqref{eq:Scaffidi} by $P_A$ gives us
\begin{align}
H_{\textrm{twist}} = &- \sum_n (-1)^{\delta_{n,n_0}} Z_{A,n} ( X_{B,n} + J ) Z_{A,n+1} \notag \\
& - \sum_n Z_{B,n-1} X_{A,n} Z_{B,n}.  \label{eq:Scaffidi_twistA}
\end{align}
Note that $Z_{A,n} X_{B,n} Z_{A,n+1}$ is a local integral of motion. We thus read off that the ground state is in the sector $Z_{A,n} X_{B,n} Z_{A,n+1} = (-1)^{\delta_{n,n_0}}$ implying a quantum number
\begin{equation}
P_B = \prod_n X_{B,n} = \prod_n Z_{A,n} X_{B,n} Z_{A,n+1} =-1. \label{eq:clustercharge}
\end{equation}
This is in contrast to a trivial decoupled case\footnote{For instance, a unitary mapping can disentangle Eq.~\eqref{eq:Scaffidi} into $H = -\sum_n \left( X_{A,n} + X_{B,n} + J Z_{A,n} Z_{A,n+1} \right)$.} where the ground state in the $P_A$-twisted sector satisfies $P_B = +1$. These quantum numbers are robust invariants on the condition that level crossings are not possible without leaving the phase. For a gapped symmetric phase, this follows from having a unique ground state in the twisted sector\footnote{This follows from symmetry fractionalization.}. For a CFT, this follows from the low-energy spectrum being universal\footnote{One should be mindful of marginal perturbations when these exist. See the discussion in Section~\ref{subsec:gaussian} where charge assignments can swap when tuning along the moduli space of $c=1$ CFTs. This does not occur for the Ising CFT.}. E.g., for the Ising criticality under consideration, for asymptotically large system sizes the finite-size energy gap in the twisted sector can be related (by the operator-state correspondence) to the difference in scaling dimension between the disorder operator ($\Delta_\mu=1/8$) and the fermion ($\Delta_\psi=1/2$), preventing a crossing (see e.g., Eq.~(7.25) in Ref.~\cite{Ginsparg88}). We thus arrive at a non-trivial topological invariant for $|J|=1$. In contrast, in the Ising phase ($|J|>1$) the ground state manifold of the twisted sector is extensively degenerate and not universal---this can be understood by noting that it traps a domain wall whose energy (relative to the untwisted sector) becomes infinite when flowing to an RG fixed-point. Thus in the symmetry-breaking case, it does not provide an invariant\footnote{E.g., $H = - \sum_n \left( Z_{A,n} Z_{A,n+1} +  X_{B,n} \left( 1-\lambda + \lambda Z_{A,n} Z_{A,n+1}\right) \right) $ is in an Ising phase for all $\lambda$ but the $P_A$-twisted sector has a crossing at $\lambda = 1/2$ where its $P_B$-charge swaps.}.

\subsection{Generalizations to two dimensions \label{subsec:higher}}

The idea that charges of twisted sectors provide topological invariants naturally generalizes to arbitrary dimensions, as has already been well-explored in the gapped case \cite{Levin12,Hung13,Chen13,Barkeshli13,Wen14,Cheng14,Zaletel14,Teo14,Else14,Kapustin14,Kapustin14b,Wang15,Tarantino16,Tiwari18}. Firstly, the notion of the symmetry twist or defect is entirely analogous to what we saw in 1D: consider a local Hamiltonian $H$ which is periodic along, say, the $y$-direction, and infinite along $x$. If $U^g = \prod_{x,y} U^g_{x,y}$ is an on-site unitary symmetry, then its twist along the $y$-direction is defined as $H_\textrm{twist} := V_{x_0} H V_{x_0}^\dagger$ with $V_{x_0} = \prod_{x \leq x_0} \prod_y U_{x,y}$. Due to $U^g$ being a symmetry and $H$ being local, we see that $H_\textrm{twist}$ only differs from $H$ for terms $x \approx x_0$. Hence, this also defines the meaning of a symmetry twist on a torus geometry (even though in that case, $H_\textrm{twist}$ is not necessarily unitarily equivalent to $H$).

\begin{figure}
	\includegraphics[scale=1]{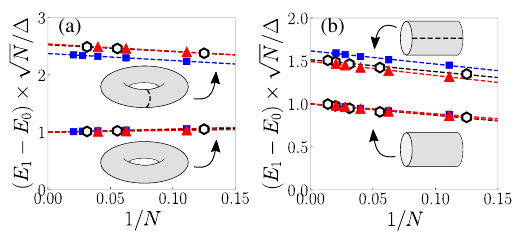}
	\caption{\textbf{Symmetry defect in the 2+1D Ising CFT.} We consider the critical ferromagnetic Ising model in a transverse field on the square, triangular and honeycomb lattices, respectively denoted by (blue) squares, (red) triangles and (white) hexagons. (a) The bottom curve shows the finite-size energy gap for periodic boundary conditions (i.e., the torus); since this is a CFT, $E_1 - E_0 \sim \Delta /\sqrt{N}$ (for total number of sites $N = L\times L$). For each lattice, we rescale the $y$-axis such that the extrapolated prefactor $\Delta$ is unity. The other data is for a torus twisted by the $\mathbb Z_2$ symmetry of the Ising model; again we extract the finite-size energy gap in this twisted sector. With the aforementioned rescaled $y$-axis, we find the universal $y$-intercepts $\approx 2.4$ (for square lattice, i.e., modular parameter $\tau = e^{i\pi/2}$) and $\approx 2.5$ (for triangular and honeycomb lattices, i.e., $\tau = e^{i\pi/3}$). (b) Similar set-up for the cylinder geometry (i.e., open in one direction, periodic in the other). Here too we find that the twisted sector has a unique ground state with the finite-size energy splitting in this rescaled $y$-axis giving the universal intercepts $\approx 1.6$ (for $\tau = e^{i\pi/2}$) and $\approx 1.5$ (for $\tau = e^{i\pi/3}$).  \label{fig:twistedIsing}}
\end{figure}

In 2D, if we have two defect lines (say, a $U^g$ twist in the $x$-direction and a $U^h$ twist in the $y$-direction) then their intersection is a point in space and it is thus natural to measure `its' charge (i.e., the charge of the ground state in this twisted sector) with respect to a third symmetry. For gapped symmetric phases of matter, the ground state in the twisted sector is unique\footnote{Indeed, symmetry fractionalization implies that the twisted sector is unitarily equivalent to the untwisted sector, and the latter is gapped by definition.}, such that its charge is a quantized topological invariant. Our contribution is to point out that this can also work for symmetry-preserving gapless systems: the low-energy spectrum (be it in a twisted sector or not) is universal, which in particular means that there can be no level crossings (except for the case where there are marginal operators, whose existence is known for a typical CFT). Hence, the charge of the ground state(s) in twisted sectors can serve as a topological invariant for gapless systems.

We note that the low-energy spectrum for 2+1D CFTs has been studied before on a torus geometry where the claim of universality was numerically confirmed \cite{Schuler16,Thomson17,Whitsitt17,Belin18,Schuler21}. We are not aware of similar numerical explorations for the case with a symmetry twist (although an epsilon-expansion was studied in Ref.~\cite{Whitsitt17}), but it is a natural extension; in particular, the symmetry twist is expected to flow to a conformal defect of the CFT.
We numerically confirm this for the transverse-field Ising model $H = - \sum_{\langle \bm n, \bm m \rangle} Z_{\bm n} Z_{\bm m} + g \sum_{\bm n} X_{\bm n}$; we set $g$ to achieve $2+1d$ Ising criticality, whose value is known for a variety of lattices \cite{Bloete02}. In Ref.~\cite{Schuler16}, the universal finite-size spectrum on a torus geometry was studied. In the present work, we rescale the spectrum such that the finite-size gap (in this untwisted sector) is asymptotically $E_1 - E_0 \sim 1/L$. We can now twist one of the periodic directions with the $\mathbb Z_2$ symmetry (this involves flipping the sign of $ Z_{\bm n} Z_{\bm m}$ on certain bonds). In Fig.~\ref{fig:twistedIsing}(a), we show the finite-size energy gap in this twisted sector for the square, triangular and honeycomb lattices. First, up to a global prefactor of $1/L$, we observe that the gap tends to a finite value, i.e., the ground state is unique also in the twisted sector. Second, its asymptotic value coincides for the triangular and honeycomb lattices: this confirms our claim of universality (since in the scaling limit, only the information of the modular parameter remains, which for both lattices is $\tau = e^{\pi i/3}$). In Fig.~\ref{fig:twistedIsing}(b) we confirm that the universality also appears for the cylinder geometry. Hence, as explained above, the charge of its ground state can serve as a robust topological invariant, as we will illustrate in the next subsection.

In fact, one can argue that the following stronger property holds: a \emph{conformal symmetry} defect or twist is automatically also a \emph{topological} defect\footnote{This is a direct consequence from noting that since it is derived from a symmetry, the defect line commutes with the stress-energy tensor; remembering how the latter encodes the dependence of the action on the metric, the claim follows.}, i.e., the universal finite-size energy spectrum is independent of deformations of the line defect. This property is not needed to define the topological invariant, but it is key in arguing that a non-trivial value of the invariant implies edge modes, as discussed in Section~\ref{subsec:2Dedge}.

\begin{figure}
	\includegraphics[scale=1]{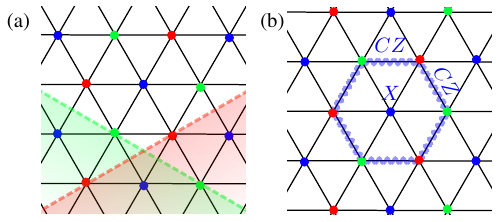}
	\caption{\textbf{Triangular lattice, twists, and the Yoshida SPT model.} (a) The triangular lattice with A (green), B (red) and C (blue) sublattices. The dashed lines denote possible ways of twisting by $P_A$ and $P_B$ symmetry (e.g., the latter corresponds to conjugating $H$ by $\prod X_{B,\bm n}$ on the shaded red area). (b)  An elementary term in the Yoshida SPT model protected by $\mathbb Z_2^3$ symmetry; the wiggly line represents a product of control-$Z$ gates. \label{fig:TL}}
\end{figure}

\subsubsection{Examples in 2D \label{subsec:2Dexample}}

Let us illustrate the above 2D topological invariant in a variety of examples. For simplicity, all of these will focus on the symmetry group $\mathbb Z_2 \times \mathbb Z_2 \times \mathbb Z_2$. As a natural setting for this symmetry, we consider the triangular lattice with the three sublattices shown in Fig.~\ref{fig:TL}(a); for each sublattice we define the $\mathbb Z_2$ symmetry operator $P_\lambda = \prod_{\bm n} X_{\lambda, \bm n}$ with $\lambda=A,B,C$. For any symmetric Hamiltonian $H$, we can twist by $P_A$ and $P_B$ along the two dashed lines in Fig.~\ref{fig:TL}(a), leading to $H_\textrm{twist}$. If the ground state of $H_\textrm{twist}$ is unique, it must have a well-defined charge under $P_C$, providing a topological invariant\footnote{Similar to the discussion in Section~\ref{subsec:reformulation}, this only works in the absence of spontaneous symmetry-breaking: if $P_C$ is broken, then there are ground states with opposite charge, whereas if $P_A$ or $P_B$ is broken, the twisted sector can have level crossings without leaving the phase.}. This can also apply in the case of degeneracy, e.g., if all ground states have the same charge under $P_C$.

Before we discuss examples of concrete Hamiltonians $H$ satisfying the above condition, let us note that topological invariants are only interesting if we can also realize a phase with a \emph{different} value of the invariant. Fortunately, we have a general result to this end: defining the SPT-entangler $U_\mathrm{2D} = \prod_{\triangle} CCZ$ (applying the control-control-$Z$ gate\footnote{This three-site gate is defined as $CCZ|\sigma_1 \sigma_2 \sigma_3 \rangle = (-1)^{\delta_{\sigma_1 \sigma_2 \sigma_3,\uparrow \uparrow \uparrow}} | \sigma_1 \sigma_2 \sigma_3 \rangle$.} on every triangle of the triangular lattice), we derive the following in Appendix~\ref{app:2D}.
\begin{shaded*}
\emph{\textbf{Different topological invariants in 2D:} let $H$ be a (gapped or gapless) model on the triangular lattice commuting with $P_A$, $P_B$ and $P_C$, and suppose it has a well-defined topological invariant as described above (e.g., all ground states in the twisted sector have the same $P_C$ charge). Then $H$ and $H' := U_\mathrm{2D} H U_\mathrm{2D}^\dagger$ have distinct topological invariants. More precisely, if $H_\textrm{twist}$ is defined as above, then its ground state charge under $P_C$ is \underline{opposite} that of $\left(H'\right)_\textrm{twist}$.}
\end{shaded*}

The simplest example is the trivial gapped paramagnet: e.g., $H = - \sum_{\bm n} \left( X_{A,\bm n} + X_{B,\bm n} + X_{C,\bm n} \right)$. In this case, $H_\textrm{twist} = H$, such that we read off that the ground state charge is $P_C = 1$ (since $X_{C,\bm n}=1$). The SPT-entangled Hamiltonian $H' = U_{\mathrm 2D} H U_{\mathrm 2D}^\dagger$ is Yoshida's $\mathbb Z_2^3$ SPT model \cite{Yoshida16,Yoshida17}. The terms in $H'$ are of the form $U_\mathrm{2D} X_{\lambda,\bm n}U_{\mathrm{2D}}^\dagger = X_{\lambda,\bm n} \times \prod_\textrm{hex} CZ$ where the product runs around the six neighboring sites as shown in Fig.~\ref{fig:TL}(b). By the above general result, we see that in the twisted sector of the Yoshida model, the ground state charge is $P_C = -1$, giving a non-trivial value of the topological invariant. (This particular case is calculated explicitly in Appendix~\ref{subsec:simple}.) This confirms that these two models cannot be adiabatically connected whilst preserving the symmetry and the universality class (in this case, the finite energy gap). Note that we even obtain a stronger condition: any critical point between the trivial and non-trivial SPT phase must have a degeneracy in the twisted sector (labeled by opposite charges $P_C = \pm 1$).

Let us discuss topological invariants in gapless examples. First note that in the above gapped example, to derive $P_C = 1$ (for the twisted sector) we did not actually use that $P_A$ and $P_B$ are gapped symmetries (i.e., that the $A$ and $B$ degrees of freedom are gapped). The argument relied only on $H$ not coupling the sublattices. Hence, the same conclusion follows as long as at least one of the three symmetries is gapped; without loss of generality, let this be $P_C$. In particular, let $H = H_C + H_{AB}$ with $H_C = - \sum_{\bm n} X_{C,\bm n}$ and $H_{AB}$ an arbitrary (symmetry-preserving) Hamiltonian on the $A$ and $B$ sublattices. Twisting by $P_A$ and $P_B$ will clearly not affect the $C$ degrees of freedom, such that the ground state satisfies $X_{\bm n} = +1$ and thus $P_C = 1$ in the twisted sector. Note that the ground state subspace is not necessarily unique: twisting by $P_A$ and $P_B$ could introduce a degeneracy, but this argument shows that $P_C =1$ for all ground states. This charge acts as a topological invariant, with $H' = U_\textrm{2D} H U_\textrm{2D}^\dagger$ giving a model with $P_C = -1$. Let us stress that although we obtained the invariant in a fine-tuned limit where the $C$ sites satisfy $X_{C,\bm n}=1$, by virtue of universality the resulting topological invariant is well-defined and robust as long as one does not leave the universality class.

This simple insight applies to a whole array of examples, including the 2D example discussed by Scaffidi, Parker and Vasseur \cite{Scaffidi17} where $H$ is decoupled such that the $A$ sites formed a gapless $U(1)$ spin liquid and the $B$ and $C$ sites were gapped. In that work, it was demonstrated that $H' =U_\textrm{2D} H U_\textrm{2D}^\dagger$ has topologically-protected edge excitations. Here, we have derived a bulk topological invariant distinguishing the models $H$ and $H'$ studied in Ref.~\cite{Scaffidi17}. The invariant is well-defined by virtue of having a gapped sector; it is not sensitive to the particular choice of gapless degrees of freedom. Another interesting byproduct is that any continuous phase transition between the trivial phase and the $\mathbb Z_2^3$ SPT phase must necessarily be gapless for \emph{all} three symmetries (otherwise its remaining gapped degrees of freedom would remember the topological invariant, preventing the required self-duality of an SPT transition).

Our topological invariant can also apply to systems with no gapped symmetries. As a concrete example, let us consider $\textrm{Ising}^3 = \textrm{Ising} \times \textrm{Ising} \times \textrm{Ising}$ criticality, where we define $H$ as having a copy of the critical Ising model on each of the three sublattices of the triangular lattice. Twisting by $P_A$ and $P_B$ again does not affect the $C$ sites. From Ref.~\cite{Schuler16}, we learn that the ground state of the critical Ising model on the torus is unique and satisfies $P_C = 1$. We can thus already conclude that $H$ has a well-defined topological invariant, with all the ground states in the twisted sector satisfying $P_C = 1$. In fact, from Fig.~\ref{fig:twistedIsing} we see that this ground state is unique (although this is strictly speaking not necessary to have a well-defined invariant). Using our general result above, we see that $H' = U_\textrm{2D} H U_\textrm{2D}^\dagger$ gives a version of the Ising${}^3$ criticality which has the opposite invariant $P_C = -1$ in the twisted sector. As long as one preserves Ising${}^3$ criticality, these two symmetry-enriched versions cannot be connected! Moreover, as we will see in the next subsection, $H'$ must host non-trivial edge phenomena, which are protected by this bulk topological invariant.

The above Ising${}^3$ example is a good proof of principle for the existence of topologically-distinct symmetry-enriched versions of the same universality class (even in the absence of gapped degrees of freedom). However, to some extent the example is academic in nature: if one does not enforce the Ising$^{3}$ criticality by hand, then energy-energy perturbations between the distinct sublattices are expected to trigger a flow to $O(3)$ criticality with cubic anisotropy\footnote{More precisely, it will flow to the cubic fixed point if we preserve the symmetry between the three sublattices, otherwise it will flow to $\textrm{Ising}\times O(2)$; nevertheless, the critical exponents of both CFTs are very similar to those of the $O(3)$ CFT, only differing beyond the third decimal \cite{Chester20}.} \cite{Aharony73,Carmona00,Hasenbusch11,Adzhemyan19,Chester20}. Fortunately, our general result above still applies: if the $O(3)$ criticality has the property that all ground states in the sector twisted by $P_A$ and $P_B$ have the same $P_C$ charge\footnote{In particular, it is sufficient to show that the ground state in the twisted sector is unique.}, then the SPT-entangler $U_\textrm{2D}$ gives a topologically non-trivial version of $O(3)$ criticality (this reasoning applies to both the standard $O(3)$ fixed point as well as its cubic perturbation). While we expect this property of $O(3)$ criticality to hold, its numerical confirmation goes beyond the scope of the present work such that we leave it to future studies. Moreover, it could be explored via $\varepsilon$-expansion using the methods of Ref.~\cite{Whitsitt17}. If true, this means that the two $O(3)$ critical lines that appeared in Ref.~\cite{Dupont21} are in fact topologically-distinct and must be separated by a multi-critical point where the twisted sector becomes degenerate. (The particular model studied in Ref.~\cite{Dupont21} instead has an intermediate symmetry-breaking phase, suggesting that such a multi-critical point will require more fine-tuning.)

\begin{figure}
	\includegraphics[scale=1]{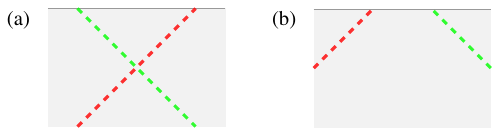}
	\caption{\textbf{Edge modes from 2D topological invariant.} (a) We start with two intersecting symmetry defects (red and green dashed lines) in a system with a boundary (solid gray line). (b) If the boundary is symmetry-preserving, the defect is topological and can be moved upward such that the intersection can be `slid off' the system. If the intersection carried a non-trivial charge (i.e., the bulk is topologically non-trivial), then consistency demands that the intersections of the defect line at the boundary are themselves non-trivial, excluding a trivial boundary condition (see main text for details). \label{fig:2Dboundary}}
\end{figure}

\subsubsection{Implications for the edge \label{subsec:2Dedge}}

In this last section, we sketch an argument for why a non-trivial value of the topological invariant (i.e., the charge in the twisted sector is different from the untwisted case) implies a non-trivial boundary theory. This generalizes the bulk-boundary correspondence to cases with a gapless bulk. The idea is to consider the above charge not as a property of the system as \emph{whole}, but rather as being associated to the \emph{intersection} of two symmetry defects. This intuitive idea is consistent with two facts: first, as discussed above, the symmetry defect is actually topological, which means the charge is unchanged when we arbitrarily deform the defect lines, and second, if one repeats the above arguments and calculations for non-intersecting defect lines, one finds that the charge remains trivial, i.e., the presence of intersecting defects is crucial in the examples we discussed.

Using this basic idea, one can infer a non-trivial constraint on the edge theory. Indeed, for a closed spatial manifold---like the torus considered above---the intersection cannot be removed by deforming the defect lines. In contrast, for a manifold with a boundary, such intersections \emph{can} be undone. For concreteness, let us consider the situation in Fig.~\ref{fig:2Dboundary}(a), where the bulk has two intersecting symmetry defects. If the boundary theory also preserves the symmetries, then the defects are topological both in the bulk and the boundary, such that we can freely move the defect lines. In particular, nothing prevents us from sliding the intersection upward, such that it eventually disappears as in Fig.~\ref{fig:2Dboundary}(b). We end up with an apparent paradox if the original intersection carried a non-trivial charge, since the final configuration has no more intersection to host a non-trivial charge (seemingly violating the topological nature of these defects). The resolution is that the localized points where the symmetry defect intersects the boundary must themselves be non-trivial. In particular, if these localized points carry a degeneracy with opposite charges, then it can freely absorb the bulk topological invariant without causing an inconsistency. Note that this non-trivial property excludes a trivial boundary condition!

Let us illustrate this for the $\textrm{Ising}^{3}$ CFT discussed in Section~\ref{subsec:2Dexample}. For the non-topological case of three decoupled copies of the Ising CFT on each of the three sublattices of the triangular lattice, we found that the intersection of the $P_A$ and $P_B$ symmetry defects hosts a trivial charge $P_C = 1$. In this case, one can freely slide off the intersection without obtaining a condition on the boundary theory. In contrast, for the SPT-entangled version, we found a charge $P_C=-1$. Hence, the above argument implies that either the boundary spontaneously breaks $P_A$ or $P_B$, or where the $P_A$ (or $P_B$) defect intersects the boundary it must host a degeneracy with charges $P_C = \pm 1$. This excludes the free boundary condition: Fig.~\ref{fig:twistedIsing}(b) shows that an Ising CFT with a symmetry defect intersecting the free boundary condition has a \emph{unique} ground state. We thus conclude that the topologically non-trivial $\mathbb Z_2^3$-enriched Ising$^{3}$ CFT cannot have a free boundary condition. Possible alternatives are symmetry-breaking or fine-tuned gapless edges; we leave a study of concrete lattice models to future work. Note that a trivial Ising CFT can also have a symmetry-breaking edge, but in that case one can drive a boundary phase transition into a free boundary condition: here that is impossible without also driving the bulk into a different universality class.

\section{Outlook}

Similar to how gapped degrees of freedom can realize a multitude of distinct phases of matter in the presence of a global symmetry $G$ \cite{Fidkowski11class,Turner11class,Chen10,Schuch11}, a given universality class has many distinct symmetry-enriched versions. This forms the concept of $G$-enriched quantum criticality. We have introduced various invariants to characterize this, the crucial one being the \emph{symmetry flux} associated to a global symmetry in the 1D case. In the simplest case, this has a well-defined charge which then serves as a topological invariant. We related this to the presence of edge modes, whose finite-size splitting depends on whether the flux is charged under a gapped symmetry. A particularly novel aspect to come out is that the Ising CFT enriched by an anti-unitary symmetry can host an edge mode whose finite-size splitting scales as $\sim 1/L^{14}$. The aforementioned invariants also allow to broach the classification of $G$-CFTs, giving a complete picture for the Ising CFT and a partial understanding for the Gaussian CFT. The latter served to discuss various subtleties, such as how the $c=1$ orbifold CFT \emph{seemingly} has a non-trivial topological invariant. We also demonstrated how the concept of symmetry-enriched quantum criticality allows to unify previous works into a single picture. Lastly, we have shown how these novel topological invariants can be generalized to higher dimensions by studying symmetry charges of CFTs in twisted sectors.

Examples of topologically non-trivial symmetry-enriched CFTs are presumably rather ubiquitous, once one knows where to look. For example, one can expect them to occur---although not exclusively---at transitions between gapped SPT phases and symmetry-breaking phases where the ground state degeneracies (for open boundaries) have compatible symmetry properties. Indeed, this guideline suggests that it should occur for the spin-1 XXZ chain, as we confirmed in Fig.~\ref{fig:xiloc}. Moreover, as we established in Section~\ref{sec:lit}, various known examples of critical systems with edge modes can be reinterpreted as phase transitions between distinct non-trivial gapped phases of matter (see also Ref.~\onlinecite{Parker18b}). It would hence be interesting to uncover more experimentally relevant examples of this sort. It is likely that these are already present---but overlooked---in known phase diagrams (the critical spin-1 XXZ chain illustrates this).

Related to finding more physical realizations of this phenomenon, is the question of what its practical use could be. The unique phenomenology of topologically non-trivial symmetry-enriched quantum criticality is that it allows for a localized topological edge modes which is apparently oblivious to long-range correlations in the bulk. Relatedly, it allows for quantum information to remain localized whilst the bulk is tuned through a quantum critical point. Such surprising stability of quantum information could conceivably prove to be useful.

On the theoretical front, an obvious---and important---challenge is the complete classification of $G$-enriched CFTs for which we have given partial results in one spatial dimension. Such a task is vast, as it contains the classification of all CFTs as a subset. A more realistic question is thus: what is the classification for $G$-CFTs where the CFT is already known and understood in the absence of additional symmetries. In this work, we provided the answer for the Ising CFT. Despite making progress for the case of the symmetry-enriched $c=1$ CFT, its classification is still an open issue for general symmetry groups, as it is for all remaining CFTs. As a first step, it is conceivable that our classification for the Ising CFT can be extended to all minimal unitary models. Moreover, in our classification we presumed a unitary group $G$ for convenience, even though we have discussed anti-unitary examples. It would hence be useful to extend this to include the case of time-reversal symmetry, and perhaps even spatial or anomalous symmetries---which we have not touched upon.
An equally exciting question is whether our topological invariants can be efficiently detected in a tensor-network representation of the ground state, such as the multi-scale entanglement renormalization group ansatz (MERA) \cite{Vidal08}.

Finally, while the last section of this paper already touched upon generalizations of these topological invariants to higher dimensions, many exciting open questions remain. The most non-trivial setting is the one where there are no gapped degrees of freedom to appeal to. An interesting case study beyond the Ising$^{3}$ discussed in the present work would be $O(3)$ criticality in $2+1d$: as discussed in Section~\ref{sec:higher}, it would admit a topological invariant if twisting the $x$-($y$-)direction by, say, a $\pi$-rotation around the internal $x$-($y$-)axis gives rise to a unique ground state\footnote{Indeed, the $\mathbb Z_2^3$ subgroup discussed in Section~\ref{sec:higher} can be identified---as a subgroup of $O(3)$---with the $\mathbb Z_2\times \mathbb Z_2$ subgroup of $\pi$-rotations and the $\mathbb Z_2$ subgroup generated by $- \mathbb I \in O(3)$.}. In fact, this would give a non-trivial topological invariant for the $O(3)$ criticality between the N\'eel phase and 2D AKLT state found in Ref.~\cite{Zhang17}. More generally, one can revisit well-studied CFTs in 2+1d from the perspective of studying their twisted sectors. As we have seen, the invariants embedded in these twisted sectors can put non-trivial constraints on the possible boundary fixed points, which also deserves further study. We suspect that the tools presented in this work are timely considering the recent foray into higher-dimensional gapless topological phases \cite{Scaffidi17,Zhang17,Weber18,Xu20,Wu20,Verresen20,Thorngren21} which have thus far been lacking a unifying framework.

\begin{acknowledgments}
The authors would like to thank Dave Aasen, Erez Berg, John Cardy, Shai Chester, Henrik Dreyer, Matthew Fisher, Tarun Grover, Arbel Haim, Duncan Haldane, Anna Keselman, Zohar Komargodski, Andreas Ludwig, Max Metlitski, Lesik Motrunich, Daniel Parker, Jonathan Ruhman, Thomas Scaffidi, Todadri Senthil, Kevin Slagle, Ady Stern, Nat Tantivasadakarn, Alexandra Thomson, Romain Vasseur, Ashvin Vishwanath, Yifan Wang and Xueda Wen for stimulating conversations.
RV is supported by the Simons Collaboration on Ultra-Quantum Matter, which is a grant from the Simons Foundation (651440, A.V.) and by the Harvard Quantum Initiative Postdoctoral Fellowship in Science and Engineering.
\newpage
RV conducted part of this research at the Max-Planck-Institute for the Physics of Complex Systems, at the Technical University of Munich, and at the KITP, supported by the Deutsche Forschungsgemeinschaft (DFG, German Research Foundation) through the Collaborative Research Center SFB 1143 and by the National Science Foundation under Grant No. NSF PHY-1748958 and by the Heising-Simons Foundation.
RT conducted part of this research at the Weizmann Institute of Science, supported by the Zuckerman Fellowship.
NGJ conducted part of this research at the School of Mathematics, University of Bristol.
FP acknowledges the support of the DFG Research Unit FOR 1807 through grants no. PO 1370/2-1, TRR80 through grant no. 107745057, the European Research Council (ERC) under the European Union's Horizon 2020 research and innovation program (grant agreement no. 771537).
\end{acknowledgments}

\bibliography{PRX_updated_Nov_2021.bbl}

\begin{thebibliography}{139}%
\makeatletter
\providecommand \@ifxundefined [1]{%
 \@ifx{#1\undefined}
}%
\providecommand \@ifnum [1]{%
 \ifnum #1\expandafter \@firstoftwo
 \else \expandafter \@secondoftwo
 \fi
}%
\providecommand \@ifx [1]{%
 \ifx #1\expandafter \@firstoftwo
 \else \expandafter \@secondoftwo
 \fi
}%
\providecommand \natexlab [1]{#1}%
\providecommand \enquote  [1]{``#1''}%
\providecommand \bibnamefont  [1]{#1}%
\providecommand \bibfnamefont [1]{#1}%
\providecommand \citenamefont [1]{#1}%
\providecommand \href@noop [0]{\@secondoftwo}%
\providecommand \href [0]{\begingroup \@sanitize@url \@href}%
\providecommand \@href[1]{\@@startlink{#1}\@@href}%
\providecommand \@@href[1]{\endgroup#1\@@endlink}%
\providecommand \@sanitize@url [0]{\catcode `\\12\catcode `\$12\catcode
  `\&12\catcode `\#12\catcode `\^12\catcode `\_12\catcode `\%12\relax}%
\providecommand \@@startlink[1]{}%
\providecommand \@@endlink[0]{}%
\providecommand \url  [0]{\begingroup\@sanitize@url \@url }%
\providecommand \@url [1]{\endgroup\@href {#1}{\urlprefix }}%
\providecommand \urlprefix  [0]{URL }%
\providecommand \Eprint [0]{\href }%
\providecommand \doibase [0]{http://dx.doi.org/}%
\providecommand \selectlanguage [0]{\@gobble}%
\providecommand \bibinfo  [0]{\@secondoftwo}%
\providecommand \bibfield  [0]{\@secondoftwo}%
\providecommand \translation [1]{[#1]}%
\providecommand \BibitemOpen [0]{}%
\providecommand \bibitemStop [0]{}%
\providecommand \bibitemNoStop [0]{.\EOS\space}%
\providecommand \EOS [0]{\spacefactor3000\relax}%
\providecommand \BibitemShut  [1]{\csname bibitem#1\endcsname}%
\let\auto@bib@innerbib\@empty
\bibitem [{\citenamefont {Fidkowski}\ and\ \citenamefont
  {Kitaev}(2011)}]{Fidkowski11class}%
  \BibitemOpen
  \bibfield  {author} {\bibinfo {author} {\bibfnamefont {Lukasz}\ \bibnamefont
  {Fidkowski}}\ and\ \bibinfo {author} {\bibfnamefont {Alexei}\ \bibnamefont
  {Kitaev}},\ }\bibfield  {title} {\enquote {\bibinfo {title} {Topological
  phases of fermions in one dimension},}\ }\href {\doibase
  10.1103/PhysRevB.83.075103} {\bibfield  {journal} {\bibinfo  {journal} {Phys.
  Rev. B}\ }\textbf {\bibinfo {volume} {83}},\ \bibinfo {pages} {075103}
  (\bibinfo {year} {2011})}\BibitemShut {NoStop}%
\bibitem [{\citenamefont {Turner}\ \emph {et~al.}(2011)\citenamefont {Turner},
  \citenamefont {Pollmann},\ and\ \citenamefont {Berg}}]{Turner11class}%
  \BibitemOpen
  \bibfield  {author} {\bibinfo {author} {\bibfnamefont {Ari~M.}\ \bibnamefont
  {Turner}}, \bibinfo {author} {\bibfnamefont {Frank}\ \bibnamefont
  {Pollmann}}, \ and\ \bibinfo {author} {\bibfnamefont {Erez}\ \bibnamefont
  {Berg}},\ }\bibfield  {title} {\enquote {\bibinfo {title} {Topological phases
  of one-dimensional fermions: An entanglement point of view},}\ }\href
  {\doibase 10.1103/PhysRevB.83.075102} {\bibfield  {journal} {\bibinfo
  {journal} {Phys. Rev. B}\ }\textbf {\bibinfo {volume} {83}},\ \bibinfo
  {pages} {075102} (\bibinfo {year} {2011})}\BibitemShut {NoStop}%
\bibitem [{\citenamefont {Chen}\ \emph {et~al.}(2011)\citenamefont {Chen},
  \citenamefont {Gu},\ and\ \citenamefont {Wen}}]{Chen10}%
  \BibitemOpen
  \bibfield  {author} {\bibinfo {author} {\bibfnamefont {Xie}\ \bibnamefont
  {Chen}}, \bibinfo {author} {\bibfnamefont {Zheng-Cheng}\ \bibnamefont {Gu}},
  \ and\ \bibinfo {author} {\bibfnamefont {Xiao-Gang}\ \bibnamefont {Wen}},\
  }\bibfield  {title} {\enquote {\bibinfo {title} {Classification of gapped
  symmetric phases in one-dimensional spin systems},}\ }\href {\doibase
  10.1103/PhysRevB.83.035107} {\bibfield  {journal} {\bibinfo  {journal} {Phys.
  Rev. B}\ }\textbf {\bibinfo {volume} {83}},\ \bibinfo {pages} {035107}
  (\bibinfo {year} {2011})}\BibitemShut {NoStop}%
\bibitem [{\citenamefont {{Schuch, N. and P\'erez-Garc\'ia, D. and Cirac, J.
  I.}}(2011)}]{Schuch11}%
  \BibitemOpen
  \bibfield  {author} {\bibinfo {author} {\bibnamefont {{Schuch, N. and
  P\'erez-Garc\'ia, D. and Cirac, J. I.}}},\ }\bibfield  {title} {\enquote
  {\bibinfo {title} {Classifying quantum phases using matrix product states and
  projected entangled pair states},}\ }\href {\doibase
  10.1103/PhysRevB.84.165139} {\bibfield  {journal} {\bibinfo  {journal} {Phys.
  Rev. B}\ }\textbf {\bibinfo {volume} {84}},\ \bibinfo {pages} {165139}
  (\bibinfo {year} {2011})}\BibitemShut {NoStop}%
\bibitem [{\citenamefont {Chen}\ \emph {et~al.}(2013)\citenamefont {Chen},
  \citenamefont {Gu}, \citenamefont {Liu},\ and\ \citenamefont {Wen}}]{Chen13}%
  \BibitemOpen
  \bibfield  {author} {\bibinfo {author} {\bibfnamefont {Xie}\ \bibnamefont
  {Chen}}, \bibinfo {author} {\bibfnamefont {Zheng-Cheng}\ \bibnamefont {Gu}},
  \bibinfo {author} {\bibfnamefont {Zheng-Xin}\ \bibnamefont {Liu}}, \ and\
  \bibinfo {author} {\bibfnamefont {Xiao-Gang}\ \bibnamefont {Wen}},\
  }\bibfield  {title} {\enquote {\bibinfo {title} {Symmetry protected
  topological orders and the group cohomology of their symmetry group},}\
  }\href {\doibase 10.1103/PhysRevB.87.155114} {\bibfield  {journal} {\bibinfo
  {journal} {Phys. Rev. B}\ }\textbf {\bibinfo {volume} {87}},\ \bibinfo
  {pages} {155114} (\bibinfo {year} {2013})}\BibitemShut {NoStop}%
\bibitem [{\citenamefont {{Kapustin}}(2014)}]{Kapustin14}%
  \BibitemOpen
  \bibfield  {author} {\bibinfo {author} {\bibfnamefont {Anton}\ \bibnamefont
  {{Kapustin}}},\ }\bibfield  {title} {\enquote {\bibinfo {title} {{Symmetry
  Protected Topological Phases, Anomalies, and Cobordisms: Beyond Group
  Cohomology}},}\ }\href@noop {} {\bibfield  {journal} {\bibinfo  {journal}
  {arXiv e-prints}\ ,\ \bibinfo {eid} {arXiv:1403.1467}} (\bibinfo {year}
  {2014})},\ \Eprint {http://arxiv.org/abs/1403.1467} {arXiv:1403.1467
  [cond-mat.str-el]} \BibitemShut {NoStop}%
\bibitem [{\citenamefont {Kapustin}\ \emph {et~al.}(2015)\citenamefont
  {Kapustin}, \citenamefont {Thorngren}, \citenamefont {Turzillo},\ and\
  \citenamefont {Wang}}]{Kapustin15}%
  \BibitemOpen
  \bibfield  {author} {\bibinfo {author} {\bibfnamefont {Anton}\ \bibnamefont
  {Kapustin}}, \bibinfo {author} {\bibfnamefont {Ryan}\ \bibnamefont
  {Thorngren}}, \bibinfo {author} {\bibfnamefont {Alex}\ \bibnamefont
  {Turzillo}}, \ and\ \bibinfo {author} {\bibfnamefont {Zitao}\ \bibnamefont
  {Wang}},\ }\bibfield  {title} {\enquote {\bibinfo {title} {Fermionic symmetry
  protected topological phases and cobordisms},}\ }\href {\doibase
  10.1007/JHEP12(2015)052} {\bibfield  {journal} {\bibinfo  {journal} {Journal
  of High Energy Physics}\ }\textbf {\bibinfo {volume} {2015}},\ \bibinfo
  {pages} {1--21} (\bibinfo {year} {2015})}\BibitemShut {NoStop}%
\bibitem [{\citenamefont {Senthil}(2015)}]{Senthil15}%
  \BibitemOpen
  \bibfield  {author} {\bibinfo {author} {\bibfnamefont {T.}~\bibnamefont
  {Senthil}},\ }\bibfield  {title} {\enquote {\bibinfo {title}
  {{Symmetry-Protected Topological Phases of Quantum Matter}},}\ }\href
  {\doibase 10.1146/annurev-conmatphys-031214-014740} {\bibfield  {journal}
  {\bibinfo  {journal} {Annual Review of Condensed Matter Physics}\ }\textbf
  {\bibinfo {volume} {6}},\ \bibinfo {pages} {299--324} (\bibinfo {year}
  {2015})}\BibitemShut {NoStop}%
\bibitem [{\citenamefont {Affleck}\ \emph {et~al.}(1988)\citenamefont
  {Affleck}, \citenamefont {Kennedy}, \citenamefont {Lieb},\ and\ \citenamefont
  {Tasaki}}]{Affleck88}%
  \BibitemOpen
  \bibfield  {author} {\bibinfo {author} {\bibfnamefont {Ian}\ \bibnamefont
  {Affleck}}, \bibinfo {author} {\bibfnamefont {Tom}\ \bibnamefont {Kennedy}},
  \bibinfo {author} {\bibfnamefont {Elliott~H.}\ \bibnamefont {Lieb}}, \ and\
  \bibinfo {author} {\bibfnamefont {Hal}\ \bibnamefont {Tasaki}},\ }\bibfield
  {title} {\enquote {\bibinfo {title} {Valence bond ground states in isotropic
  quantum antiferromagnets},}\ }\href
  {https://projecteuclid.org:443/euclid.cmp/1104161001} {\bibfield  {journal}
  {\bibinfo  {journal} {Comm. Math. Phys.}\ }\textbf {\bibinfo {volume}
  {115}},\ \bibinfo {pages} {477--528} (\bibinfo {year} {1988})}\BibitemShut
  {NoStop}%
\bibitem [{\citenamefont {Kennedy}(1990)}]{Kennedy90}%
  \BibitemOpen
  \bibfield  {author} {\bibinfo {author} {\bibfnamefont {T}~\bibnamefont
  {Kennedy}},\ }\bibfield  {title} {\enquote {\bibinfo {title} {Exact
  diagonalisations of open spin-1 chains},}\ }\href {\doibase
  10.1088/0953-8984/2/26/010} {\bibfield  {journal} {\bibinfo  {journal}
  {Journal of Physics: Condensed Matter}\ }\textbf {\bibinfo {volume} {2}},\
  \bibinfo {pages} {5737--5745} (\bibinfo {year} {1990})}\BibitemShut {NoStop}%
\bibitem [{\citenamefont {Kestner}\ \emph {et~al.}(2011)\citenamefont
  {Kestner}, \citenamefont {Wang}, \citenamefont {Sau},\ and\ \citenamefont
  {Das~Sarma}}]{Kestner11}%
  \BibitemOpen
  \bibfield  {author} {\bibinfo {author} {\bibfnamefont {J.~P.}\ \bibnamefont
  {Kestner}}, \bibinfo {author} {\bibfnamefont {Bin}\ \bibnamefont {Wang}},
  \bibinfo {author} {\bibfnamefont {Jay~D.}\ \bibnamefont {Sau}}, \ and\
  \bibinfo {author} {\bibfnamefont {S.}~\bibnamefont {Das~Sarma}},\ }\bibfield
  {title} {\enquote {\bibinfo {title} {{Prediction of a gapless topological
  Haldane liquid phase in a one-dimensional cold polar molecular lattice}},}\
  }\href {\doibase 10.1103/PhysRevB.83.174409} {\bibfield  {journal} {\bibinfo
  {journal} {Phys. Rev. B}\ }\textbf {\bibinfo {volume} {83}},\ \bibinfo
  {pages} {174409} (\bibinfo {year} {2011})}\BibitemShut {NoStop}%
\bibitem [{\citenamefont {Cheng}\ and\ \citenamefont {Tu}(2011)}]{Cheng11}%
  \BibitemOpen
  \bibfield  {author} {\bibinfo {author} {\bibfnamefont {Meng}\ \bibnamefont
  {Cheng}}\ and\ \bibinfo {author} {\bibfnamefont {Hong-Hao}\ \bibnamefont
  {Tu}},\ }\bibfield  {title} {\enquote {\bibinfo {title} {Majorana edge states
  in interacting two-chain ladders of fermions},}\ }\href {\doibase
  10.1103/PhysRevB.84.094503} {\bibfield  {journal} {\bibinfo  {journal} {Phys.
  Rev. B}\ }\textbf {\bibinfo {volume} {84}},\ \bibinfo {pages} {094503}
  (\bibinfo {year} {2011})}\BibitemShut {NoStop}%
\bibitem [{\citenamefont {Fidkowski}\ \emph {et~al.}(2011)\citenamefont
  {Fidkowski}, \citenamefont {Lutchyn}, \citenamefont {Nayak},\ and\
  \citenamefont {Fisher}}]{Fidkowski11longrange}%
  \BibitemOpen
  \bibfield  {author} {\bibinfo {author} {\bibfnamefont {Lukasz}\ \bibnamefont
  {Fidkowski}}, \bibinfo {author} {\bibfnamefont {Roman~M.}\ \bibnamefont
  {Lutchyn}}, \bibinfo {author} {\bibfnamefont {Chetan}\ \bibnamefont {Nayak}},
  \ and\ \bibinfo {author} {\bibfnamefont {Matthew P.~A.}\ \bibnamefont
  {Fisher}},\ }\bibfield  {title} {\enquote {\bibinfo {title} {Majorana zero
  modes in one-dimensional quantum wires without long-ranged superconducting
  order},}\ }\href {\doibase 10.1103/PhysRevB.84.195436} {\bibfield  {journal}
  {\bibinfo  {journal} {Phys. Rev. B}\ }\textbf {\bibinfo {volume} {84}},\
  \bibinfo {pages} {195436} (\bibinfo {year} {2011})}\BibitemShut {NoStop}%
\bibitem [{\citenamefont {Sau}\ \emph {et~al.}(2011)\citenamefont {Sau},
  \citenamefont {Halperin}, \citenamefont {Flensberg},\ and\ \citenamefont
  {Das~Sarma}}]{Sau11}%
  \BibitemOpen
  \bibfield  {author} {\bibinfo {author} {\bibfnamefont {Jay~D.}\ \bibnamefont
  {Sau}}, \bibinfo {author} {\bibfnamefont {B.~I.}\ \bibnamefont {Halperin}},
  \bibinfo {author} {\bibfnamefont {K.}~\bibnamefont {Flensberg}}, \ and\
  \bibinfo {author} {\bibfnamefont {S.}~\bibnamefont {Das~Sarma}},\ }\bibfield
  {title} {\enquote {\bibinfo {title} {Number conserving theory for
  topologically protected degeneracy in one-dimensional fermions},}\ }\href
  {\doibase 10.1103/PhysRevB.84.144509} {\bibfield  {journal} {\bibinfo
  {journal} {Phys. Rev. B}\ }\textbf {\bibinfo {volume} {84}},\ \bibinfo
  {pages} {144509} (\bibinfo {year} {2011})}\BibitemShut {NoStop}%
\bibitem [{\citenamefont {Ruhman}\ \emph {et~al.}(2012)\citenamefont {Ruhman},
  \citenamefont {Dalla~Torre}, \citenamefont {Huber},\ and\ \citenamefont
  {Altman}}]{Ruhman12}%
  \BibitemOpen
  \bibfield  {author} {\bibinfo {author} {\bibfnamefont {J.}~\bibnamefont
  {Ruhman}}, \bibinfo {author} {\bibfnamefont {E.~G.}\ \bibnamefont
  {Dalla~Torre}}, \bibinfo {author} {\bibfnamefont {S.~D.}\ \bibnamefont
  {Huber}}, \ and\ \bibinfo {author} {\bibfnamefont {E.}~\bibnamefont
  {Altman}},\ }\bibfield  {title} {\enquote {\bibinfo {title} {Nonlocal order
  in elongated dipolar gases},}\ }\href {\doibase 10.1103/PhysRevB.85.125121}
  {\bibfield  {journal} {\bibinfo  {journal} {Phys. Rev. B}\ }\textbf {\bibinfo
  {volume} {85}},\ \bibinfo {pages} {125121} (\bibinfo {year}
  {2012})}\BibitemShut {NoStop}%
\bibitem [{\citenamefont {{Grover}}\ and\ \citenamefont
  {{Vishwanath}}(2012)}]{Grover12}%
  \BibitemOpen
  \bibfield  {author} {\bibinfo {author} {\bibfnamefont {Tarun}\ \bibnamefont
  {{Grover}}}\ and\ \bibinfo {author} {\bibfnamefont {Ashvin}\ \bibnamefont
  {{Vishwanath}}},\ }\bibfield  {title} {\enquote {\bibinfo {title} {{Quantum
  Criticality in Topological Insulators and Superconductors: Emergence of
  Strongly Coupled Majoranas and Supersymmetry}},}\ }\href@noop {} {\bibfield
  {journal} {\bibinfo  {journal} {arXiv e-prints}\ ,\ \bibinfo {eid}
  {arXiv:1206.1332}} (\bibinfo {year} {2012})},\ \Eprint
  {http://arxiv.org/abs/1206.1332} {arXiv:1206.1332 [cond-mat.str-el]}
  \BibitemShut {NoStop}%
\bibitem [{\citenamefont {Kraus}\ \emph {et~al.}(2013)\citenamefont {Kraus},
  \citenamefont {Dalmonte}, \citenamefont {Baranov}, \citenamefont
  {L\"auchli},\ and\ \citenamefont {Zoller}}]{Kraus13}%
  \BibitemOpen
  \bibfield  {author} {\bibinfo {author} {\bibfnamefont {Christina~V.}\
  \bibnamefont {Kraus}}, \bibinfo {author} {\bibfnamefont {Marcello}\
  \bibnamefont {Dalmonte}}, \bibinfo {author} {\bibfnamefont {Mikhail~A.}\
  \bibnamefont {Baranov}}, \bibinfo {author} {\bibfnamefont {Andreas~M.}\
  \bibnamefont {L\"auchli}}, \ and\ \bibinfo {author} {\bibfnamefont
  {P.}~\bibnamefont {Zoller}},\ }\bibfield  {title} {\enquote {\bibinfo {title}
  {{Majorana Edge States in Atomic Wires Coupled by Pair Hopping}},}\ }\href
  {\doibase 10.1103/PhysRevLett.111.173004} {\bibfield  {journal} {\bibinfo
  {journal} {Phys. Rev. Lett.}\ }\textbf {\bibinfo {volume} {111}},\ \bibinfo
  {pages} {173004} (\bibinfo {year} {2013})}\BibitemShut {NoStop}%
\bibitem [{\citenamefont {Ortiz}\ \emph {et~al.}(2014)\citenamefont {Ortiz},
  \citenamefont {Dukelsky}, \citenamefont {Cobanera}, \citenamefont {Esebbag},\
  and\ \citenamefont {Beenakker}}]{Ortiz14}%
  \BibitemOpen
  \bibfield  {author} {\bibinfo {author} {\bibfnamefont {Gerardo}\ \bibnamefont
  {Ortiz}}, \bibinfo {author} {\bibfnamefont {Jorge}\ \bibnamefont {Dukelsky}},
  \bibinfo {author} {\bibfnamefont {Emilio}\ \bibnamefont {Cobanera}}, \bibinfo
  {author} {\bibfnamefont {Carlos}\ \bibnamefont {Esebbag}}, \ and\ \bibinfo
  {author} {\bibfnamefont {Carlo}\ \bibnamefont {Beenakker}},\ }\bibfield
  {title} {\enquote {\bibinfo {title} {{Many-Body Characterization of
  Particle-Conserving Topological Superfluids}},}\ }\href {\doibase
  10.1103/PhysRevLett.113.267002} {\bibfield  {journal} {\bibinfo  {journal}
  {Phys. Rev. Lett.}\ }\textbf {\bibinfo {volume} {113}},\ \bibinfo {pages}
  {267002} (\bibinfo {year} {2014})}\BibitemShut {NoStop}%
\bibitem [{\citenamefont {Keselman}\ and\ \citenamefont
  {Berg}(2015)}]{Keselman15}%
  \BibitemOpen
  \bibfield  {author} {\bibinfo {author} {\bibfnamefont {Anna}\ \bibnamefont
  {Keselman}}\ and\ \bibinfo {author} {\bibfnamefont {Erez}\ \bibnamefont
  {Berg}},\ }\bibfield  {title} {\enquote {\bibinfo {title} {Gapless
  symmetry-protected topological phase of fermions in one dimension},}\ }\href
  {\doibase 10.1103/PhysRevB.91.235309} {\bibfield  {journal} {\bibinfo
  {journal} {Phys. Rev. B}\ }\textbf {\bibinfo {volume} {91}},\ \bibinfo
  {pages} {235309} (\bibinfo {year} {2015})}\BibitemShut {NoStop}%
\bibitem [{\citenamefont {Ruhman}\ \emph {et~al.}(2015)\citenamefont {Ruhman},
  \citenamefont {Berg},\ and\ \citenamefont {Altman}}]{Ruhman15}%
  \BibitemOpen
  \bibfield  {author} {\bibinfo {author} {\bibfnamefont {Jonathan}\
  \bibnamefont {Ruhman}}, \bibinfo {author} {\bibfnamefont {Erez}\ \bibnamefont
  {Berg}}, \ and\ \bibinfo {author} {\bibfnamefont {Ehud}\ \bibnamefont
  {Altman}},\ }\bibfield  {title} {\enquote {\bibinfo {title} {{Topological
  States in a One-Dimensional Fermi Gas with Attractive Interaction}},}\ }\href
  {\doibase 10.1103/PhysRevLett.114.100401} {\bibfield  {journal} {\bibinfo
  {journal} {Phys. Rev. Lett.}\ }\textbf {\bibinfo {volume} {114}},\ \bibinfo
  {pages} {100401} (\bibinfo {year} {2015})}\BibitemShut {NoStop}%
\bibitem [{\citenamefont {Kainaris}\ and\ \citenamefont
  {Carr}(2015)}]{Kainaris15}%
  \BibitemOpen
  \bibfield  {author} {\bibinfo {author} {\bibfnamefont {Nikolaos}\
  \bibnamefont {Kainaris}}\ and\ \bibinfo {author} {\bibfnamefont {Sam~T.}\
  \bibnamefont {Carr}},\ }\bibfield  {title} {\enquote {\bibinfo {title}
  {Emergent topological properties in interacting one-dimensional systems with
  spin-orbit coupling},}\ }\href {\doibase 10.1103/PhysRevB.92.035139}
  {\bibfield  {journal} {\bibinfo  {journal} {Phys. Rev. B}\ }\textbf {\bibinfo
  {volume} {92}},\ \bibinfo {pages} {035139} (\bibinfo {year}
  {2015})}\BibitemShut {NoStop}%
\bibitem [{\citenamefont {Iemini}\ \emph {et~al.}(2015)\citenamefont {Iemini},
  \citenamefont {Mazza}, \citenamefont {Rossini}, \citenamefont {Fazio},\ and\
  \citenamefont {Diehl}}]{Iemini15}%
  \BibitemOpen
  \bibfield  {author} {\bibinfo {author} {\bibfnamefont {Fernando}\
  \bibnamefont {Iemini}}, \bibinfo {author} {\bibfnamefont {Leonardo}\
  \bibnamefont {Mazza}}, \bibinfo {author} {\bibfnamefont {Davide}\
  \bibnamefont {Rossini}}, \bibinfo {author} {\bibfnamefont {Rosario}\
  \bibnamefont {Fazio}}, \ and\ \bibinfo {author} {\bibfnamefont {Sebastian}\
  \bibnamefont {Diehl}},\ }\bibfield  {title} {\enquote {\bibinfo {title}
  {{Localized Majorana-Like Modes in a Number-Conserving Setting: An Exactly
  Solvable Model}},}\ }\href {\doibase 10.1103/PhysRevLett.115.156402}
  {\bibfield  {journal} {\bibinfo  {journal} {Phys. Rev. Lett.}\ }\textbf
  {\bibinfo {volume} {115}},\ \bibinfo {pages} {156402} (\bibinfo {year}
  {2015})}\BibitemShut {NoStop}%
\bibitem [{\citenamefont {Lang}\ and\ \citenamefont
  {B\"uchler}(2015)}]{Lang15}%
  \BibitemOpen
  \bibfield  {author} {\bibinfo {author} {\bibfnamefont {Nicolai}\ \bibnamefont
  {Lang}}\ and\ \bibinfo {author} {\bibfnamefont {Hans~Peter}\ \bibnamefont
  {B\"uchler}},\ }\bibfield  {title} {\enquote {\bibinfo {title} {Topological
  states in a microscopic model of interacting fermions},}\ }\href {\doibase
  10.1103/PhysRevB.92.041118} {\bibfield  {journal} {\bibinfo  {journal} {Phys.
  Rev. B}\ }\textbf {\bibinfo {volume} {92}},\ \bibinfo {pages} {041118(R)}
  (\bibinfo {year} {2015})}\BibitemShut {NoStop}%
\bibitem [{\citenamefont {Ortiz}\ and\ \citenamefont
  {Cobanera}(2016)}]{Ortiz16}%
  \BibitemOpen
  \bibfield  {author} {\bibinfo {author} {\bibfnamefont {Gerardo}\ \bibnamefont
  {Ortiz}}\ and\ \bibinfo {author} {\bibfnamefont {Emilio}\ \bibnamefont
  {Cobanera}},\ }\bibfield  {title} {\enquote {\bibinfo {title} {{What is a
  particle-conserving Topological Superfluid? The fate of Majorana modes beyond
  mean-field theory}},}\ }\href {\doibase
  https://doi.org/10.1016/j.aop.2016.05.020} {\bibfield  {journal} {\bibinfo
  {journal} {Annals of Physics}\ }\textbf {\bibinfo {volume} {372}},\ \bibinfo
  {pages} {357 -- 374} (\bibinfo {year} {2016})}\BibitemShut {NoStop}%
\bibitem [{\citenamefont {Montorsi}\ \emph {et~al.}(2017)\citenamefont
  {Montorsi}, \citenamefont {Dolcini}, \citenamefont {Iotti},\ and\
  \citenamefont {Rossi}}]{Montorsi17}%
  \BibitemOpen
  \bibfield  {author} {\bibinfo {author} {\bibfnamefont {Arianna}\ \bibnamefont
  {Montorsi}}, \bibinfo {author} {\bibfnamefont {Fabrizio}\ \bibnamefont
  {Dolcini}}, \bibinfo {author} {\bibfnamefont {Rita~C.}\ \bibnamefont
  {Iotti}}, \ and\ \bibinfo {author} {\bibfnamefont {Fausto}\ \bibnamefont
  {Rossi}},\ }\bibfield  {title} {\enquote {\bibinfo {title}
  {Symmetry-protected topological phases of one-dimensional interacting
  fermions with spin-charge separation},}\ }\href {\doibase
  10.1103/PhysRevB.95.245108} {\bibfield  {journal} {\bibinfo  {journal} {Phys.
  Rev. B}\ }\textbf {\bibinfo {volume} {95}},\ \bibinfo {pages} {245108}
  (\bibinfo {year} {2017})}\BibitemShut {NoStop}%
\bibitem [{\citenamefont {Wang}\ \emph {et~al.}(2017)\citenamefont {Wang},
  \citenamefont {Xu}, \citenamefont {Pu},\ and\ \citenamefont
  {Hazzard}}]{Wang17}%
  \BibitemOpen
  \bibfield  {author} {\bibinfo {author} {\bibfnamefont {Zhiyuan}\ \bibnamefont
  {Wang}}, \bibinfo {author} {\bibfnamefont {Youjiang}\ \bibnamefont {Xu}},
  \bibinfo {author} {\bibfnamefont {Han}\ \bibnamefont {Pu}}, \ and\ \bibinfo
  {author} {\bibfnamefont {Kaden R.~A.}\ \bibnamefont {Hazzard}},\ }\bibfield
  {title} {\enquote {\bibinfo {title} {Number-conserving interacting fermion
  models with exact topological superconducting ground states},}\ }\href
  {\doibase 10.1103/PhysRevB.96.115110} {\bibfield  {journal} {\bibinfo
  {journal} {Phys. Rev. B}\ }\textbf {\bibinfo {volume} {96}},\ \bibinfo
  {pages} {115110} (\bibinfo {year} {2017})}\BibitemShut {NoStop}%
\bibitem [{\citenamefont {Ruhman}\ and\ \citenamefont
  {Altman}(2017)}]{Ruhman17}%
  \BibitemOpen
  \bibfield  {author} {\bibinfo {author} {\bibfnamefont {Jonathan}\
  \bibnamefont {Ruhman}}\ and\ \bibinfo {author} {\bibfnamefont {Ehud}\
  \bibnamefont {Altman}},\ }\bibfield  {title} {\enquote {\bibinfo {title}
  {Topological degeneracy and pairing in a one-dimensional gas of spinless
  fermions},}\ }\href {\doibase 10.1103/PhysRevB.96.085133} {\bibfield
  {journal} {\bibinfo  {journal} {Phys. Rev. B}\ }\textbf {\bibinfo {volume}
  {96}},\ \bibinfo {pages} {085133} (\bibinfo {year} {2017})}\BibitemShut
  {NoStop}%
\bibitem [{\citenamefont {Scaffidi}\ \emph {et~al.}(2017)\citenamefont
  {Scaffidi}, \citenamefont {Parker},\ and\ \citenamefont
  {Vasseur}}]{Scaffidi17}%
  \BibitemOpen
  \bibfield  {author} {\bibinfo {author} {\bibfnamefont {Thomas}\ \bibnamefont
  {Scaffidi}}, \bibinfo {author} {\bibfnamefont {Daniel~E.}\ \bibnamefont
  {Parker}}, \ and\ \bibinfo {author} {\bibfnamefont {Romain}\ \bibnamefont
  {Vasseur}},\ }\bibfield  {title} {\enquote {\bibinfo {title} {{Gapless
  Symmetry-Protected Topological Order}},}\ }\href {\doibase
  10.1103/PhysRevX.7.041048} {\bibfield  {journal} {\bibinfo  {journal} {Phys.
  Rev. X}\ }\textbf {\bibinfo {volume} {7}},\ \bibinfo {pages} {041048}
  (\bibinfo {year} {2017})}\BibitemShut {NoStop}%
\bibitem [{\citenamefont {Guther}\ \emph {et~al.}(2017)\citenamefont {Guther},
  \citenamefont {Lang},\ and\ \citenamefont {B\"uchler}}]{Guther17}%
  \BibitemOpen
  \bibfield  {author} {\bibinfo {author} {\bibfnamefont {K.}~\bibnamefont
  {Guther}}, \bibinfo {author} {\bibfnamefont {N.}~\bibnamefont {Lang}}, \ and\
  \bibinfo {author} {\bibfnamefont {H.~P.}\ \bibnamefont {B\"uchler}},\
  }\bibfield  {title} {\enquote {\bibinfo {title} {{Ising anyonic topological
  phase of interacting fermions in one dimension}},}\ }\href {\doibase
  10.1103/PhysRevB.96.121109} {\bibfield  {journal} {\bibinfo  {journal} {Phys.
  Rev. B}\ }\textbf {\bibinfo {volume} {96}},\ \bibinfo {pages} {121109}
  (\bibinfo {year} {2017})}\BibitemShut {NoStop}%
\bibitem [{\citenamefont {Kainaris}\ \emph {et~al.}(2017)\citenamefont
  {Kainaris}, \citenamefont {Santos}, \citenamefont {Gutman},\ and\
  \citenamefont {Carr}}]{Kainaris17}%
  \BibitemOpen
  \bibfield  {author} {\bibinfo {author} {\bibfnamefont {Nikolaos}\
  \bibnamefont {Kainaris}}, \bibinfo {author} {\bibfnamefont {Raul~A.}\
  \bibnamefont {Santos}}, \bibinfo {author} {\bibfnamefont {D.~B.}\
  \bibnamefont {Gutman}}, \ and\ \bibinfo {author} {\bibfnamefont {Sam~T.}\
  \bibnamefont {Carr}},\ }\bibfield  {title} {\enquote {\bibinfo {title}
  {Interaction induced topological protection in one-dimensional conductors},}\
  }\href {\doibase 10.1002/prop.201600054} {\bibfield  {journal} {\bibinfo
  {journal} {Fortschritte der Physik}\ }\textbf {\bibinfo {volume} {65}},\
  \bibinfo {pages} {1600054} (\bibinfo {year} {2017})}\BibitemShut {NoStop}%
\bibitem [{\citenamefont {Jiang}\ \emph {et~al.}(2018)\citenamefont {Jiang},
  \citenamefont {Li}, \citenamefont {Seidel},\ and\ \citenamefont
  {Lee}}]{Jiang18}%
  \BibitemOpen
  \bibfield  {author} {\bibinfo {author} {\bibfnamefont {Hong-Chen}\
  \bibnamefont {Jiang}}, \bibinfo {author} {\bibfnamefont {Zi-Xiang}\
  \bibnamefont {Li}}, \bibinfo {author} {\bibfnamefont {Alexander}\
  \bibnamefont {Seidel}}, \ and\ \bibinfo {author} {\bibfnamefont {Dung-Hai}\
  \bibnamefont {Lee}},\ }\bibfield  {title} {\enquote {\bibinfo {title}
  {{Symmetry protected topological Luttinger liquids and the phase transition
  between them}},}\ }\href {\doibase
  https://doi.org/10.1016/j.scib.2018.05.010} {\bibfield  {journal} {\bibinfo
  {journal} {Science Bulletin}\ }\textbf {\bibinfo {volume} {63}},\ \bibinfo
  {pages} {753 -- 758} (\bibinfo {year} {2018})}\BibitemShut {NoStop}%
\bibitem [{\citenamefont {Zhang}\ and\ \citenamefont {Liu}(2018)}]{Zhang18}%
  \BibitemOpen
  \bibfield  {author} {\bibinfo {author} {\bibfnamefont {Rui-Xing}\
  \bibnamefont {Zhang}}\ and\ \bibinfo {author} {\bibfnamefont {Chao-Xing}\
  \bibnamefont {Liu}},\ }\bibfield  {title} {\enquote {\bibinfo {title}
  {{Crystalline Symmetry-Protected Majorana Mode in Number-Conserving Dirac
  Semimetal Nanowires}},}\ }\href {\doibase 10.1103/PhysRevLett.120.156802}
  {\bibfield  {journal} {\bibinfo  {journal} {Phys. Rev. Lett.}\ }\textbf
  {\bibinfo {volume} {120}},\ \bibinfo {pages} {156802} (\bibinfo {year}
  {2018})}\BibitemShut {NoStop}%
\bibitem [{\citenamefont {Verresen}\ \emph {et~al.}(2018)\citenamefont
  {Verresen}, \citenamefont {Jones},\ and\ \citenamefont
  {Pollmann}}]{Verresen18}%
  \BibitemOpen
  \bibfield  {author} {\bibinfo {author} {\bibfnamefont {Ruben}\ \bibnamefont
  {Verresen}}, \bibinfo {author} {\bibfnamefont {Nick~G.}\ \bibnamefont
  {Jones}}, \ and\ \bibinfo {author} {\bibfnamefont {Frank}\ \bibnamefont
  {Pollmann}},\ }\bibfield  {title} {\enquote {\bibinfo {title} {{Topology and
  Edge Modes in Quantum Critical Chains}},}\ }\href {\doibase
  10.1103/PhysRevLett.120.057001} {\bibfield  {journal} {\bibinfo  {journal}
  {Phys. Rev. Lett.}\ }\textbf {\bibinfo {volume} {120}},\ \bibinfo {pages}
  {057001} (\bibinfo {year} {2018})}\BibitemShut {NoStop}%
\bibitem [{\citenamefont {Parker}\ \emph {et~al.}(2018)\citenamefont {Parker},
  \citenamefont {Scaffidi},\ and\ \citenamefont {Vasseur}}]{Parker18}%
  \BibitemOpen
  \bibfield  {author} {\bibinfo {author} {\bibfnamefont {Daniel~E.}\
  \bibnamefont {Parker}}, \bibinfo {author} {\bibfnamefont {Thomas}\
  \bibnamefont {Scaffidi}}, \ and\ \bibinfo {author} {\bibfnamefont {Romain}\
  \bibnamefont {Vasseur}},\ }\bibfield  {title} {\enquote {\bibinfo {title}
  {{Topological Luttinger liquids from decorated domain walls}},}\ }\href
  {\doibase 10.1103/PhysRevB.97.165114} {\bibfield  {journal} {\bibinfo
  {journal} {Phys. Rev. B}\ }\textbf {\bibinfo {volume} {97}},\ \bibinfo
  {pages} {165114} (\bibinfo {year} {2018})}\BibitemShut {NoStop}%
\bibitem [{\citenamefont {Keselman}\ \emph {et~al.}(2018)\citenamefont
  {Keselman}, \citenamefont {Berg},\ and\ \citenamefont {Azaria}}]{Keselman18}%
  \BibitemOpen
  \bibfield  {author} {\bibinfo {author} {\bibfnamefont {Anna}\ \bibnamefont
  {Keselman}}, \bibinfo {author} {\bibfnamefont {Erez}\ \bibnamefont {Berg}}, \
  and\ \bibinfo {author} {\bibfnamefont {Patrick}\ \bibnamefont {Azaria}},\
  }\bibfield  {title} {\enquote {\bibinfo {title} {{From one-dimensional charge
  conserving superconductors to the gapless Haldane phase}},}\ }\href {\doibase
  10.1103/PhysRevB.98.214501} {\bibfield  {journal} {\bibinfo  {journal} {Phys.
  Rev. B}\ }\textbf {\bibinfo {volume} {98}},\ \bibinfo {pages} {214501}
  (\bibinfo {year} {2018})}\BibitemShut {NoStop}%
\bibitem [{\citenamefont {Chen}\ \emph {et~al.}(2018)\citenamefont {Chen},
  \citenamefont {Yan}, \citenamefont {Ting}, \citenamefont {Chen},\ and\
  \citenamefont {Burnell}}]{Chen18}%
  \BibitemOpen
  \bibfield  {author} {\bibinfo {author} {\bibfnamefont {Chun}\ \bibnamefont
  {Chen}}, \bibinfo {author} {\bibfnamefont {Wei}\ \bibnamefont {Yan}},
  \bibinfo {author} {\bibfnamefont {C.~S.}\ \bibnamefont {Ting}}, \bibinfo
  {author} {\bibfnamefont {Yan}\ \bibnamefont {Chen}}, \ and\ \bibinfo {author}
  {\bibfnamefont {F.~J.}\ \bibnamefont {Burnell}},\ }\bibfield  {title}
  {\enquote {\bibinfo {title} {{Flux-stabilized Majorana zero modes in coupled
  one-dimensional Fermi wires}},}\ }\href {\doibase 10.1103/PhysRevB.98.161106}
  {\bibfield  {journal} {\bibinfo  {journal} {Phys. Rev. B}\ }\textbf {\bibinfo
  {volume} {98}},\ \bibinfo {pages} {161106} (\bibinfo {year}
  {2018})}\BibitemShut {NoStop}%
\bibitem [{\citenamefont {Haldane}(1983)}]{Haldane83}%
  \BibitemOpen
  \bibfield  {author} {\bibinfo {author} {\bibfnamefont {F.~D.~M.}\
  \bibnamefont {Haldane}},\ }\bibfield  {title} {\enquote {\bibinfo {title}
  {{Continuum dynamics of the 1-D Heisenberg antiferromagnet: Identification
  with the $O(3)$ nonlinear sigma model}},}\ }\href {\doibase
  10.1016/0375-9601(83)90631-X} {\bibfield  {journal} {\bibinfo  {journal}
  {Phys. Lett. A}\ }\textbf {\bibinfo {volume} {93}},\ \bibinfo {pages} {464}
  (\bibinfo {year} {1983})}\BibitemShut {NoStop}%
\bibitem [{\citenamefont {Kennedy}\ and\ \citenamefont
  {Tasaki}(1992)}]{Kennedy92}%
  \BibitemOpen
  \bibfield  {author} {\bibinfo {author} {\bibfnamefont {Tom}\ \bibnamefont
  {Kennedy}}\ and\ \bibinfo {author} {\bibfnamefont {Hal}\ \bibnamefont
  {Tasaki}},\ }\bibfield  {title} {\enquote {\bibinfo {title} {{Hidden symmetry
  breaking and the Haldane phase in $S=1$ quantum spin chains}},}\ }\href
  {https://projecteuclid.org:443/euclid.cmp/1104250747} {\bibfield  {journal}
  {\bibinfo  {journal} {Comm. Math. Phys.}\ }\textbf {\bibinfo {volume}
  {147}},\ \bibinfo {pages} {431--484} (\bibinfo {year} {1992})}\BibitemShut
  {NoStop}%
\bibitem [{\citenamefont {Gu}\ and\ \citenamefont {Wen}(2009)}]{Gu09}%
  \BibitemOpen
  \bibfield  {author} {\bibinfo {author} {\bibfnamefont {Zheng-Cheng}\
  \bibnamefont {Gu}}\ and\ \bibinfo {author} {\bibfnamefont {Xiao-Gang}\
  \bibnamefont {Wen}},\ }\bibfield  {title} {\enquote {\bibinfo {title}
  {Tensor-entanglement-filtering renormalization approach and
  symmetry-protected topological order},}\ }\href {\doibase
  10.1103/PhysRevB.80.155131} {\bibfield  {journal} {\bibinfo  {journal} {Phys.
  Rev. B}\ }\textbf {\bibinfo {volume} {80}},\ \bibinfo {pages} {155131}
  (\bibinfo {year} {2009})}\BibitemShut {NoStop}%
\bibitem [{\citenamefont {Pollmann}\ \emph {et~al.}(2010)\citenamefont
  {Pollmann}, \citenamefont {Turner}, \citenamefont {Berg},\ and\ \citenamefont
  {Oshikawa}}]{Pollmann10}%
  \BibitemOpen
  \bibfield  {author} {\bibinfo {author} {\bibfnamefont {Frank}\ \bibnamefont
  {Pollmann}}, \bibinfo {author} {\bibfnamefont {Ari~M.}\ \bibnamefont
  {Turner}}, \bibinfo {author} {\bibfnamefont {Erez}\ \bibnamefont {Berg}}, \
  and\ \bibinfo {author} {\bibfnamefont {Masaki}\ \bibnamefont {Oshikawa}},\
  }\bibfield  {title} {\enquote {\bibinfo {title} {Entanglement spectrum of a
  topological phase in one dimension},}\ }\href {\doibase
  10.1103/PhysRevB.81.064439} {\bibfield  {journal} {\bibinfo  {journal} {Phys.
  Rev. B}\ }\textbf {\bibinfo {volume} {81}},\ \bibinfo {pages} {064439}
  (\bibinfo {year} {2010})}\BibitemShut {NoStop}%
\bibitem [{\citenamefont {Chen}\ \emph {et~al.}(2003)\citenamefont {Chen},
  \citenamefont {Hida},\ and\ \citenamefont {Sanctuary}}]{Chen03}%
  \BibitemOpen
  \bibfield  {author} {\bibinfo {author} {\bibfnamefont {Wei}\ \bibnamefont
  {Chen}}, \bibinfo {author} {\bibfnamefont {Kazuo}\ \bibnamefont {Hida}}, \
  and\ \bibinfo {author} {\bibfnamefont {B.~C.}\ \bibnamefont {Sanctuary}},\
  }\bibfield  {title} {\enquote {\bibinfo {title} {{Ground-state phase diagram
  of $S=1$ $\mathrm{XXZ}$ chains with uniaxial single-ion-type anisotropy}},}\
  }\href {\doibase 10.1103/PhysRevB.67.104401} {\bibfield  {journal} {\bibinfo
  {journal} {Phys. Rev. B}\ }\textbf {\bibinfo {volume} {67}},\ \bibinfo
  {pages} {104401} (\bibinfo {year} {2003})}\BibitemShut {NoStop}%
\bibitem [{\citenamefont {Vafa}(1986)}]{Vafa86}%
  \BibitemOpen
  \bibfield  {author} {\bibinfo {author} {\bibfnamefont {Cumrun}\ \bibnamefont
  {Vafa}},\ }\bibfield  {title} {\enquote {\bibinfo {title} {Modular invariance
  and discrete torsion on orbifolds},}\ }\href {\doibase
  https://doi.org/10.1016/0550-3213(86)90379-2} {\bibfield  {journal} {\bibinfo
   {journal} {Nuclear Physics B}\ }\textbf {\bibinfo {volume} {273}},\ \bibinfo
  {pages} {592--606} (\bibinfo {year} {1986})}\BibitemShut {NoStop}%
\bibitem [{\citenamefont {Vafa}\ and\ \citenamefont {Witten}(1995)}]{Vafa95}%
  \BibitemOpen
  \bibfield  {author} {\bibinfo {author} {\bibfnamefont {Cumrun}\ \bibnamefont
  {Vafa}}\ and\ \bibinfo {author} {\bibfnamefont {Edward}\ \bibnamefont
  {Witten}},\ }\bibfield  {title} {\enquote {\bibinfo {title} {On orbifolds
  with discrete torsion},}\ }\href
  {http://www.sciencedirect.com/science/article/pii/0393044094000489}
  {\bibfield  {journal} {\bibinfo  {journal} {Journal of Geometry and Physics}\
  }\textbf {\bibinfo {volume} {15}},\ \bibinfo {pages} {189 -- 214} (\bibinfo
  {year} {1995})}\BibitemShut {NoStop}%
\bibitem [{\citenamefont {{Douglas}}(1998)}]{Douglas98}%
  \BibitemOpen
  \bibfield  {author} {\bibinfo {author} {\bibfnamefont {Michael~R.}\
  \bibnamefont {{Douglas}}},\ }\bibfield  {title} {\enquote {\bibinfo {title}
  {{D-branes and Discrete Torsion}},}\ }\href@noop {} {\bibfield  {journal}
  {\bibinfo  {journal} {arXiv e-prints}\ ,\ \bibinfo {eid} {hep-th/9807235}}
  (\bibinfo {year} {1998})},\ \Eprint {http://arxiv.org/abs/hep-th/9807235}
  {arXiv:hep-th/9807235 [hep-th]} \BibitemShut {NoStop}%
\bibitem [{\citenamefont {Sharpe}(2003)}]{Sharpe03}%
  \BibitemOpen
  \bibfield  {author} {\bibinfo {author} {\bibfnamefont {Eric}\ \bibnamefont
  {Sharpe}},\ }\bibfield  {title} {\enquote {\bibinfo {title} {Discrete
  torsion},}\ }\href {\doibase 10.1103/PhysRevD.68.126003} {\bibfield
  {journal} {\bibinfo  {journal} {Phys. Rev. D}\ }\textbf {\bibinfo {volume}
  {68}},\ \bibinfo {pages} {126003} (\bibinfo {year} {2003})}\BibitemShut
  {NoStop}%
\bibitem [{\citenamefont {Senthil}\ \emph {et~al.}(2004)\citenamefont
  {Senthil}, \citenamefont {Vishwanath}, \citenamefont {Balents}, \citenamefont
  {Sachdev},\ and\ \citenamefont {Fisher}}]{Senthil04}%
  \BibitemOpen
  \bibfield  {author} {\bibinfo {author} {\bibfnamefont {T.}~\bibnamefont
  {Senthil}}, \bibinfo {author} {\bibfnamefont {Ashvin}\ \bibnamefont
  {Vishwanath}}, \bibinfo {author} {\bibfnamefont {Leon}\ \bibnamefont
  {Balents}}, \bibinfo {author} {\bibfnamefont {Subir}\ \bibnamefont
  {Sachdev}}, \ and\ \bibinfo {author} {\bibfnamefont {Matthew P.~A.}\
  \bibnamefont {Fisher}},\ }\bibfield  {title} {\enquote {\bibinfo {title}
  {{Deconfined Quantum Critical Points}},}\ }\href {\doibase
  10.1126/science.1091806} {\bibfield  {journal} {\bibinfo  {journal}
  {Science}\ }\textbf {\bibinfo {volume} {303}},\ \bibinfo {pages} {1490--1494}
  (\bibinfo {year} {2004})},\ \Eprint
  {http://arxiv.org/abs/https://science.sciencemag.org/content/303/5663/1490.full.pdf}
  {https://science.sciencemag.org/content/303/5663/1490.full.pdf} \BibitemShut
  {NoStop}%
\bibitem [{\citenamefont {Jiang}\ and\ \citenamefont
  {Motrunich}(2019)}]{Jiang19}%
  \BibitemOpen
  \bibfield  {author} {\bibinfo {author} {\bibfnamefont {Shenghan}\
  \bibnamefont {Jiang}}\ and\ \bibinfo {author} {\bibfnamefont {Olexei}\
  \bibnamefont {Motrunich}},\ }\bibfield  {title} {\enquote {\bibinfo {title}
  {{Ising ferromagnet to valence bond solid transition in a one-dimensional
  spin chain: Analogies to deconfined quantum critical points}},}\ }\href
  {\doibase 10.1103/PhysRevB.99.075103} {\bibfield  {journal} {\bibinfo
  {journal} {Phys. Rev. B}\ }\textbf {\bibinfo {volume} {99}},\ \bibinfo
  {pages} {075103} (\bibinfo {year} {2019})}\BibitemShut {NoStop}%
\bibitem [{\citenamefont {Roberts}\ \emph {et~al.}(2019)\citenamefont
  {Roberts}, \citenamefont {Jiang},\ and\ \citenamefont
  {Motrunich}}]{Roberts19}%
  \BibitemOpen
  \bibfield  {author} {\bibinfo {author} {\bibfnamefont {Brenden}\ \bibnamefont
  {Roberts}}, \bibinfo {author} {\bibfnamefont {Shenghan}\ \bibnamefont
  {Jiang}}, \ and\ \bibinfo {author} {\bibfnamefont {Olexei~I.}\ \bibnamefont
  {Motrunich}},\ }\bibfield  {title} {\enquote {\bibinfo {title} {Deconfined
  quantum critical point in one dimension},}\ }\href {\doibase
  10.1103/PhysRevB.99.165143} {\bibfield  {journal} {\bibinfo  {journal} {Phys.
  Rev. B}\ }\textbf {\bibinfo {volume} {99}},\ \bibinfo {pages} {165143}
  (\bibinfo {year} {2019})}\BibitemShut {NoStop}%
\bibitem [{\citenamefont {Pollmann}\ and\ \citenamefont
  {Turner}(2012)}]{Pollmann12b}%
  \BibitemOpen
  \bibfield  {author} {\bibinfo {author} {\bibfnamefont {Frank}\ \bibnamefont
  {Pollmann}}\ and\ \bibinfo {author} {\bibfnamefont {Ari~M.}\ \bibnamefont
  {Turner}},\ }\bibfield  {title} {\enquote {\bibinfo {title} {Detection of
  symmetry-protected topological phases in one dimension},}\ }\href {\doibase
  10.1103/PhysRevB.86.125441} {\bibfield  {journal} {\bibinfo  {journal} {Phys.
  Rev. B}\ }\textbf {\bibinfo {volume} {86}},\ \bibinfo {pages} {125441}
  (\bibinfo {year} {2012})}\BibitemShut {NoStop}%
\bibitem [{\citenamefont {Suzuki}(1971)}]{Suzuki71}%
  \BibitemOpen
  \bibfield  {author} {\bibinfo {author} {\bibfnamefont {Masuo}\ \bibnamefont
  {Suzuki}},\ }\bibfield  {title} {\enquote {\bibinfo {title} {{Relationship
  among Exactly Soluble Models of Critical Phenomena. I*)2D Ising Model, Dimer
  Problem and the Generalized XY-Model}},}\ }\href {\doibase
  10.1143/PTP.46.1337} {\bibfield  {journal} {\bibinfo  {journal} {Progress of
  Theoretical Physics}\ }\textbf {\bibinfo {volume} {46}},\ \bibinfo {pages}
  {1337} (\bibinfo {year} {1971})}\BibitemShut {NoStop}%
\bibitem [{\citenamefont {Raussendorf}\ and\ \citenamefont
  {Briegel}(2001)}]{Raussendorf01}%
  \BibitemOpen
  \bibfield  {author} {\bibinfo {author} {\bibfnamefont {Robert}\ \bibnamefont
  {Raussendorf}}\ and\ \bibinfo {author} {\bibfnamefont {Hans~J.}\ \bibnamefont
  {Briegel}},\ }\bibfield  {title} {\enquote {\bibinfo {title} {{A One-Way
  Quantum Computer}},}\ }\href {\doibase 10.1103/PhysRevLett.86.5188}
  {\bibfield  {journal} {\bibinfo  {journal} {Phys. Rev. Lett.}\ }\textbf
  {\bibinfo {volume} {86}},\ \bibinfo {pages} {5188--5191} (\bibinfo {year}
  {2001})}\BibitemShut {NoStop}%
\bibitem [{\citenamefont {Keating}\ and\ \citenamefont
  {Mezzadri}(2004)}]{Keating04}%
  \BibitemOpen
  \bibfield  {author} {\bibinfo {author} {\bibfnamefont {J.P.}\ \bibnamefont
  {Keating}}\ and\ \bibinfo {author} {\bibfnamefont {F.}~\bibnamefont
  {Mezzadri}},\ }\bibfield  {title} {\enquote {\bibinfo {title} {Random matrix
  theory and entanglement in quantum spin chains},}\ }\href {\doibase
  10.1007/s00220-004-1188-2} {\bibfield  {journal} {\bibinfo  {journal}
  {Communications in Mathematical Physics}\ }\textbf {\bibinfo {volume}
  {252}},\ \bibinfo {pages} {543--579} (\bibinfo {year} {2004})}\BibitemShut
  {NoStop}%
\bibitem [{\citenamefont {Son}\ \emph {et~al.}(2011)\citenamefont {Son},
  \citenamefont {Amico}, \citenamefont {Fazio}, \citenamefont {Hamma},
  \citenamefont {Pascazio},\ and\ \citenamefont {Vedral}}]{Son11}%
  \BibitemOpen
  \bibfield  {author} {\bibinfo {author} {\bibfnamefont {W.}~\bibnamefont
  {Son}}, \bibinfo {author} {\bibfnamefont {L.}~\bibnamefont {Amico}}, \bibinfo
  {author} {\bibfnamefont {R.}~\bibnamefont {Fazio}}, \bibinfo {author}
  {\bibfnamefont {A.}~\bibnamefont {Hamma}}, \bibinfo {author} {\bibfnamefont
  {S.}~\bibnamefont {Pascazio}}, \ and\ \bibinfo {author} {\bibfnamefont
  {V.}~\bibnamefont {Vedral}},\ }\bibfield  {title} {\enquote {\bibinfo {title}
  {Quantum phase transition between cluster and antiferromagnetic states},}\
  }\href {\doibase 10.1209/0295-5075/95/50001} {\bibfield  {journal} {\bibinfo
  {journal} {{EPL} (Europhysics Letters)}\ }\textbf {\bibinfo {volume} {95}},\
  \bibinfo {pages} {50001} (\bibinfo {year} {2011})}\BibitemShut {NoStop}%
\bibitem [{\citenamefont {Verresen}\ \emph {et~al.}(2017)\citenamefont
  {Verresen}, \citenamefont {Moessner},\ and\ \citenamefont
  {Pollmann}}]{Verresen17}%
  \BibitemOpen
  \bibfield  {author} {\bibinfo {author} {\bibfnamefont {Ruben}\ \bibnamefont
  {Verresen}}, \bibinfo {author} {\bibfnamefont {Roderich}\ \bibnamefont
  {Moessner}}, \ and\ \bibinfo {author} {\bibfnamefont {Frank}\ \bibnamefont
  {Pollmann}},\ }\bibfield  {title} {\enquote {\bibinfo {title}
  {One-dimensional symmetry protected topological phases and their
  transitions},}\ }\href {\doibase 10.1103/PhysRevB.96.165124} {\bibfield
  {journal} {\bibinfo  {journal} {Phys. Rev. B}\ }\textbf {\bibinfo {volume}
  {96}},\ \bibinfo {pages} {165124} (\bibinfo {year} {2017})}\BibitemShut
  {NoStop}%
\bibitem [{\citenamefont {Else}\ \emph {et~al.}(2013)\citenamefont {Else},
  \citenamefont {Bartlett},\ and\ \citenamefont {Doherty}}]{Else13}%
  \BibitemOpen
  \bibfield  {author} {\bibinfo {author} {\bibfnamefont {Dominic~V.}\
  \bibnamefont {Else}}, \bibinfo {author} {\bibfnamefont {Stephen~D.}\
  \bibnamefont {Bartlett}}, \ and\ \bibinfo {author} {\bibfnamefont
  {Andrew~C.}\ \bibnamefont {Doherty}},\ }\bibfield  {title} {\enquote
  {\bibinfo {title} {Hidden symmetry-breaking picture of symmetry-protected
  topological order},}\ }\href {\doibase 10.1103/PhysRevB.88.085114} {\bibfield
   {journal} {\bibinfo  {journal} {Phys. Rev. B}\ }\textbf {\bibinfo {volume}
  {88}},\ \bibinfo {pages} {085114} (\bibinfo {year} {2013})}\BibitemShut
  {NoStop}%
\bibitem [{\citenamefont {Levin}\ and\ \citenamefont {Gu}(2012)}]{Levin12}%
  \BibitemOpen
  \bibfield  {author} {\bibinfo {author} {\bibfnamefont {Michael}\ \bibnamefont
  {Levin}}\ and\ \bibinfo {author} {\bibfnamefont {Zheng-Cheng}\ \bibnamefont
  {Gu}},\ }\bibfield  {title} {\enquote {\bibinfo {title} {Braiding statistics
  approach to symmetry-protected topological phases},}\ }\href {\doibase
  10.1103/PhysRevB.86.115109} {\bibfield  {journal} {\bibinfo  {journal} {Phys.
  Rev. B}\ }\textbf {\bibinfo {volume} {86}},\ \bibinfo {pages} {115109}
  (\bibinfo {year} {2012})}\BibitemShut {NoStop}%
\bibitem [{\citenamefont {Wang}\ and\ \citenamefont {Levin}(2015)}]{Wang15}%
  \BibitemOpen
  \bibfield  {author} {\bibinfo {author} {\bibfnamefont {Chenjie}\ \bibnamefont
  {Wang}}\ and\ \bibinfo {author} {\bibfnamefont {Michael}\ \bibnamefont
  {Levin}},\ }\bibfield  {title} {\enquote {\bibinfo {title} {Topological
  invariants for gauge theories and symmetry-protected topological phases},}\
  }\href {\doibase 10.1103/PhysRevB.91.165119} {\bibfield  {journal} {\bibinfo
  {journal} {Phys. Rev. B}\ }\textbf {\bibinfo {volume} {91}},\ \bibinfo
  {pages} {165119} (\bibinfo {year} {2015})}\BibitemShut {NoStop}%
\bibitem [{\citenamefont {Di~Francesco}\ \emph {et~al.}(1997)\citenamefont
  {Di~Francesco}, \citenamefont {Mathieu},\ and\ \citenamefont
  {Senechal}}]{DiFrancesco97}%
  \BibitemOpen
  \bibfield  {author} {\bibinfo {author} {\bibfnamefont {P.}~\bibnamefont
  {Di~Francesco}}, \bibinfo {author} {\bibfnamefont {P.}~\bibnamefont
  {Mathieu}}, \ and\ \bibinfo {author} {\bibfnamefont {D.}~\bibnamefont
  {Senechal}},\ }\href {\doibase 10.1007/978-1-4612-2256-9} {\emph {\bibinfo
  {title} {{Conformal Field Theory}}}},\ Graduate Texts in Contemporary
  Physics\ (\bibinfo  {publisher} {Springer-Verlag},\ \bibinfo {address} {New
  York},\ \bibinfo {year} {1997})\BibitemShut {NoStop}%
\bibitem [{\citenamefont {Jones}\ and\ \citenamefont
  {Verresen}(2019)}]{Jones19}%
  \BibitemOpen
  \bibfield  {author} {\bibinfo {author} {\bibfnamefont {N.~G.}\ \bibnamefont
  {Jones}}\ and\ \bibinfo {author} {\bibfnamefont {R.}~\bibnamefont
  {Verresen}},\ }\bibfield  {title} {\enquote {\bibinfo {title} {{Asymptotic
  Correlations in Gapped and Critical Topological Phases of 1D Quantum
  Systems}},}\ }\href {\doibase 10.1007/s10955-019-02257-9} {\bibfield
  {journal} {\bibinfo  {journal} {Journal of Statistical Physics}\ }\textbf
  {\bibinfo {volume} {175}},\ \bibinfo {pages} {1164--1213} (\bibinfo {year}
  {2019})}\BibitemShut {NoStop}%
\bibitem [{\citenamefont {Watts}(2001)}]{Watts01}%
  \BibitemOpen
  \bibfield  {author} {\bibinfo {author} {\bibfnamefont {G.M.T.}\ \bibnamefont
  {Watts}},\ }\bibfield  {title} {\enquote {\bibinfo {title} {{On the boundary
  Ising model with disorder operators}},}\ }\href {\doibase
  10.1016/S0550-3213(00)00720-3} {\bibfield  {journal} {\bibinfo  {journal}
  {Nuclear Physics B}\ }\textbf {\bibinfo {volume} {596}},\ \bibinfo {pages}
  {513 -- 524} (\bibinfo {year} {2001})}\BibitemShut {NoStop}%
\bibitem [{\citenamefont {Cardy}(1986{\natexlab{a}})}]{Cardy86}%
  \BibitemOpen
  \bibfield  {author} {\bibinfo {author} {\bibfnamefont {John~L.}\ \bibnamefont
  {Cardy}},\ }\bibfield  {title} {\enquote {\bibinfo {title} {Effect of
  boundary conditions on the operator content of two-dimensional conformally
  invariant theories},}\ }\href {\doibase 10.1016/0550-3213(86)90596-1}
  {\bibfield  {journal} {\bibinfo  {journal} {Nuclear Physics B}\ }\textbf
  {\bibinfo {volume} {275}},\ \bibinfo {pages} {200 -- 218} (\bibinfo {year}
  {1986}{\natexlab{a}})}\BibitemShut {NoStop}%
\bibitem [{\citenamefont {Vodola}\ \emph {et~al.}(2014)\citenamefont {Vodola},
  \citenamefont {Lepori}, \citenamefont {Ercolessi}, \citenamefont {Gorshkov},\
  and\ \citenamefont {Pupillo}}]{Vodola14}%
  \BibitemOpen
  \bibfield  {author} {\bibinfo {author} {\bibfnamefont {Davide}\ \bibnamefont
  {Vodola}}, \bibinfo {author} {\bibfnamefont {Luca}\ \bibnamefont {Lepori}},
  \bibinfo {author} {\bibfnamefont {Elisa}\ \bibnamefont {Ercolessi}}, \bibinfo
  {author} {\bibfnamefont {Alexey~V.}\ \bibnamefont {Gorshkov}}, \ and\
  \bibinfo {author} {\bibfnamefont {Guido}\ \bibnamefont {Pupillo}},\
  }\bibfield  {title} {\enquote {\bibinfo {title} {{Kitaev Chains with
  Long-Range Pairing}},}\ }\href {\doibase 10.1103/PhysRevLett.113.156402}
  {\bibfield  {journal} {\bibinfo  {journal} {Phys. Rev. Lett.}\ }\textbf
  {\bibinfo {volume} {113}},\ \bibinfo {pages} {156402} (\bibinfo {year}
  {2014})}\BibitemShut {NoStop}%
\bibitem [{\citenamefont {Vodola}\ \emph {et~al.}(2015)\citenamefont {Vodola},
  \citenamefont {Lepori}, \citenamefont {Ercolessi},\ and\ \citenamefont
  {Pupillo}}]{Vodola15}%
  \BibitemOpen
  \bibfield  {author} {\bibinfo {author} {\bibfnamefont {Davide}\ \bibnamefont
  {Vodola}}, \bibinfo {author} {\bibfnamefont {Luca}\ \bibnamefont {Lepori}},
  \bibinfo {author} {\bibfnamefont {Elisa}\ \bibnamefont {Ercolessi}}, \ and\
  \bibinfo {author} {\bibfnamefont {Guido}\ \bibnamefont {Pupillo}},\
  }\bibfield  {title} {\enquote {\bibinfo {title} {{Long-range Ising and Kitaev
  models: phases, correlations and edge modes}},}\ }\href {\doibase
  10.1088/1367-2630/18/1/015001} {\bibfield  {journal} {\bibinfo  {journal}
  {New Journal of Physics}\ }\textbf {\bibinfo {volume} {18}},\ \bibinfo
  {pages} {015001} (\bibinfo {year} {2015})}\BibitemShut {NoStop}%
\bibitem [{\citenamefont {Patrick}\ \emph {et~al.}(2017)\citenamefont
  {Patrick}, \citenamefont {Neupert},\ and\ \citenamefont
  {Pachos}}]{Patrick17}%
  \BibitemOpen
  \bibfield  {author} {\bibinfo {author} {\bibfnamefont {Kristian}\
  \bibnamefont {Patrick}}, \bibinfo {author} {\bibfnamefont {Titus}\
  \bibnamefont {Neupert}}, \ and\ \bibinfo {author} {\bibfnamefont
  {Jiannis~K.}\ \bibnamefont {Pachos}},\ }\bibfield  {title} {\enquote
  {\bibinfo {title} {{Topological Quantum Liquids with Long-Range
  Couplings}},}\ }\href {\doibase 10.1103/PhysRevLett.118.267002} {\bibfield
  {journal} {\bibinfo  {journal} {Phys. Rev. Lett.}\ }\textbf {\bibinfo
  {volume} {118}},\ \bibinfo {pages} {267002} (\bibinfo {year}
  {2017})}\BibitemShut {NoStop}%
\bibitem [{\citenamefont {J\"ager}\ \emph {et~al.}(2020)\citenamefont
  {J\"ager}, \citenamefont {Dell'Anna},\ and\ \citenamefont
  {Morigi}}]{Jaeger20}%
  \BibitemOpen
  \bibfield  {author} {\bibinfo {author} {\bibfnamefont {Simon~B.}\
  \bibnamefont {J\"ager}}, \bibinfo {author} {\bibfnamefont {Luca}\
  \bibnamefont {Dell'Anna}}, \ and\ \bibinfo {author} {\bibfnamefont
  {Giovanna}\ \bibnamefont {Morigi}},\ }\bibfield  {title} {\enquote {\bibinfo
  {title} {{Edge states of the long-range Kitaev chain: An analytical
  study}},}\ }\href {\doibase 10.1103/PhysRevB.102.035152} {\bibfield
  {journal} {\bibinfo  {journal} {Phys. Rev. B}\ }\textbf {\bibinfo {volume}
  {102}},\ \bibinfo {pages} {035152} (\bibinfo {year} {2020})}\BibitemShut
  {NoStop}%
\bibitem [{\citenamefont {Kitaev}(2001)}]{Kitaev01}%
  \BibitemOpen
  \bibfield  {author} {\bibinfo {author} {\bibfnamefont {A}~\bibnamefont
  {Kitaev}},\ }\bibfield  {title} {\enquote {\bibinfo {title} {{Unpaired
  Majorana fermions in quantum wires}},}\ }\href@noop {} {\bibfield  {journal}
  {\bibinfo  {journal} {Physics-Uspekhi}\ }\textbf {\bibinfo {volume} {44}},\
  \bibinfo {pages} {131} (\bibinfo {year} {2001})}\BibitemShut {NoStop}%
\bibitem [{\citenamefont {Parker}\ \emph {et~al.}(2019)\citenamefont {Parker},
  \citenamefont {Vasseur},\ and\ \citenamefont {Scaffidi}}]{Parker18b}%
  \BibitemOpen
  \bibfield  {author} {\bibinfo {author} {\bibfnamefont {Daniel~E.}\
  \bibnamefont {Parker}}, \bibinfo {author} {\bibfnamefont {Romain}\
  \bibnamefont {Vasseur}}, \ and\ \bibinfo {author} {\bibfnamefont {Thomas}\
  \bibnamefont {Scaffidi}},\ }\bibfield  {title} {\enquote {\bibinfo {title}
  {{Topologically Protected Long Edge Coherence Times in Symmetry-Broken
  Phases}},}\ }\href {\doibase 10.1103/PhysRevLett.122.240605} {\bibfield
  {journal} {\bibinfo  {journal} {Phys. Rev. Lett.}\ }\textbf {\bibinfo
  {volume} {122}},\ \bibinfo {pages} {240605} (\bibinfo {year}
  {2019})}\BibitemShut {NoStop}%
\bibitem [{\citenamefont {Hung}\ and\ \citenamefont {Wen}(2013)}]{Hung13}%
  \BibitemOpen
  \bibfield  {author} {\bibinfo {author} {\bibfnamefont {Ling-Yan}\
  \bibnamefont {Hung}}\ and\ \bibinfo {author} {\bibfnamefont {Xiao-Gang}\
  \bibnamefont {Wen}},\ }\bibfield  {title} {\enquote {\bibinfo {title}
  {Quantized topological terms in weak-coupling gauge theories with a global
  symmetry and their connection to symmetry-enriched topological phases},}\
  }\href {\doibase 10.1103/PhysRevB.87.165107} {\bibfield  {journal} {\bibinfo
  {journal} {Phys. Rev. B}\ }\textbf {\bibinfo {volume} {87}},\ \bibinfo
  {pages} {165107} (\bibinfo {year} {2013})}\BibitemShut {NoStop}%
\bibitem [{\citenamefont {Barkeshli}\ \emph {et~al.}(2013)\citenamefont
  {Barkeshli}, \citenamefont {Jian},\ and\ \citenamefont {Qi}}]{Barkeshli13}%
  \BibitemOpen
  \bibfield  {author} {\bibinfo {author} {\bibfnamefont {Maissam}\ \bibnamefont
  {Barkeshli}}, \bibinfo {author} {\bibfnamefont {Chao-Ming}\ \bibnamefont
  {Jian}}, \ and\ \bibinfo {author} {\bibfnamefont {Xiao-Liang}\ \bibnamefont
  {Qi}},\ }\bibfield  {title} {\enquote {\bibinfo {title} {{Theory of defects
  in Abelian topological states}},}\ }\href {\doibase
  10.1103/PhysRevB.88.235103} {\bibfield  {journal} {\bibinfo  {journal} {Phys.
  Rev. B}\ }\textbf {\bibinfo {volume} {88}},\ \bibinfo {pages} {235103}
  (\bibinfo {year} {2013})}\BibitemShut {NoStop}%
\bibitem [{\citenamefont {Wen}(2014)}]{Wen14}%
  \BibitemOpen
  \bibfield  {author} {\bibinfo {author} {\bibfnamefont {Xiao-Gang}\
  \bibnamefont {Wen}},\ }\bibfield  {title} {\enquote {\bibinfo {title}
  {Symmetry-protected topological invariants of symmetry-protected topological
  phases of interacting bosons and fermions},}\ }\href {\doibase
  10.1103/PhysRevB.89.035147} {\bibfield  {journal} {\bibinfo  {journal} {Phys.
  Rev. B}\ }\textbf {\bibinfo {volume} {89}},\ \bibinfo {pages} {035147}
  (\bibinfo {year} {2014})}\BibitemShut {NoStop}%
\bibitem [{\citenamefont {Cheng}\ and\ \citenamefont {Gu}(2014)}]{Cheng14}%
  \BibitemOpen
  \bibfield  {author} {\bibinfo {author} {\bibfnamefont {Meng}\ \bibnamefont
  {Cheng}}\ and\ \bibinfo {author} {\bibfnamefont {Zheng-Cheng}\ \bibnamefont
  {Gu}},\ }\bibfield  {title} {\enquote {\bibinfo {title} {{Topological
  Response Theory of Abelian Symmetry-Protected Topological Phases in Two
  Dimensions}},}\ }\href {\doibase 10.1103/PhysRevLett.112.141602} {\bibfield
  {journal} {\bibinfo  {journal} {Phys. Rev. Lett.}\ }\textbf {\bibinfo
  {volume} {112}},\ \bibinfo {pages} {141602} (\bibinfo {year}
  {2014})}\BibitemShut {NoStop}%
\bibitem [{\citenamefont {Zaletel}(2014)}]{Zaletel14}%
  \BibitemOpen
  \bibfield  {author} {\bibinfo {author} {\bibfnamefont {Michael~P.}\
  \bibnamefont {Zaletel}},\ }\bibfield  {title} {\enquote {\bibinfo {title}
  {Detecting two-dimensional symmetry-protected topological order in a
  ground-state wave function},}\ }\href {\doibase 10.1103/PhysRevB.90.235113}
  {\bibfield  {journal} {\bibinfo  {journal} {Phys. Rev. B}\ }\textbf {\bibinfo
  {volume} {90}},\ \bibinfo {pages} {235113} (\bibinfo {year}
  {2014})}\BibitemShut {NoStop}%
\bibitem [{\citenamefont {Teo}\ \emph {et~al.}(2014)\citenamefont {Teo},
  \citenamefont {Roy},\ and\ \citenamefont {Chen}}]{Teo14}%
  \BibitemOpen
  \bibfield  {author} {\bibinfo {author} {\bibfnamefont {Jeffrey C.~Y.}\
  \bibnamefont {Teo}}, \bibinfo {author} {\bibfnamefont {Abhishek}\
  \bibnamefont {Roy}}, \ and\ \bibinfo {author} {\bibfnamefont {Xiao}\
  \bibnamefont {Chen}},\ }\bibfield  {title} {\enquote {\bibinfo {title}
  {Unconventional fusion and braiding of topological defects in a lattice
  model},}\ }\href {\doibase 10.1103/PhysRevB.90.115118} {\bibfield  {journal}
  {\bibinfo  {journal} {Phys. Rev. B}\ }\textbf {\bibinfo {volume} {90}},\
  \bibinfo {pages} {115118} (\bibinfo {year} {2014})}\BibitemShut {NoStop}%
\bibitem [{\citenamefont {Else}\ and\ \citenamefont {Nayak}(2014)}]{Else14}%
  \BibitemOpen
  \bibfield  {author} {\bibinfo {author} {\bibfnamefont {Dominic~V.}\
  \bibnamefont {Else}}\ and\ \bibinfo {author} {\bibfnamefont {Chetan}\
  \bibnamefont {Nayak}},\ }\bibfield  {title} {\enquote {\bibinfo {title}
  {Classifying symmetry-protected topological phases through the anomalous
  action of the symmetry on the edge},}\ }\href {\doibase
  10.1103/PhysRevB.90.235137} {\bibfield  {journal} {\bibinfo  {journal} {Phys.
  Rev. B}\ }\textbf {\bibinfo {volume} {90}},\ \bibinfo {pages} {235137}
  (\bibinfo {year} {2014})}\BibitemShut {NoStop}%
\bibitem [{\citenamefont {Kapustin}(2014)}]{Kapustin14b}%
  \BibitemOpen
  \bibfield  {author} {\bibinfo {author} {\bibfnamefont {Anton}\ \bibnamefont
  {Kapustin}},\ }\href@noop {} {\enquote {\bibinfo {title} {{Bosonic
  Topological Insulators and Paramagnets: a view from cobordisms}},}\ }
  (\bibinfo {year} {2014}),\ \Eprint {http://arxiv.org/abs/1404.6659}
  {arXiv:1404.6659 [cond-mat.str-el]} \BibitemShut {NoStop}%
\bibitem [{\citenamefont {Tarantino}\ \emph {et~al.}(2016)\citenamefont
  {Tarantino}, \citenamefont {Lindner},\ and\ \citenamefont
  {Fidkowski}}]{Tarantino16}%
  \BibitemOpen
  \bibfield  {author} {\bibinfo {author} {\bibfnamefont {Nicolas}\ \bibnamefont
  {Tarantino}}, \bibinfo {author} {\bibfnamefont {Netanel~H}\ \bibnamefont
  {Lindner}}, \ and\ \bibinfo {author} {\bibfnamefont {Lukasz}\ \bibnamefont
  {Fidkowski}},\ }\bibfield  {title} {\enquote {\bibinfo {title} {{Symmetry
  fractionalization and twist defects}},}\ }\href {\doibase
  10.1088/1367-2630/18/3/035006} {\bibfield  {journal} {\bibinfo  {journal}
  {New Journal of Physics}\ }\textbf {\bibinfo {volume} {18}},\ \bibinfo
  {pages} {035006} (\bibinfo {year} {2016})}\BibitemShut {NoStop}%
\bibitem [{\citenamefont {Tiwari}\ \emph {et~al.}(2018)\citenamefont {Tiwari},
  \citenamefont {Chen}, \citenamefont {Shiozaki},\ and\ \citenamefont
  {Ryu}}]{Tiwari18}%
  \BibitemOpen
  \bibfield  {author} {\bibinfo {author} {\bibfnamefont {Apoorv}\ \bibnamefont
  {Tiwari}}, \bibinfo {author} {\bibfnamefont {Xiao}\ \bibnamefont {Chen}},
  \bibinfo {author} {\bibfnamefont {Ken}\ \bibnamefont {Shiozaki}}, \ and\
  \bibinfo {author} {\bibfnamefont {Shinsei}\ \bibnamefont {Ryu}},\ }\bibfield
  {title} {\enquote {\bibinfo {title} {{Bosonic topological phases of matter:
  Bulk-boundary correspondence, symmetry protected topological invariants, and
  gauging}},}\ }\href {\doibase 10.1103/PhysRevB.97.245133} {\bibfield
  {journal} {\bibinfo  {journal} {Phys. Rev. B}\ }\textbf {\bibinfo {volume}
  {97}},\ \bibinfo {pages} {245133} (\bibinfo {year} {2018})}\BibitemShut
  {NoStop}%
\bibitem [{\citenamefont {Ruelle}\ and\ \citenamefont
  {Verhoeven}(1998)}]{Ruelle98}%
  \BibitemOpen
  \bibfield  {author} {\bibinfo {author} {\bibfnamefont {P.}~\bibnamefont
  {Ruelle}}\ and\ \bibinfo {author} {\bibfnamefont {O.}~\bibnamefont
  {Verhoeven}},\ }\bibfield  {title} {\enquote {\bibinfo {title} {Discrete
  symmetries of unitary minimal conformal theories},}\ }\href {\doibase
  10.1016/S0550-3213(98)00639-7} {\bibfield  {journal} {\bibinfo  {journal}
  {Nuclear Physics B}\ }\textbf {\bibinfo {volume} {535}},\ \bibinfo {pages}
  {650 -- 680} (\bibinfo {year} {1998})}\BibitemShut {NoStop}%
\bibitem [{\citenamefont {White}(1992)}]{White92}%
  \BibitemOpen
  \bibfield  {author} {\bibinfo {author} {\bibfnamefont {Steven~R.}\
  \bibnamefont {White}},\ }\bibfield  {title} {\enquote {\bibinfo {title}
  {Density matrix formulation for quantum renormalization groups},}\ }\href
  {\doibase 10.1103/PhysRevLett.69.2863} {\bibfield  {journal} {\bibinfo
  {journal} {Phys. Rev. Lett.}\ }\textbf {\bibinfo {volume} {69}},\ \bibinfo
  {pages} {2863--2866} (\bibinfo {year} {1992})}\BibitemShut {NoStop}%
\bibitem [{\citenamefont {Kj\"all}\ \emph {et~al.}(2013)\citenamefont
  {Kj\"all}, \citenamefont {Zaletel}, \citenamefont {Mong}, \citenamefont
  {Bardarson},\ and\ \citenamefont {Pollmann}}]{Kjall13}%
  \BibitemOpen
  \bibfield  {author} {\bibinfo {author} {\bibfnamefont {Jonas~A.}\
  \bibnamefont {Kj\"all}}, \bibinfo {author} {\bibfnamefont {Michael~P.}\
  \bibnamefont {Zaletel}}, \bibinfo {author} {\bibfnamefont {Roger S.~K.}\
  \bibnamefont {Mong}}, \bibinfo {author} {\bibfnamefont {Jens~H.}\
  \bibnamefont {Bardarson}}, \ and\ \bibinfo {author} {\bibfnamefont {Frank}\
  \bibnamefont {Pollmann}},\ }\bibfield  {title} {\enquote {\bibinfo {title}
  {{Phase diagram of the anisotropic spin-2 XXZ model: Infinite-system density
  matrix renormalization group study}},}\ }\href {\doibase
  10.1103/PhysRevB.87.235106} {\bibfield  {journal} {\bibinfo  {journal} {Phys.
  Rev. B}\ }\textbf {\bibinfo {volume} {87}},\ \bibinfo {pages} {235106}
  (\bibinfo {year} {2013})}\BibitemShut {NoStop}%
\bibitem [{\citenamefont {Yamanaka}\ \emph {et~al.}(1993)\citenamefont
  {Yamanaka}, \citenamefont {Hatsugai},\ and\ \citenamefont
  {Kohmoto}}]{Yamanaka93}%
  \BibitemOpen
  \bibfield  {author} {\bibinfo {author} {\bibfnamefont {M.}~\bibnamefont
  {Yamanaka}}, \bibinfo {author} {\bibfnamefont {Y.}~\bibnamefont {Hatsugai}},
  \ and\ \bibinfo {author} {\bibfnamefont {M.}~\bibnamefont {Kohmoto}},\
  }\bibfield  {title} {\enquote {\bibinfo {title} {{Phase diagram of the S=1/2
  quantum spin chain with bond alternation}},}\ }\href {\doibase
  10.1103/PhysRevB.48.9555} {\bibfield  {journal} {\bibinfo  {journal} {Phys.
  Rev. B}\ }\textbf {\bibinfo {volume} {48}},\ \bibinfo {pages} {9555--9563}
  (\bibinfo {year} {1993})}\BibitemShut {NoStop}%
\bibitem [{\citenamefont {Ejima}\ \emph {et~al.}(2018)\citenamefont {Ejima},
  \citenamefont {Yamaguchi}, \citenamefont {Essler}, \citenamefont {Lange},
  \citenamefont {Ohta},\ and\ \citenamefont {Fehske}}]{Satoshi18}%
  \BibitemOpen
  \bibfield  {author} {\bibinfo {author} {\bibfnamefont {Satoshi}\ \bibnamefont
  {Ejima}}, \bibinfo {author} {\bibfnamefont {Tomoki}\ \bibnamefont
  {Yamaguchi}}, \bibinfo {author} {\bibfnamefont {Fabian H.~L.}\ \bibnamefont
  {Essler}}, \bibinfo {author} {\bibfnamefont {Florian}\ \bibnamefont {Lange}},
  \bibinfo {author} {\bibfnamefont {Yukinori}\ \bibnamefont {Ohta}}, \ and\
  \bibinfo {author} {\bibfnamefont {Holger}\ \bibnamefont {Fehske}},\
  }\bibfield  {title} {\enquote {\bibinfo {title} {{Exotic criticality in the
  dimerized spin-1 $XXZ$ chain with single-ion anisotropy}},}\ }\href {\doibase
  10.21468/SciPostPhys.5.6.059} {\bibfield  {journal} {\bibinfo  {journal}
  {SciPost Phys.}\ }\textbf {\bibinfo {volume} {5}},\ \bibinfo {pages} {59}
  (\bibinfo {year} {2018})}\BibitemShut {NoStop}%
\bibitem [{\citenamefont {Cardy}(1989)}]{Cardy89}%
  \BibitemOpen
  \bibfield  {author} {\bibinfo {author} {\bibfnamefont {John~L.}\ \bibnamefont
  {Cardy}},\ }\bibfield  {title} {\enquote {\bibinfo {title} {{Boundary
  conditions, fusion rules and the Verlinde formula}},}\ }\href {\doibase
  10.1016/0550-3213(89)90521-X} {\bibfield  {journal} {\bibinfo  {journal}
  {Nuclear Physics B}\ }\textbf {\bibinfo {volume} {324}},\ \bibinfo {pages}
  {581 -- 596} (\bibinfo {year} {1989})}\BibitemShut {NoStop}%
\bibitem [{\citenamefont {Blumenhagen}\ and\ \citenamefont
  {Plauschinn}(2009)}]{Blumenhagen09}%
  \BibitemOpen
  \bibfield  {author} {\bibinfo {author} {\bibfnamefont {R.}~\bibnamefont
  {Blumenhagen}}\ and\ \bibinfo {author} {\bibfnamefont {E.}~\bibnamefont
  {Plauschinn}},\ }\href {https://www.springer.com/gb/book/9783642004490}
  {\emph {\bibinfo {title} {{Introduction to Conformal Field Theory: With
  Applications to String Theory}}}},\ Lecture Notes in Physics\ (\bibinfo
  {publisher} {Springer Berlin Heidelberg},\ \bibinfo {year}
  {2009})\BibitemShut {NoStop}%
\bibitem [{\citenamefont {Hida}(1992)}]{Hida92}%
  \BibitemOpen
  \bibfield  {author} {\bibinfo {author} {\bibfnamefont {Kazuo}\ \bibnamefont
  {Hida}},\ }\bibfield  {title} {\enquote {\bibinfo {title} {{Crossover between
  the Haldane-gap phase and the dimer phase in the spin-1/2 alternating
  Heisenberg chain}},}\ }\href {\doibase 10.1103/PhysRevB.45.2207} {\bibfield
  {journal} {\bibinfo  {journal} {Phys. Rev. B}\ }\textbf {\bibinfo {volume}
  {45}},\ \bibinfo {pages} {2207--2212} (\bibinfo {year} {1992})}\BibitemShut
  {NoStop}%
\bibitem [{\citenamefont {Ginsparg}(1990)}]{Ginsparg88}%
  \BibitemOpen
  \bibfield  {author} {\bibinfo {author} {\bibfnamefont {P}~\bibnamefont
  {Ginsparg}},\ }\bibfield  {title} {\enquote {\bibinfo {title} {Applied
  conformal field theory.}}\ }in\ \href@noop {} {\emph {\bibinfo {booktitle}
  {Les Houches, Session XLIX, 1988,Fields,Strings and Critical Phenomena}}},\
  \bibinfo {editor} {edited by\ \bibinfo {editor} {\bibfnamefont
  {E.}~\bibnamefont {Brezin}}\ and\ \bibinfo {editor} {\bibfnamefont
  {J.}~\bibnamefont {Zinn-Justin}}}\ (\bibinfo  {publisher} {Elsevier},\
  \bibinfo {year} {1990})\BibitemShut {NoStop}%
\bibitem [{\citenamefont {Affleck}(1990)}]{Affleck88b}%
  \BibitemOpen
  \bibfield  {author} {\bibinfo {author} {\bibfnamefont {Ian}\ \bibnamefont
  {Affleck}},\ }\bibfield  {title} {\enquote {\bibinfo {title} {Field theory
  methods and quantum critical phenomena},}\ }in\ \href@noop {} {\emph
  {\bibinfo {booktitle} {Les Houches, Session XLIX, 1988,Fields,Strings and
  Critical Phenomena}}},\ \bibinfo {editor} {edited by\ \bibinfo {editor}
  {\bibfnamefont {E.}~\bibnamefont {Brezin}}\ and\ \bibinfo {editor}
  {\bibfnamefont {J.}~\bibnamefont {Zinn-Justin}}}\ (\bibinfo  {publisher}
  {Elsevier},\ \bibinfo {year} {1990})\BibitemShut {NoStop}%
\bibitem [{\citenamefont {Sachdev}(2001)}]{Sachdev01}%
  \BibitemOpen
  \bibfield  {author} {\bibinfo {author} {\bibfnamefont {S.}~\bibnamefont
  {Sachdev}},\ }\href {\doibase 10.1017/CBO9780511973765} {\emph {\bibinfo
  {title} {{Quantum Phase Transitions}}}}\ (\bibinfo  {publisher} {Cambridge
  University Press},\ \bibinfo {year} {2001})\BibitemShut {NoStop}%
\bibitem [{\citenamefont {Giamarchi}(2004)}]{Giamarchi04}%
  \BibitemOpen
  \bibfield  {author} {\bibinfo {author} {\bibfnamefont {T.}~\bibnamefont
  {Giamarchi}},\ }\href {\doibase 10.1093/acprof:oso/9780198525004.001.0001}
  {\emph {\bibinfo {title} {{Quantum Physics in One Dimension}}}},\
  International Series of Monographs on Physics\ (\bibinfo  {publisher}
  {Clarendon Press},\ \bibinfo {year} {2004})\BibitemShut {NoStop}%
\bibitem [{\citenamefont {{Geraedts}}\ and\ \citenamefont
  {{Motrunich}}(2014)}]{Geraedts14}%
  \BibitemOpen
  \bibfield  {author} {\bibinfo {author} {\bibfnamefont {Scott~D.}\
  \bibnamefont {{Geraedts}}}\ and\ \bibinfo {author} {\bibfnamefont
  {Olexei~I.}\ \bibnamefont {{Motrunich}}},\ }\bibfield  {title} {\enquote
  {\bibinfo {title} {{Exact Models for Symmetry-Protected Topological Phases in
  One Dimension}},}\ }\href@noop {} {\bibfield  {journal} {\bibinfo  {journal}
  {arXiv e-prints}\ ,\ \bibinfo {eid} {arXiv:1410.1580}} (\bibinfo {year}
  {2014})},\ \Eprint {http://arxiv.org/abs/1410.1580} {arXiv:1410.1580
  [cond-mat.stat-mech]} \BibitemShut {NoStop}%
\bibitem [{\citenamefont {Santos}(2015)}]{Santos15}%
  \BibitemOpen
  \bibfield  {author} {\bibinfo {author} {\bibfnamefont {Luiz~H.}\ \bibnamefont
  {Santos}},\ }\bibfield  {title} {\enquote {\bibinfo {title}
  {{Rokhsar-Kivelson models of bosonic symmetry-protected topological
  states}},}\ }\href {\doibase 10.1103/PhysRevB.91.155150} {\bibfield
  {journal} {\bibinfo  {journal} {Phys. Rev. B}\ }\textbf {\bibinfo {volume}
  {91}},\ \bibinfo {pages} {155150} (\bibinfo {year} {2015})}\BibitemShut
  {NoStop}%
\bibitem [{\citenamefont {Zamolodchikov}\ and\ \citenamefont
  {Fateev}(1985)}]{Zamolodchikov85}%
  \BibitemOpen
  \bibfield  {author} {\bibinfo {author} {\bibfnamefont {A~B}\ \bibnamefont
  {Zamolodchikov}}\ and\ \bibinfo {author} {\bibfnamefont {V~A}\ \bibnamefont
  {Fateev}},\ }\bibfield  {title} {\enquote {\bibinfo {title} {{Nonlocal
  (parafermion) currents in two-dimensional conformal quantum field theory and
  self-dual critical points in $Z_N$-symmetric statistical systems}},}\ }\href
  {https://www.osti.gov/biblio/5929972} {\bibfield  {journal} {\bibinfo
  {journal} {Sov. Phys. - JETP (Engl. Transl.); (United States)}\ }\textbf
  {\bibinfo {volume} {62}} (\bibinfo {year} {1985})}\BibitemShut {NoStop}%
\bibitem [{\citenamefont {von Gehlen}\ and\ \citenamefont
  {Rittenberg}(1986)}]{Gehlen86}%
  \BibitemOpen
  \bibfield  {author} {\bibinfo {author} {\bibfnamefont {G}~\bibnamefont {von
  Gehlen}}\ and\ \bibinfo {author} {\bibfnamefont {V}~\bibnamefont
  {Rittenberg}},\ }\bibfield  {title} {\enquote {\bibinfo {title} {{Operator
  content of the three-state Potts quantum chain}},}\ }\href {\doibase
  10.1088/0305-4470/19/10/013} {\bibfield  {journal} {\bibinfo  {journal}
  {Journal of Physics A: Mathematical and General}\ }\textbf {\bibinfo {volume}
  {19}},\ \bibinfo {pages} {L625--L629} (\bibinfo {year} {1986})}\BibitemShut
  {NoStop}%
\bibitem [{\citenamefont {Tsui}\ \emph {et~al.}(2017)\citenamefont {Tsui},
  \citenamefont {Huang}, \citenamefont {Jiang},\ and\ \citenamefont
  {Lee}}]{Tsui17}%
  \BibitemOpen
  \bibfield  {author} {\bibinfo {author} {\bibfnamefont {Lokman}\ \bibnamefont
  {Tsui}}, \bibinfo {author} {\bibfnamefont {Yen-Ta}\ \bibnamefont {Huang}},
  \bibinfo {author} {\bibfnamefont {Hong-Chen}\ \bibnamefont {Jiang}}, \ and\
  \bibinfo {author} {\bibfnamefont {Dung-Hai}\ \bibnamefont {Lee}},\ }\bibfield
   {title} {\enquote {\bibinfo {title} {{The phase transitions between $Z_n
  \times Z_n$ bosonic topological phases in 1+1D, and a constraint on the
  central charge for the critical points between bosonic symmetry protected
  topological phases}},}\ }\href {\doibase
  https://doi.org/10.1016/j.nuclphysb.2017.03.021} {\bibfield  {journal}
  {\bibinfo  {journal} {Nuclear Physics B}\ }\textbf {\bibinfo {volume}
  {919}},\ \bibinfo {pages} {470 -- 503} (\bibinfo {year} {2017})}\BibitemShut
  {NoStop}%
\bibitem [{\citenamefont {Motruk}\ \emph {et~al.}(2013)\citenamefont {Motruk},
  \citenamefont {Berg}, \citenamefont {Turner},\ and\ \citenamefont
  {Pollmann}}]{Motruk13}%
  \BibitemOpen
  \bibfield  {author} {\bibinfo {author} {\bibfnamefont {Johannes}\
  \bibnamefont {Motruk}}, \bibinfo {author} {\bibfnamefont {Erez}\ \bibnamefont
  {Berg}}, \bibinfo {author} {\bibfnamefont {Ari~M.}\ \bibnamefont {Turner}}, \
  and\ \bibinfo {author} {\bibfnamefont {Frank}\ \bibnamefont {Pollmann}},\
  }\bibfield  {title} {\enquote {\bibinfo {title} {Topological phases in gapped
  edges of fractionalized systems},}\ }\href {\doibase
  10.1103/PhysRevB.88.085115} {\bibfield  {journal} {\bibinfo  {journal} {Phys.
  Rev. B}\ }\textbf {\bibinfo {volume} {88}},\ \bibinfo {pages} {085115}
  (\bibinfo {year} {2013})}\BibitemShut {NoStop}%
\bibitem [{\citenamefont {Meidan}\ \emph {et~al.}(2017)\citenamefont {Meidan},
  \citenamefont {Berg},\ and\ \citenamefont {Stern}}]{Meidan17}%
  \BibitemOpen
  \bibfield  {author} {\bibinfo {author} {\bibfnamefont {D.}~\bibnamefont
  {Meidan}}, \bibinfo {author} {\bibfnamefont {E.}~\bibnamefont {Berg}}, \ and\
  \bibinfo {author} {\bibfnamefont {Ady}\ \bibnamefont {Stern}},\ }\bibfield
  {title} {\enquote {\bibinfo {title} {Classification of topological phases of
  parafermionic chains with symmetries},}\ }\href {\doibase
  10.1103/PhysRevB.95.205104} {\bibfield  {journal} {\bibinfo  {journal} {Phys.
  Rev. B}\ }\textbf {\bibinfo {volume} {95}},\ \bibinfo {pages} {205104}
  (\bibinfo {year} {2017})}\BibitemShut {NoStop}%
\bibitem [{\citenamefont {{Chen}}\ \emph {et~al.}(2014)\citenamefont {{Chen}},
  \citenamefont {{Lu}},\ and\ \citenamefont {{Vishwanath}}}]{Chen14}%
  \BibitemOpen
  \bibfield  {author} {\bibinfo {author} {\bibfnamefont {Xie}\ \bibnamefont
  {{Chen}}}, \bibinfo {author} {\bibfnamefont {Yuan-Ming}\ \bibnamefont
  {{Lu}}}, \ and\ \bibinfo {author} {\bibfnamefont {Ashvin}\ \bibnamefont
  {{Vishwanath}}},\ }\bibfield  {title} {\enquote {\bibinfo {title}
  {{Symmetry-protected topological phases from decorated domain walls}},}\
  }\href {\doibase 10.1038/ncomms4507} {\bibfield  {journal} {\bibinfo
  {journal} {Nature Communications}\ }\textbf {\bibinfo {volume} {5}},\
  \bibinfo {eid} {3507} (\bibinfo {year} {2014})}\BibitemShut {NoStop}%
\bibitem [{\citenamefont {McAvity}\ and\ \citenamefont
  {Osborn}(1993)}]{McAvity93}%
  \BibitemOpen
  \bibfield  {author} {\bibinfo {author} {\bibfnamefont {D.M.}\ \bibnamefont
  {McAvity}}\ and\ \bibinfo {author} {\bibfnamefont {H.}~\bibnamefont
  {Osborn}},\ }\bibfield  {title} {\enquote {\bibinfo {title} {Energy-momentum
  tensor in conformal field theories near a boundary},}\ }\href {\doibase
  https://doi.org/10.1016/0550-3213(93)90005-A} {\bibfield  {journal} {\bibinfo
   {journal} {Nuclear Physics B}\ }\textbf {\bibinfo {volume} {406}},\ \bibinfo
  {pages} {655--680} (\bibinfo {year} {1993})}\BibitemShut {NoStop}%
\bibitem [{\citenamefont {McAvity}\ and\ \citenamefont
  {Osborn}(1995)}]{McAvity95}%
  \BibitemOpen
  \bibfield  {author} {\bibinfo {author} {\bibfnamefont {D.M.}\ \bibnamefont
  {McAvity}}\ and\ \bibinfo {author} {\bibfnamefont {H.}~\bibnamefont
  {Osborn}},\ }\bibfield  {title} {\enquote {\bibinfo {title} {Conformal field
  theories near a boundary in general dimensions},}\ }\href {\doibase
  https://doi.org/10.1016/0550-3213(95)00476-9} {\bibfield  {journal} {\bibinfo
   {journal} {Nuclear Physics B}\ }\textbf {\bibinfo {volume} {455}},\ \bibinfo
  {pages} {522--576} (\bibinfo {year} {1995})}\BibitemShut {NoStop}%
\bibitem [{\citenamefont {Liendo}\ \emph {et~al.}(2013)\citenamefont {Liendo},
  \citenamefont {Rastelli},\ and\ \citenamefont {van Rees}}]{Liendo13}%
  \BibitemOpen
  \bibfield  {author} {\bibinfo {author} {\bibfnamefont {Pedro}\ \bibnamefont
  {Liendo}}, \bibinfo {author} {\bibfnamefont {Leonardo}\ \bibnamefont
  {Rastelli}}, \ and\ \bibinfo {author} {\bibfnamefont {Balt~C.}\ \bibnamefont
  {van Rees}},\ }\bibfield  {title} {\enquote {\bibinfo {title} {The bootstrap
  program for boundary {CFT}$_d$},}\ }\href {\doibase 10.1007/jhep07(2013)113}
  {\bibfield  {journal} {\bibinfo  {journal} {Journal of High Energy Physics}\
  }\textbf {\bibinfo {volume} {2013}} (\bibinfo {year} {2013}),\
  10.1007/jhep07(2013)113}\BibitemShut {NoStop}%
\bibitem [{\citenamefont {Billó}\ \emph {et~al.}(2013)\citenamefont {Billó},
  \citenamefont {Caselle}, \citenamefont {Gaiotto}, \citenamefont {Gliozzi},
  \citenamefont {Meineri},\ and\ \citenamefont {Pellegrini}}]{Billo13}%
  \BibitemOpen
  \bibfield  {author} {\bibinfo {author} {\bibfnamefont {M.}~\bibnamefont
  {Billó}}, \bibinfo {author} {\bibfnamefont {M.}~\bibnamefont {Caselle}},
  \bibinfo {author} {\bibfnamefont {D.}~\bibnamefont {Gaiotto}}, \bibinfo
  {author} {\bibfnamefont {F.}~\bibnamefont {Gliozzi}}, \bibinfo {author}
  {\bibfnamefont {M.}~\bibnamefont {Meineri}}, \ and\ \bibinfo {author}
  {\bibfnamefont {R.}~\bibnamefont {Pellegrini}},\ }\href@noop {} {\enquote
  {\bibinfo {title} {{Line defects in the 3d Ising model}},}\ } (\bibinfo
  {year} {2013}),\ \Eprint {http://arxiv.org/abs/1304.4110} {arXiv:1304.4110
  [hep-th]} \BibitemShut {NoStop}%
\bibitem [{\citenamefont {Gaiotto}\ \emph {et~al.}(2014)\citenamefont
  {Gaiotto}, \citenamefont {Mazac},\ and\ \citenamefont {Paulos}}]{Gaiotto14}%
  \BibitemOpen
  \bibfield  {author} {\bibinfo {author} {\bibfnamefont {Davide}\ \bibnamefont
  {Gaiotto}}, \bibinfo {author} {\bibfnamefont {Dalimil}\ \bibnamefont
  {Mazac}}, \ and\ \bibinfo {author} {\bibfnamefont {Miguel~F.}\ \bibnamefont
  {Paulos}},\ }\bibfield  {title} {\enquote {\bibinfo {title} {{Bootstrapping
  the 3d Ising twist defect}},}\ }\href {\doibase 10.1007/jhep03(2014)100}
  {\bibfield  {journal} {\bibinfo  {journal} {Journal of High Energy Physics}\
  }\textbf {\bibinfo {volume} {2014}} (\bibinfo {year} {2014}),\
  10.1007/jhep03(2014)100}\BibitemShut {NoStop}%
\bibitem [{\citenamefont {Gliozzi}\ \emph {et~al.}(2015)\citenamefont
  {Gliozzi}, \citenamefont {Liendo}, \citenamefont {Meineri},\ and\
  \citenamefont {Rago}}]{Gliozzi15}%
  \BibitemOpen
  \bibfield  {author} {\bibinfo {author} {\bibfnamefont {Ferdinando}\
  \bibnamefont {Gliozzi}}, \bibinfo {author} {\bibfnamefont {Pedro}\
  \bibnamefont {Liendo}}, \bibinfo {author} {\bibfnamefont {Marco}\
  \bibnamefont {Meineri}}, \ and\ \bibinfo {author} {\bibfnamefont {Antonio}\
  \bibnamefont {Rago}},\ }\bibfield  {title} {\enquote {\bibinfo {title}
  {{Boundary and interface CFTs from the conformal bootstrap}},}\ }\href
  {\doibase 10.1007/jhep05(2015)036} {\bibfield  {journal} {\bibinfo  {journal}
  {Journal of High Energy Physics}\ }\textbf {\bibinfo {volume} {2015}}
  (\bibinfo {year} {2015}),\ 10.1007/jhep05(2015)036}\BibitemShut {NoStop}%
\bibitem [{\citenamefont {{Bill{\`o}}}\ \emph {et~al.}(2016)\citenamefont
  {{Bill{\`o}}}, \citenamefont {{Gon{\c{c}}alves}}, \citenamefont {{Lauria}},\
  and\ \citenamefont {{Meineri}}}]{Billo16}%
  \BibitemOpen
  \bibfield  {author} {\bibinfo {author} {\bibfnamefont {Marco}\ \bibnamefont
  {{Bill{\`o}}}}, \bibinfo {author} {\bibfnamefont {Vasco}\ \bibnamefont
  {{Gon{\c{c}}alves}}}, \bibinfo {author} {\bibfnamefont {Edoardo}\
  \bibnamefont {{Lauria}}}, \ and\ \bibinfo {author} {\bibfnamefont {Marco}\
  \bibnamefont {{Meineri}}},\ }\bibfield  {title} {\enquote {\bibinfo {title}
  {{Defects in conformal field theory}},}\ }\href {\doibase
  10.1007/JHEP04(2016)091} {\bibfield  {journal} {\bibinfo  {journal} {Journal
  of High Energy Physics}\ }\textbf {\bibinfo {volume} {2016}},\ \bibinfo {eid}
  {91} (\bibinfo {year} {2016})},\ \Eprint {http://arxiv.org/abs/1601.02883}
  {arXiv:1601.02883 [hep-th]} \BibitemShut {NoStop}%
\bibitem [{\citenamefont {{Fukuda}}\ \emph {et~al.}(2018)\citenamefont
  {{Fukuda}}, \citenamefont {{Kobayashi}},\ and\ \citenamefont
  {{Nishioka}}}]{Fukuda18}%
  \BibitemOpen
  \bibfield  {author} {\bibinfo {author} {\bibfnamefont {Masayuki}\
  \bibnamefont {{Fukuda}}}, \bibinfo {author} {\bibfnamefont {Nozomu}\
  \bibnamefont {{Kobayashi}}}, \ and\ \bibinfo {author} {\bibfnamefont
  {Tatsuma}\ \bibnamefont {{Nishioka}}},\ }\bibfield  {title} {\enquote
  {\bibinfo {title} {{Operator product expansion for conformal defects}},}\
  }\href {\doibase 10.1007/JHEP01(2018)013} {\bibfield  {journal} {\bibinfo
  {journal} {Journal of High Energy Physics}\ }\textbf {\bibinfo {volume}
  {2018}},\ \bibinfo {eid} {13} (\bibinfo {year} {2018})},\ \Eprint
  {http://arxiv.org/abs/1710.11165} {arXiv:1710.11165 [hep-th]} \BibitemShut
  {NoStop}%
\bibitem [{\citenamefont {Lauria}\ \emph {et~al.}(2019)\citenamefont {Lauria},
  \citenamefont {Meineri},\ and\ \citenamefont {Trevisani}}]{Lauria19}%
  \BibitemOpen
  \bibfield  {author} {\bibinfo {author} {\bibfnamefont {Edoardo}\ \bibnamefont
  {Lauria}}, \bibinfo {author} {\bibfnamefont {Marco}\ \bibnamefont {Meineri}},
  \ and\ \bibinfo {author} {\bibfnamefont {Emilio}\ \bibnamefont {Trevisani}},\
  }\bibfield  {title} {\enquote {\bibinfo {title} {Spinning operators and
  defects in conformal field theory},}\ }\href {\doibase
  10.1007/jhep08(2019)066} {\bibfield  {journal} {\bibinfo  {journal} {Journal
  of High Energy Physics}\ }\textbf {\bibinfo {volume} {2019}} (\bibinfo {year}
  {2019}),\ 10.1007/jhep08(2019)066}\BibitemShut {NoStop}%
\bibitem [{\citenamefont {Metlitski}(2020)}]{Metlitski20}%
  \BibitemOpen
  \bibfield  {author} {\bibinfo {author} {\bibfnamefont {Max~A.}\ \bibnamefont
  {Metlitski}},\ }\href@noop {} {\enquote {\bibinfo {title} {{Boundary
  criticality of the $O(N)$ model in $d = 3$ critically revisited}},}\ }
  (\bibinfo {year} {2020}),\ \Eprint {http://arxiv.org/abs/2009.05119}
  {arXiv:2009.05119 [cond-mat.str-el]} \BibitemShut {NoStop}%
\bibitem [{\citenamefont {{Dey}}\ and\ \citenamefont
  {{S{\"o}derberg}}(2020)}]{Dey20}%
  \BibitemOpen
  \bibfield  {author} {\bibinfo {author} {\bibfnamefont {Parijat}\ \bibnamefont
  {{Dey}}}\ and\ \bibinfo {author} {\bibfnamefont {Alexander}\ \bibnamefont
  {{S{\"o}derberg}}},\ }\bibfield  {title} {\enquote {\bibinfo {title} {{On
  Analytic Bootstrap for Interface and Boundary CFT}},}\ }\href@noop {}
  {\bibfield  {journal} {\bibinfo  {journal} {arXiv e-prints}\ ,\ \bibinfo
  {eid} {arXiv:2012.11344}} (\bibinfo {year} {2020})},\ \Eprint
  {http://arxiv.org/abs/2012.11344} {arXiv:2012.11344 [hep-th]} \BibitemShut
  {NoStop}%
\bibitem [{\citenamefont {{Antunes}}(2021)}]{Antunes21}%
  \BibitemOpen
  \bibfield  {author} {\bibinfo {author} {\bibfnamefont {Ant{\'o}nio}\
  \bibnamefont {{Antunes}}},\ }\bibfield  {title} {\enquote {\bibinfo {title}
  {{Conformal Bootstrap near the edge}},}\ }\href@noop {} {\bibfield  {journal}
  {\bibinfo  {journal} {arXiv e-prints}\ ,\ \bibinfo {eid} {arXiv:2103.03132}}
  (\bibinfo {year} {2021})},\ \Eprint {http://arxiv.org/abs/2103.03132}
  {arXiv:2103.03132 [hep-th]} \BibitemShut {NoStop}%
\bibitem [{\citenamefont {Lauria}\ \emph {et~al.}(2021)\citenamefont {Lauria},
  \citenamefont {Liendo}, \citenamefont {van Rees},\ and\ \citenamefont
  {Zhao}}]{Lauria21}%
  \BibitemOpen
  \bibfield  {author} {\bibinfo {author} {\bibfnamefont {Edoardo}\ \bibnamefont
  {Lauria}}, \bibinfo {author} {\bibfnamefont {Pedro}\ \bibnamefont {Liendo}},
  \bibinfo {author} {\bibfnamefont {Balt~C.}\ \bibnamefont {van Rees}}, \ and\
  \bibinfo {author} {\bibfnamefont {Xiang}\ \bibnamefont {Zhao}},\ }\bibfield
  {title} {\enquote {\bibinfo {title} {Line and surface defects for the free
  scalar field},}\ }\href {\doibase 10.1007/jhep01(2021)060} {\bibfield
  {journal} {\bibinfo  {journal} {Journal of High Energy Physics}\ }\textbf
  {\bibinfo {volume} {2021}} (\bibinfo {year} {2021}),\
  10.1007/jhep01(2021)060}\BibitemShut {NoStop}%
\bibitem [{\citenamefont {Herzog}\ and\ \citenamefont
  {Shrestha}(2021)}]{Herzog21}%
  \BibitemOpen
  \bibfield  {author} {\bibinfo {author} {\bibfnamefont {Christopher~P.}\
  \bibnamefont {Herzog}}\ and\ \bibinfo {author} {\bibfnamefont {Abhay}\
  \bibnamefont {Shrestha}},\ }\bibfield  {title} {\enquote {\bibinfo {title}
  {{Two point functions in defect CFTs}},}\ }\href {\doibase
  10.1007/jhep04(2021)226} {\bibfield  {journal} {\bibinfo  {journal} {Journal
  of High Energy Physics}\ }\textbf {\bibinfo {volume} {2021}} (\bibinfo {year}
  {2021}),\ 10.1007/jhep04(2021)226}\BibitemShut {NoStop}%
\bibitem [{\citenamefont {Zhao}\ \emph {et~al.}(2021)\citenamefont {Zhao},
  \citenamefont {Yan}, \citenamefont {Cheng},\ and\ \citenamefont
  {Meng}}]{Zhao21}%
  \BibitemOpen
  \bibfield  {author} {\bibinfo {author} {\bibfnamefont {Jiarui}\ \bibnamefont
  {Zhao}}, \bibinfo {author} {\bibfnamefont {Zheng}\ \bibnamefont {Yan}},
  \bibinfo {author} {\bibfnamefont {Meng}\ \bibnamefont {Cheng}}, \ and\
  \bibinfo {author} {\bibfnamefont {Zi~Yang}\ \bibnamefont {Meng}},\
  }\href@noop {} {\enquote {\bibinfo {title} {{Higher-form symmetry breaking at
  Ising transitions}},}\ } (\bibinfo {year} {2021}),\ \Eprint
  {http://arxiv.org/abs/2011.12543} {arXiv:2011.12543 [cond-mat.str-el]}
  \BibitemShut {NoStop}%
\bibitem [{\citenamefont {Wu}\ \emph {et~al.}(2021)\citenamefont {Wu},
  \citenamefont {Jian},\ and\ \citenamefont {Xu}}]{Wu21}%
  \BibitemOpen
  \bibfield  {author} {\bibinfo {author} {\bibfnamefont {Xiao-Chuan}\
  \bibnamefont {Wu}}, \bibinfo {author} {\bibfnamefont {Chao-Ming}\
  \bibnamefont {Jian}}, \ and\ \bibinfo {author} {\bibfnamefont {Cenke}\
  \bibnamefont {Xu}},\ }\href@noop {} {\enquote {\bibinfo {title} {{Universal
  Features of Higher-Form Symmetries at Phase Transitions}},}\ } (\bibinfo
  {year} {2021}),\ \Eprint {http://arxiv.org/abs/2101.10342} {arXiv:2101.10342
  [cond-mat.str-el]} \BibitemShut {NoStop}%
\bibitem [{\citenamefont {Cardy}(1986{\natexlab{b}})}]{Cardy86operatorcontent}%
  \BibitemOpen
  \bibfield  {author} {\bibinfo {author} {\bibfnamefont {John~L.}\ \bibnamefont
  {Cardy}},\ }\bibfield  {title} {\enquote {\bibinfo {title} {Operator content
  of two-dimensional conformally invariant theories},}\ }\href {\doibase
  https://doi.org/10.1016/0550-3213(86)90552-3} {\bibfield  {journal} {\bibinfo
   {journal} {Nuclear Physics B}\ }\textbf {\bibinfo {volume} {270}},\ \bibinfo
  {pages} {186--204} (\bibinfo {year} {1986}{\natexlab{b}})}\BibitemShut
  {NoStop}%
\bibitem [{\citenamefont {Schuler}\ \emph {et~al.}(2016)\citenamefont
  {Schuler}, \citenamefont {Whitsitt}, \citenamefont {Henry}, \citenamefont
  {Sachdev},\ and\ \citenamefont {L\"auchli}}]{Schuler16}%
  \BibitemOpen
  \bibfield  {author} {\bibinfo {author} {\bibfnamefont {Michael}\ \bibnamefont
  {Schuler}}, \bibinfo {author} {\bibfnamefont {Seth}\ \bibnamefont
  {Whitsitt}}, \bibinfo {author} {\bibfnamefont {Louis-Paul}\ \bibnamefont
  {Henry}}, \bibinfo {author} {\bibfnamefont {Subir}\ \bibnamefont {Sachdev}},
  \ and\ \bibinfo {author} {\bibfnamefont {Andreas~M.}\ \bibnamefont
  {L\"auchli}},\ }\bibfield  {title} {\enquote {\bibinfo {title} {{Universal
  Signatures of Quantum Critical Points from Finite-Size Torus Spectra: A
  Window into the Operator Content of Higher-Dimensional Conformal Field
  Theories}},}\ }\href {\doibase 10.1103/PhysRevLett.117.210401} {\bibfield
  {journal} {\bibinfo  {journal} {Phys. Rev. Lett.}\ }\textbf {\bibinfo
  {volume} {117}},\ \bibinfo {pages} {210401} (\bibinfo {year}
  {2016})}\BibitemShut {NoStop}%
\bibitem [{\citenamefont {Thomson}\ and\ \citenamefont
  {Sachdev}(2017)}]{Thomson17}%
  \BibitemOpen
  \bibfield  {author} {\bibinfo {author} {\bibfnamefont {Alex}\ \bibnamefont
  {Thomson}}\ and\ \bibinfo {author} {\bibfnamefont {Subir}\ \bibnamefont
  {Sachdev}},\ }\bibfield  {title} {\enquote {\bibinfo {title} {Spectrum of
  conformal gauge theories on a torus},}\ }\href {\doibase
  10.1103/PhysRevB.95.205128} {\bibfield  {journal} {\bibinfo  {journal} {Phys.
  Rev. B}\ }\textbf {\bibinfo {volume} {95}},\ \bibinfo {pages} {205128}
  (\bibinfo {year} {2017})}\BibitemShut {NoStop}%
\bibitem [{\citenamefont {Whitsitt}\ \emph {et~al.}(2017)\citenamefont
  {Whitsitt}, \citenamefont {Schuler}, \citenamefont {Henry}, \citenamefont
  {L\"auchli},\ and\ \citenamefont {Sachdev}}]{Whitsitt17}%
  \BibitemOpen
  \bibfield  {author} {\bibinfo {author} {\bibfnamefont {Seth}\ \bibnamefont
  {Whitsitt}}, \bibinfo {author} {\bibfnamefont {Michael}\ \bibnamefont
  {Schuler}}, \bibinfo {author} {\bibfnamefont {Louis-Paul}\ \bibnamefont
  {Henry}}, \bibinfo {author} {\bibfnamefont {Andreas~M.}\ \bibnamefont
  {L\"auchli}}, \ and\ \bibinfo {author} {\bibfnamefont {Subir}\ \bibnamefont
  {Sachdev}},\ }\bibfield  {title} {\enquote {\bibinfo {title} {{Spectrum of
  the Wilson-Fisher conformal field theory on the torus}},}\ }\href {\doibase
  10.1103/PhysRevB.96.035142} {\bibfield  {journal} {\bibinfo  {journal} {Phys.
  Rev. B}\ }\textbf {\bibinfo {volume} {96}},\ \bibinfo {pages} {035142}
  (\bibinfo {year} {2017})}\BibitemShut {NoStop}%
\bibitem [{\citenamefont {Belin}\ \emph {et~al.}(2018)\citenamefont {Belin},
  \citenamefont {de~Boer},\ and\ \citenamefont {Kruthoff}}]{Belin18}%
  \BibitemOpen
  \bibfield  {author} {\bibinfo {author} {\bibfnamefont {Alexandre}\
  \bibnamefont {Belin}}, \bibinfo {author} {\bibfnamefont {Jan}\ \bibnamefont
  {de~Boer}}, \ and\ \bibinfo {author} {\bibfnamefont {Jorrit}\ \bibnamefont
  {Kruthoff}},\ }\bibfield  {title} {\enquote {\bibinfo {title} {{Comments on a
  state-operator correspondence for the torus}},}\ }\href {\doibase
  10.21468/SciPostPhys.5.6.060} {\bibfield  {journal} {\bibinfo  {journal}
  {SciPost Phys.}\ }\textbf {\bibinfo {volume} {5}},\ \bibinfo {pages} {60}
  (\bibinfo {year} {2018})}\BibitemShut {NoStop}%
\bibitem [{\citenamefont {Schuler}\ \emph {et~al.}(2021)\citenamefont
  {Schuler}, \citenamefont {Hesselmann}, \citenamefont {Whitsitt},
  \citenamefont {Lang}, \citenamefont {Wessel},\ and\ \citenamefont
  {L\"auchli}}]{Schuler21}%
  \BibitemOpen
  \bibfield  {author} {\bibinfo {author} {\bibfnamefont {Michael}\ \bibnamefont
  {Schuler}}, \bibinfo {author} {\bibfnamefont {Stephan}\ \bibnamefont
  {Hesselmann}}, \bibinfo {author} {\bibfnamefont {Seth}\ \bibnamefont
  {Whitsitt}}, \bibinfo {author} {\bibfnamefont {Thomas~C.}\ \bibnamefont
  {Lang}}, \bibinfo {author} {\bibfnamefont {Stefan}\ \bibnamefont {Wessel}}, \
  and\ \bibinfo {author} {\bibfnamefont {Andreas~M.}\ \bibnamefont
  {L\"auchli}},\ }\bibfield  {title} {\enquote {\bibinfo {title} {{Torus
  spectroscopy of the Gross-Neveu-Yukawa quantum field theory: Free Dirac
  versus chiral Ising fixed point}},}\ }\href {\doibase
  10.1103/PhysRevB.103.125128} {\bibfield  {journal} {\bibinfo  {journal}
  {Phys. Rev. B}\ }\textbf {\bibinfo {volume} {103}},\ \bibinfo {pages}
  {125128} (\bibinfo {year} {2021})}\BibitemShut {NoStop}%
\bibitem [{\citenamefont {Bl\"ote}\ and\ \citenamefont
  {Deng}(2002)}]{Bloete02}%
  \BibitemOpen
  \bibfield  {author} {\bibinfo {author} {\bibfnamefont {Henk W.~J.}\
  \bibnamefont {Bl\"ote}}\ and\ \bibinfo {author} {\bibfnamefont {Youjin}\
  \bibnamefont {Deng}},\ }\bibfield  {title} {\enquote {\bibinfo {title}
  {{Cluster Monte Carlo simulation of the transverse Ising model}},}\ }\href
  {\doibase 10.1103/PhysRevE.66.066110} {\bibfield  {journal} {\bibinfo
  {journal} {Phys. Rev. E}\ }\textbf {\bibinfo {volume} {66}},\ \bibinfo
  {pages} {066110} (\bibinfo {year} {2002})}\BibitemShut {NoStop}%
\bibitem [{\citenamefont {Yoshida}(2016)}]{Yoshida16}%
  \BibitemOpen
  \bibfield  {author} {\bibinfo {author} {\bibfnamefont {Beni}\ \bibnamefont
  {Yoshida}},\ }\bibfield  {title} {\enquote {\bibinfo {title} {Topological
  phases with generalized global symmetries},}\ }\href {\doibase
  10.1103/PhysRevB.93.155131} {\bibfield  {journal} {\bibinfo  {journal} {Phys.
  Rev. B}\ }\textbf {\bibinfo {volume} {93}},\ \bibinfo {pages} {155131}
  (\bibinfo {year} {2016})}\BibitemShut {NoStop}%
\bibitem [{\citenamefont {Yoshida}(2017)}]{Yoshida17}%
  \BibitemOpen
  \bibfield  {author} {\bibinfo {author} {\bibfnamefont {Beni}\ \bibnamefont
  {Yoshida}},\ }\bibfield  {title} {\enquote {\bibinfo {title} {Gapped
  boundaries, group cohomology and fault-tolerant logical gates},}\ }\href
  {\doibase https://doi.org/10.1016/j.aop.2016.12.014} {\bibfield  {journal}
  {\bibinfo  {journal} {Annals of Physics}\ }\textbf {\bibinfo {volume}
  {377}},\ \bibinfo {pages} {387--413} (\bibinfo {year} {2017})}\BibitemShut
  {NoStop}%
\bibitem [{\citenamefont {Chester}\ \emph {et~al.}(2020)\citenamefont
  {Chester}, \citenamefont {Landry}, \citenamefont {Liu}, \citenamefont
  {Poland}, \citenamefont {Simmons-Duffin}, \citenamefont {Su},\ and\
  \citenamefont {Vichi}}]{Chester20}%
  \BibitemOpen
  \bibfield  {author} {\bibinfo {author} {\bibfnamefont {Shai~M.}\ \bibnamefont
  {Chester}}, \bibinfo {author} {\bibfnamefont {Walter}\ \bibnamefont
  {Landry}}, \bibinfo {author} {\bibfnamefont {Junyu}\ \bibnamefont {Liu}},
  \bibinfo {author} {\bibfnamefont {David}\ \bibnamefont {Poland}}, \bibinfo
  {author} {\bibfnamefont {David}\ \bibnamefont {Simmons-Duffin}}, \bibinfo
  {author} {\bibfnamefont {Ning}\ \bibnamefont {Su}}, \ and\ \bibinfo {author}
  {\bibfnamefont {Alessandro}\ \bibnamefont {Vichi}},\ }\href@noop {} {\enquote
  {\bibinfo {title} {{Bootstrapping Heisenberg Magnets and their Cubic
  Instability}},}\ } (\bibinfo {year} {2020}),\ \Eprint
  {http://arxiv.org/abs/2011.14647} {arXiv:2011.14647 [hep-th]} \BibitemShut
  {NoStop}%
\bibitem [{\citenamefont {Aharony}(1973)}]{Aharony73}%
  \BibitemOpen
  \bibfield  {author} {\bibinfo {author} {\bibfnamefont {Amnon}\ \bibnamefont
  {Aharony}},\ }\bibfield  {title} {\enquote {\bibinfo {title} {{Critical
  Behavior of Anisotropic Cubic Systems}},}\ }\href {\doibase
  10.1103/PhysRevB.8.4270} {\bibfield  {journal} {\bibinfo  {journal} {Phys.
  Rev. B}\ }\textbf {\bibinfo {volume} {8}},\ \bibinfo {pages} {4270--4273}
  (\bibinfo {year} {1973})}\BibitemShut {NoStop}%
\bibitem [{\citenamefont {Manuel~Carmona}\ \emph {et~al.}(2000)\citenamefont
  {Manuel~Carmona}, \citenamefont {Pelissetto},\ and\ \citenamefont
  {Vicari}}]{Carmona00}%
  \BibitemOpen
  \bibfield  {author} {\bibinfo {author} {\bibfnamefont {Jos\'e}\ \bibnamefont
  {Manuel~Carmona}}, \bibinfo {author} {\bibfnamefont {Andrea}\ \bibnamefont
  {Pelissetto}}, \ and\ \bibinfo {author} {\bibfnamefont {Ettore}\ \bibnamefont
  {Vicari}},\ }\bibfield  {title} {\enquote {\bibinfo {title} {{$N$-component
  Ginzburg-Landau Hamiltonian with cubic anisotropy: A six-loop study}},}\
  }\href {\doibase 10.1103/PhysRevB.61.15136} {\bibfield  {journal} {\bibinfo
  {journal} {Phys. Rev. B}\ }\textbf {\bibinfo {volume} {61}},\ \bibinfo
  {pages} {15136--15151} (\bibinfo {year} {2000})}\BibitemShut {NoStop}%
\bibitem [{\citenamefont {Hasenbusch}\ and\ \citenamefont
  {Vicari}(2011)}]{Hasenbusch11}%
  \BibitemOpen
  \bibfield  {author} {\bibinfo {author} {\bibfnamefont {Martin}\ \bibnamefont
  {Hasenbusch}}\ and\ \bibinfo {author} {\bibfnamefont {Ettore}\ \bibnamefont
  {Vicari}},\ }\bibfield  {title} {\enquote {\bibinfo {title} {Anisotropic
  perturbations in three-dimensional o($n$)-symmetric vector models},}\ }\href
  {\doibase 10.1103/PhysRevB.84.125136} {\bibfield  {journal} {\bibinfo
  {journal} {Phys. Rev. B}\ }\textbf {\bibinfo {volume} {84}},\ \bibinfo
  {pages} {125136} (\bibinfo {year} {2011})}\BibitemShut {NoStop}%
\bibitem [{\citenamefont {Adzhemyan}\ \emph {et~al.}(2019)\citenamefont
  {Adzhemyan}, \citenamefont {Ivanova}, \citenamefont {Kompaniets},
  \citenamefont {Kudlis},\ and\ \citenamefont {Sokolov}}]{Adzhemyan19}%
  \BibitemOpen
  \bibfield  {author} {\bibinfo {author} {\bibfnamefont {Loran~Ts.}\
  \bibnamefont {Adzhemyan}}, \bibinfo {author} {\bibfnamefont {Ella~V.}\
  \bibnamefont {Ivanova}}, \bibinfo {author} {\bibfnamefont {Mikhail~V.}\
  \bibnamefont {Kompaniets}}, \bibinfo {author} {\bibfnamefont {Andrey}\
  \bibnamefont {Kudlis}}, \ and\ \bibinfo {author} {\bibfnamefont
  {Aleksandr~I.}\ \bibnamefont {Sokolov}},\ }\bibfield  {title} {\enquote
  {\bibinfo {title} {Six-loop epsilon expansion study of three-dimensional
  n-vector model with cubic anisotropy},}\ }\href {\doibase
  https://doi.org/10.1016/j.nuclphysb.2019.02.001} {\bibfield  {journal}
  {\bibinfo  {journal} {Nuclear Physics B}\ }\textbf {\bibinfo {volume}
  {940}},\ \bibinfo {pages} {332--350} (\bibinfo {year} {2019})}\BibitemShut
  {NoStop}%
\bibitem [{\citenamefont {Dupont}\ \emph {et~al.}(2021)\citenamefont {Dupont},
  \citenamefont {Gazit},\ and\ \citenamefont {Scaffidi}}]{Dupont21}%
  \BibitemOpen
  \bibfield  {author} {\bibinfo {author} {\bibfnamefont {Maxime}\ \bibnamefont
  {Dupont}}, \bibinfo {author} {\bibfnamefont {Snir}\ \bibnamefont {Gazit}}, \
  and\ \bibinfo {author} {\bibfnamefont {Thomas}\ \bibnamefont {Scaffidi}},\
  }\bibfield  {title} {\enquote {\bibinfo {title} {{From trivial to topological
  paramagnets: The case of ${\mathbb{Z}}_{2}$ and ${\mathbb{Z}}_{2}^{3}$
  symmetries in two dimensions}},}\ }\href {\doibase
  10.1103/PhysRevB.103.144437} {\bibfield  {journal} {\bibinfo  {journal}
  {Phys. Rev. B}\ }\textbf {\bibinfo {volume} {103}},\ \bibinfo {pages}
  {144437} (\bibinfo {year} {2021})}\BibitemShut {NoStop}%
\bibitem [{\citenamefont {Vidal}(2008)}]{Vidal08}%
  \BibitemOpen
  \bibfield  {author} {\bibinfo {author} {\bibfnamefont {G.}~\bibnamefont
  {Vidal}},\ }\bibfield  {title} {\enquote {\bibinfo {title} {{Class of Quantum
  Many-Body States That Can Be Efficiently Simulated}},}\ }\href {\doibase
  10.1103/PhysRevLett.101.110501} {\bibfield  {journal} {\bibinfo  {journal}
  {Phys. Rev. Lett.}\ }\textbf {\bibinfo {volume} {101}},\ \bibinfo {pages}
  {110501} (\bibinfo {year} {2008})}\BibitemShut {NoStop}%
\bibitem [{\citenamefont {Zhang}\ and\ \citenamefont {Wang}(2017)}]{Zhang17}%
  \BibitemOpen
  \bibfield  {author} {\bibinfo {author} {\bibfnamefont {Long}\ \bibnamefont
  {Zhang}}\ and\ \bibinfo {author} {\bibfnamefont {Fa}~\bibnamefont {Wang}},\
  }\bibfield  {title} {\enquote {\bibinfo {title} {{Unconventional Surface
  Critical Behavior Induced by a Quantum Phase Transition from the
  Two-Dimensional Affleck-Kennedy-Lieb-Tasaki Phase to a N\'eel-Ordered
  Phase}},}\ }\href {\doibase 10.1103/PhysRevLett.118.087201} {\bibfield
  {journal} {\bibinfo  {journal} {Phys. Rev. Lett.}\ }\textbf {\bibinfo
  {volume} {118}},\ \bibinfo {pages} {087201} (\bibinfo {year}
  {2017})}\BibitemShut {NoStop}%
\bibitem [{\citenamefont {Weber}\ \emph {et~al.}(2018)\citenamefont {Weber},
  \citenamefont {Parisen~Toldin},\ and\ \citenamefont {Wessel}}]{Weber18}%
  \BibitemOpen
  \bibfield  {author} {\bibinfo {author} {\bibfnamefont {Lukas}\ \bibnamefont
  {Weber}}, \bibinfo {author} {\bibfnamefont {Francesco}\ \bibnamefont
  {Parisen~Toldin}}, \ and\ \bibinfo {author} {\bibfnamefont {Stefan}\
  \bibnamefont {Wessel}},\ }\bibfield  {title} {\enquote {\bibinfo {title}
  {Nonordinary edge criticality of two-dimensional quantum critical magnets},}\
  }\href {\doibase 10.1103/PhysRevB.98.140403} {\bibfield  {journal} {\bibinfo
  {journal} {Phys. Rev. B}\ }\textbf {\bibinfo {volume} {98}},\ \bibinfo
  {pages} {140403} (\bibinfo {year} {2018})}\BibitemShut {NoStop}%
\bibitem [{\citenamefont {Xu}\ \emph {et~al.}(2020)\citenamefont {Xu},
  \citenamefont {Wu}, \citenamefont {Jian},\ and\ \citenamefont {Xu}}]{Xu20}%
  \BibitemOpen
  \bibfield  {author} {\bibinfo {author} {\bibfnamefont {Yichen}\ \bibnamefont
  {Xu}}, \bibinfo {author} {\bibfnamefont {Xiao-Chuan}\ \bibnamefont {Wu}},
  \bibinfo {author} {\bibfnamefont {Chao-Ming}\ \bibnamefont {Jian}}, \ and\
  \bibinfo {author} {\bibfnamefont {Cenke}\ \bibnamefont {Xu}},\ }\bibfield
  {title} {\enquote {\bibinfo {title} {Topological edge and interface states at
  bulk disorder-to-order quantum critical points},}\ }\href {\doibase
  10.1103/PhysRevB.101.184419} {\bibfield  {journal} {\bibinfo  {journal}
  {Phys. Rev. B}\ }\textbf {\bibinfo {volume} {101}},\ \bibinfo {pages}
  {184419} (\bibinfo {year} {2020})}\BibitemShut {NoStop}%
\bibitem [{\citenamefont {Wu}\ \emph {et~al.}(2020)\citenamefont {Wu},
  \citenamefont {Xu}, \citenamefont {Geng}, \citenamefont {Jian},\ and\
  \citenamefont {Xu}}]{Wu20}%
  \BibitemOpen
  \bibfield  {author} {\bibinfo {author} {\bibfnamefont {Xiao-Chuan}\
  \bibnamefont {Wu}}, \bibinfo {author} {\bibfnamefont {Yichen}\ \bibnamefont
  {Xu}}, \bibinfo {author} {\bibfnamefont {Hao}\ \bibnamefont {Geng}}, \bibinfo
  {author} {\bibfnamefont {Chao-Ming}\ \bibnamefont {Jian}}, \ and\ \bibinfo
  {author} {\bibfnamefont {Cenke}\ \bibnamefont {Xu}},\ }\bibfield  {title}
  {\enquote {\bibinfo {title} {Boundary criticality of topological quantum
  phase transitions in two-dimensional systems},}\ }\href {\doibase
  10.1103/PhysRevB.101.174406} {\bibfield  {journal} {\bibinfo  {journal}
  {Phys. Rev. B}\ }\textbf {\bibinfo {volume} {101}},\ \bibinfo {pages}
  {174406} (\bibinfo {year} {2020})}\BibitemShut {NoStop}%
\bibitem [{\citenamefont {Verresen}(2020)}]{Verresen20}%
  \BibitemOpen
  \bibfield  {author} {\bibinfo {author} {\bibfnamefont {Ruben}\ \bibnamefont
  {Verresen}},\ }\href@noop {} {\enquote {\bibinfo {title} {Topology and edge
  states survive quantum criticality between topological insulators},}\ }
  (\bibinfo {year} {2020}),\ \Eprint {http://arxiv.org/abs/2003.05453}
  {arXiv:2003.05453 [cond-mat.str-el]} \BibitemShut {NoStop}%
\bibitem [{\citenamefont {Thorngren}\ \emph {et~al.}(2021)\citenamefont
  {Thorngren}, \citenamefont {Vishwanath},\ and\ \citenamefont
  {Verresen}}]{Thorngren21}%
  \BibitemOpen
  \bibfield  {author} {\bibinfo {author} {\bibfnamefont {Ryan}\ \bibnamefont
  {Thorngren}}, \bibinfo {author} {\bibfnamefont {Ashvin}\ \bibnamefont
  {Vishwanath}}, \ and\ \bibinfo {author} {\bibfnamefont {Ruben}\ \bibnamefont
  {Verresen}},\ }\bibfield  {title} {\enquote {\bibinfo {title} {Intrinsically
  gapless topological phases},}\ }\href {\doibase 10.1103/PhysRevB.104.075132}
  {\bibfield  {journal} {\bibinfo  {journal} {Phys. Rev. B}\ }\textbf {\bibinfo
  {volume} {104}},\ \bibinfo {pages} {075132} (\bibinfo {year}
  {2021})}\BibitemShut {NoStop}%
\bibitem [{\citenamefont {Anand}\ \emph {et~al.}(2017)\citenamefont {Anand},
  \citenamefont {Genest}, \citenamefont {Katz}, \citenamefont {Khandker},\ and\
  \citenamefont {Walters}}]{Anand17}%
  \BibitemOpen
  \bibfield  {author} {\bibinfo {author} {\bibfnamefont {Nikhil}\ \bibnamefont
  {Anand}}, \bibinfo {author} {\bibfnamefont {Vincent~X.}\ \bibnamefont
  {Genest}}, \bibinfo {author} {\bibfnamefont {Emanuel}\ \bibnamefont {Katz}},
  \bibinfo {author} {\bibfnamefont {Zuhair~U.}\ \bibnamefont {Khandker}}, \
  and\ \bibinfo {author} {\bibfnamefont {Matthew~T.}\ \bibnamefont {Walters}},\
  }\bibfield  {title} {\enquote {\bibinfo {title} {{RG flow from $\varphi^4$
  theory to the 2D Ising model}},}\ }\href
  {https://doi.org/10.1007/JHEP08(2017)056} {\bibfield  {journal} {\bibinfo
  {journal} {Journal of High Energy Physics}\ }\textbf {\bibinfo {volume}
  {2017}},\ \bibinfo {pages} {56} (\bibinfo {year} {2017})}\BibitemShut
  {NoStop}%
\bibitem [{\citenamefont {Dijkgraaf}\ \emph {et~al.}(1988)\citenamefont
  {Dijkgraaf}, \citenamefont {Verlinde},\ and\ \citenamefont
  {Verlinde}}]{Dijkgraaf88}%
  \BibitemOpen
  \bibfield  {author} {\bibinfo {author} {\bibfnamefont {Robbert}\ \bibnamefont
  {Dijkgraaf}}, \bibinfo {author} {\bibfnamefont {Erik}\ \bibnamefont
  {Verlinde}}, \ and\ \bibinfo {author} {\bibfnamefont {Herman}\ \bibnamefont
  {Verlinde}},\ }\bibfield  {title} {\enquote {\bibinfo {title} {{$c=1$
  conformal field theories on Riemann surfaces}},}\ }\href
  {https://projecteuclid.org:443/euclid.cmp/1104161089} {\bibfield  {journal}
  {\bibinfo  {journal} {Comm. Math. Phys.}\ }\textbf {\bibinfo {volume}
  {115}},\ \bibinfo {pages} {649--690} (\bibinfo {year} {1988})}\BibitemShut
  {NoStop}%
\bibitem [{\citenamefont {Ginsparg}(1988)}]{Ginsparg88b}%
  \BibitemOpen
  \bibfield  {author} {\bibinfo {author} {\bibfnamefont {P.}~\bibnamefont
  {Ginsparg}},\ }\bibfield  {title} {\enquote {\bibinfo {title} {Curiosities at
  $c = 1$},}\ }\href {\doibase https://doi.org/10.1016/0550-3213(88)90249-0}
  {\bibfield  {journal} {\bibinfo  {journal} {Nuclear Physics B}\ }\textbf
  {\bibinfo {volume} {295}},\ \bibinfo {pages} {153 -- 170} (\bibinfo {year}
  {1988})}\BibitemShut {NoStop}%
\bibitem [{\citenamefont {Kiritsis}(1989)}]{Kiritsis89}%
  \BibitemOpen
  \bibfield  {author} {\bibinfo {author} {\bibfnamefont {Elias~B.}\
  \bibnamefont {Kiritsis}},\ }\bibfield  {title} {\enquote {\bibinfo {title}
  {Proof of the completeness of the classification of rational conformal
  theories with c=1},}\ }\href {\doibase
  https://doi.org/10.1016/0370-2693(89)90073-7} {\bibfield  {journal} {\bibinfo
   {journal} {Physics Letters B}\ }\textbf {\bibinfo {volume} {217}},\ \bibinfo
  {pages} {427 -- 430} (\bibinfo {year} {1989})}\BibitemShut {NoStop}%
\end{thebibliography}%


\widetext

\appendix
\setcounter{equation}{0}
\setcounter{figure}{0}
\setcounter{table}{0}
\renewcommand{\thefigure}{\thesection\arabic{figure}}

\setcounter{figure}{0}
\section{Symmetry fluxes and their charges \label{app:symflux}}

\subsection{Symmetry properties of unique symmetry fluxes}

For any $g \in G$ and $h \in C(g)$, we define $\chi_g(h)$ through $U^h \mathcal S^g U^{h\dagger} = \chi_g(h) \mathcal S^g$ (presuming we have chosen the endpoint operator $\mathcal O^g$ such that $C(g)$ acts nicely on it; see the discussion in Section~\ref{sec:symflux}). From this definition, it directly follows that $\chi_g(hk) = \chi_g(h) \chi_g(k)$, i.e., $\chi_g: C(g) \to U(1)$ is a one-dimensional representation. Other useful properties---which hold for bosonic systems which are gapped or described by a CFT---are $\chi_g(h) = \chi_h(g)^{-1}$ and $\chi_g(g) =1 $. In the gapped case, these can be derived from the concept of symmetry fractionalization. More generally, these can be argued based on modular invariance of the partition function. Note that $\chi_g(g)=1$ need not be true for fermionic systems: the Kitaev chain is a paradigmatic example where the symmetry flux of $P$ is charged under itself!

\subsection{From abelian charges to cocycles \label{app:chargetococycle}}

Here we show how specifying the above charge $\chi_g$ for any $g \in G$ is equivalent to specifying a projective representation of $G$ if $G$ is abelian. 

Due to the structure theorem, we have $G \cong \mathbb Z_{r_1} \times \cdots \mathbb Z_{r_n}$ (for convenience we take $G$ to be finite). Let $g_1$, ..., $g_n$ be a set of generators. We now define a central extension of $G$, which is a group $\tilde G$ generated by the symbols $\hat g_1$, ..., $\hat g_n$, and complex phases. To define the relations between these generating elements, it is useful to introduce the shorthand $[a,b] \equiv a b a^{-1} b^{-1}$. The relations of $\tilde G$ are then $\hat g_j^{r_j} \equiv 1$ and $[\hat g_j, \hat g_k] \equiv \chi_{g_k}(g_j)$. A priori, it is not trivial that this definition is consistent, since there are non-trivial relationships between commutators. In particular: $[a,b] = [b,a]^{-1}$ and $[a,bc] = [a,b] [b,[a,c]] [a,c]$ and $[g,g^{-1}] = 1$. However, the properties of $\chi_g$ mentioned in the previous subsection indeed show that the consistency relations are satisfied.

We have thus defined a central extension $U(1) \to \hat G \to G$. This short exact sequence simply means that $G \cong \tilde G / U(1)$, as one can readily verify. This is equivalent to defining a projective representation of $G$. The latter is often characterized in terms of a cocycle $\omega(g,h)$. The standard way of obtaining this from the extension, is by first defining a section $s: G \to \hat G$, i.e., an embedding of the original group into the extended one. It is sufficient to define this on the products of the generating elements: $s(g_1^{k_1} \cdots g_n^{k_n}) \equiv \hat g_1^{k_1} \cdots \hat g_n^{k_n}$. The cocycle is then determined via $\omega(g,h) = s(g) s(h) s(gh)^{-1}$.

\subsection{Gapped symmetries \label{app:gappedsym}}

In this subsection, we focus on symmetries which act only on gapped degrees of freedom.

\subsubsection{Symmetry flux from symmetry fractionalization: string order parameter}

For gapped symmetries, there is the notion of symmetry fractionalization \cite{Fidkowski11class,Turner11class}. This says that if one acts with the symmetry operator on a finite but large region, it effectively only acts non-trivially near the edges: $U^g_m U^g_{m+1} \cdots U^g_{n-1} = U^g_L U^g_R$. These obey a \emph{projective} representation $U^g_R U^h_R = e^{i\omega(g,h)} U^{gh}_R$; here $\omega(\cdot,\cdot)$ characterizes the so-called second group cohomology class. These fractional symmetries, $U^g_L$ and $U^g_R$, might have non-trivial symmetry properties which force their expectation value to be zero. A string order parameter is then usually defined by finding an operator $\mathcal O^g$ that cancels these symmetry properties such that $\langle \mathcal O^{g\dagger}_m  U^g_L \rangle \neq 0$ and $\langle U^g_R \mathcal O^g_n \rangle \neq 0$. The resulting string order parameter is then $\mathcal O^{g\dagger}_m  U^g_{m} U^g_{m+1} \cdots U^g_{n-1} \mathcal O^g_n$ which has long-range order \emph{by construction}. Note that this exactly satisfies the condition for the symmetry flux of $g$ as defined in Section~\ref{subsec:symfluxdef}.

\subsubsection{Uniqueness of symmetry flux}

Suppose one has a second operator $\tilde{\mathcal O}$ that satisfies the same properties. In particular, $\langle \tilde{\mathcal O}^{g\dagger}_m  U^g_L \rangle \neq 0$ and $\langle U^g_R \tilde{\mathcal O}^g_n \rangle \neq 0$. Then the linear combination $\mathcal T^g_n := \langle U^g_R \tilde{\mathcal O}^g_n \rangle \times \mathcal S^g_n - \langle U^g_R {\mathcal O}^g_n \rangle \times \tilde{\mathcal S}^g_n$ no longer has long-range order, and thus by the definition of the equivalence class in Section~\ref{sec:symflux}, the symmetry fluxes $\mathcal S^g$ and $\tilde{\mathcal S^g}$ generate the same class, i.e., the symmetry flux is unique. To prove the above claim:
\begin{align}
\langle \mathcal T^{g\dagger}_m \mathcal T^{g\vphantom\dagger}_n \rangle &= \left\langle \big( \langle U^g_R \tilde{\mathcal O}^g_n \rangle \mathcal O^g_m - \langle U^g_R {\mathcal O}^g_n \rangle  \tilde{\mathcal O}^g_m \big)^\dagger U^g_m U^g_{m+1} \cdots U^g_{n-1}\big( \langle U^g_R \tilde{\mathcal O}^g_n \rangle \mathcal O^g_n - \langle U^g_R {\mathcal O}^g_n \rangle  \tilde{\mathcal O}^g_n \big) \right\rangle \\
&= \langle \cdots \rangle \times \left\langle U^g_R \big( \langle U^g_R \tilde{\mathcal O}^g_n \rangle \mathcal O^g_n - \langle U^g_R {\mathcal O}^g_n \rangle  \tilde{\mathcal O}^g_n \big) \right\rangle =  \langle \cdots \rangle \times \left( \langle U^g_R \tilde{\mathcal O}^g_n \rangle  \langle U^g_R \mathcal O^g_n \rangle - \langle U^g_R {\mathcal O}^g_n \rangle  \langle U^g_R \tilde{\mathcal O}^g_n \rangle \right) =0.
\end{align}

\subsubsection{Charges from the cocycle}

Having shown the uniqueness of the symmetry flux of $g \in G$, we can now consider its symmetry properties. For any $h \in C(g)$ (i.e., the elements of commuting with $g$), we can consider $U^h \mathcal S^g \left( U^h \right)^\dagger = \chi_g(h) \mathcal S^g$ with $\chi_g(h) \in U(1)$. As noted before, the function $\chi_g: C(g) \to U(1): h \mapsto \chi_g(h)$ is a one-dimensional representation of the stabilizer of $g$. We now show that $\chi_g(h) = e^{-i(\omega(h,g) + \omega(hg,h^{-1}))}$. By definition of $\mathcal O^g$, it has the opposite symmetry property of $U_R^g$ (indeed, otherwise $\mathcal O^g U^g_R$ could not have a nonzero expectation value), hence $U^h U_R^g U^{h\dagger} = \chi_g^*(h) U_R^g$. The left-hand side equals $U_R^h U_R^g U^{h\dagger}_R = e^{i\omega(h,g)} U_R^{hg} U^{h^{-1}}_R = e^{i(\omega(h,g) + \omega(hg,h^{-1}))} U^g_R$. QED.

Note that in Section~\ref{app:chargetococycle} we proved that this relationship can be inverted if $G$ is abelian.

\setcounter{figure}{0}
\section{Symmetry-enriched Ising CFT in an exactly solvable model \label{app:freeferm}}

Here we solve the Hamiltonian $H = J_1 H_1 + J_\textrm{Hal} H_\textrm{Hal} + J_x H_x$ by mapping it to free fermions. We use the formalism of Ref.~\onlinecite{Verresen18} to obtain a simple solution that straightforwardly allows us to extract bulk correlation lengths and edge mode localization lengths from zeros of holomorphic functions.

After a Jordan-Wigner transformation, we have a quadratic fermionic chain. More generally, when there is a two-site unit cell, it is useful to write the Hamiltonian as follows:
\begin{equation}
H = -i \sum_{n \in \textrm{sites}} \sum_{\alpha \in \mathbb Z} \left( \tilde \gamma_{2n-1}, \tilde \gamma_{2n} \right) T_\alpha \left( \begin{array}{cc} \gamma_{2(n+\alpha)-1} \\ \gamma_{2(n+\alpha)} \end{array} \right).
\end{equation}

In particular, for the above Hamiltonian, we obtain
\begin{equation}
T_{-1} = \left( \begin{array}{cc} 0 & J_H \\ 0 & 0 \end{array} \right),
\qquad 
T_0 = \left( \begin{array}{cc} 0 & J_1 + J_x \\ J_1 & 0 \end{array} \right),
\qquad 
T_1 = \left( \begin{array}{cc} 0 & 0 \\ J_H + J_x & 0 \end{array} \right).
\end{equation}

As in Ref.~\onlinecite{Verresen18}, it is useful to consider
\begin{equation}
F(z) := \sum_\alpha T_\alpha z^\alpha = \left( \begin{array}{cc} 0 & J_1 + J_x + J_H/z \\ (J_H + J_x )z + J_1 & 0 \end{array} \right) = \left( \begin{array}{cc} 0 & h(z) \\ g(z) & 0 \end{array} \right),
\end{equation}
where we have defined $h(z) = \frac{J_1 + J_x}{z} \left( z + \frac{J_H}{J_1 + J_x}  \right)$ and $g(z) = (J_H + J_x ) \left( z + \frac{J_1}{J_H + J_x } \right)$ (written in such a way that their zeros can be read off). In the style of Ref.~\onlinecite{Verresen18}, we can associate a correlation length to both functions (the extra factor of two in the numerator is due to the two-site unit cell):
\begin{equation}
\xi_h = \frac{2}{\left| \ln \left| \frac{J_H}{J_1 + J_x} \right| \right|}
\qquad \textrm{and} \qquad
\xi_g = \frac{2}{\left| \ln \left| \frac{J_1}{J_H + J_x} \right| \right|}. \label{eq:ximodel}
\end{equation}
`The' correlation length of the system is then given by $\xi = \max \{\xi_h,\xi_g\}$.

\begin{figure}[h]
\includegraphics[scale=1]{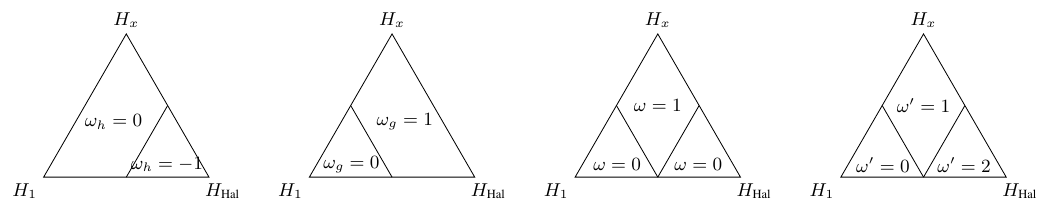}
\caption{The first two plots show the topological invariant associated to $h(z)$ and $g(z)$. The third plot shows the BDI topological invariant which can be expressed as $\omega = \omega_h + \omega_g$. In the last plot, we show $\omega' := |\omega_h| + |\omega_g|$, which is the one to consider in this case. \label{fig:omegas}}
\end{figure}

For edge modes and topological invariants, it is useful to consider the winding number $\omega = N_z - N_p$ associated to $h(z)$ and $g(z)$, denoted by $\omega_h$ and $\omega_g$ respectively. Their values are shown in the phase diagrams in Fig.~\ref{fig:omegas}. This figure clarifies that the topological phase transition is driven by $h(z)$, whereas the topology of $g(z)$ is stable throughout that same region. It is hence natural to conclude that $\xi_\textrm{loc} = \xi_g$. (This has also been confirmed with numerics.)

In Fig.~\ref{fig:phasediagram}(b) of the main text, the red path is parametrized by $J_x = 2\lambda J_1$ and $J_H = 2(1-\lambda) J_1$ (for $\lambda \in [0,1]$). Hence, $\frac{J_1}{J_H + J_x} = \frac{1}{2}$ and $\frac{J_H}{J_1+J_x} = \frac{1-\lambda}{\sfrac{1}{2}+\lambda}$ From Eq.~\eqref{eq:ximodel}, we conclude that the edge mode has constant localization length $\xi_\textrm{loc} = \sfrac{2}{\ln 2}$, whereas the bulk correlation length is
\begin{equation}
\xi = \left\{ \begin{array}{ccl}
\frac{2}{ \left| \ln\left( \frac{1-\lambda}{\sfrac{1}{2}+\lambda} \right) \right|} & & \textrm{for } \lambda \in [0,\sfrac{1}{2}] \\
\frac{2}{ \ln 2 } & & \textrm{for } \lambda \in [\sfrac{1}{2},1] .
\end{array}
\right.
\end{equation}

\setcounter{figure}{0}
\section{Duality mapping \texorpdfstring{$\bm{ (I_x,I_y,I_z) \leftrightarrow (1,\textrm{Hal},0) }$}{} \label{app:duality}}

One can define the following unitary mapping (ignoring boundary conditions):
\begin{align}
X_{2n+1} &= (-1)^n \tilde Z_1 \cdots \tilde Z_{2n-1} \tilde Z_{2n} \\
Y_{2n+1} &= (-1)^n \tilde Z_1 \cdots \tilde Z_{2n-1} \tilde Y_{2n} \tilde Y_{2n+1} \\
Z_{2n+1} &= -\tilde X_{2n} \tilde Y_{2n+1} \\
X_{2n+2} &= (-1)^n \tilde Z_1 \cdots \tilde Z_{2n-1} \tilde Z_{2n} \tilde X_{2n+1} \tilde X_{2n+2} \\
Y_{2n+2} &= (-1)^{n+1} \tilde Z_1 \cdots \tilde Z_{2n-1} \tilde Z_{2n} \tilde Z_{2n+1} \\
Z_{2n+2} &= \tilde Y_{2n+1} \tilde X_{2n+2}.
\end{align}
One can check that as thus defined, the operators satisfy the relevant algebra: operators on different sites commute, and on the same site they form a representation of the Pauli algebra.

From the above correspondences, one can derive:
\begin{align}
X_{2n} X_{2n+1} &= \tilde Y_{2n-1} \tilde Y_{2n} \\
Y_{2n} Y_{2n+1} &= \tilde Y_{2n} \tilde Y_{2n+1} \\
X_{2n+1} X_{2n+2} &= \tilde X_{2n+1} \tilde X_{2n+2} \\
Y_{2n+1} Y_{2n+2} &= \tilde X_{2n} \tilde X_{2n+1}.
\end{align}
This directly implies that $H_x \leftrightarrow H_1$ and $H_y \leftrightarrow H_\textrm{Hal}$.

Similarly, one can check:
\begin{align}
Z_{2n} Z_{2n+1} &= - \tilde Y_{2n-1} \tilde Y_{2n+1} \\
Z_{2n+1} Z_{2n+2} &= - \tilde X_{2n} \tilde X_{2n+2} \\
-Y_{2n} Y_{2n+2} &= \tilde Z_{2n} \tilde Z_{2n+1} \\
-X_{2n+1} X_{2n+3} &= \tilde Z_{2n+1} \tilde Z_{2n+2}.
\end{align}
Hence, $H_z \leftrightarrow H_0$. (Caveat: depending on which direction of the mapping one takes, $H_0$ has $XX$ couplings on even/odd sites. However, since all five other Hamiltonians are inversion symmetric (when inverting along bonds \emph{between} two-site unit cells), we can always concatenate with spatial inversion to obtain the desired variant.)

\setcounter{figure}{0}
\section{The bait-and-switch lemma \label{app:baitandswitch}}

Consider two $G$-enriched Ising CFTs, which we refer to as the A and B systems. Suppose that each has the same charge for their $\sigma$ operator. We now prove that if we stack the A-system which has been perturbed into its \emph{gapped} symmetry-preserving phase on top of the \emph{critical} B-system, then we can \emph{smoothly} connect this to the B-system in its \emph{gapped} symmetry-preserving phase stacked on top of the \emph{critical} A-system. Conceptually, this says that all (non-symmetry-breaking) $G$-enriched Ising CFTs with the same charges for local operators can be realized by stacking gapped SPT phases on top of a reference Ising CFT. (Note that in the presence of symmetry-breaking, one can apply this lemma to the remaining symmetry group.) This lemma can be seen as a generalization of Corollary 1 in the Appendix of \cite{Verresen18} to the interacting case.

It is convenient to use the representation of the Ising CFT as a $\phi^4$ theory. In particular, for a decoupled stack of the above two critical Ising CFTs, the Lagrangian would be
\begin{equation}
\mathcal L_0 = \sum_{i=A,B} \left( (\partial \phi_i)^2 - \phi_i^4 - m_c^2 \phi_i^2 \right).
\end{equation}
The parameter $m_c$ is taken such that we are at the Ising fixed point. (This is roughly $m_c^2 \approx 0.5$ according to Ref.~\onlinecite{Anand17}, but its precise value is not important to the argument.) We now show that the situation where (only) the A system is gapped, i.e. $\mathcal L = \mathcal L_0 - m^2 \phi_A^2$, can be smoothly connected to where (only) the B system is gapped, i.e. $\mathcal L = \mathcal L_0 - m^2 \phi_B^2$, preserving both the Ising universality class and the $G$ symmetry throughout.

Since by assumption the $\phi_A$ and $\phi_B$ fields have the same charges under each element of $G$, we can consider the following symmetric coupling (where $\theta \in [0,\pi/2]$ is a free parameter):
\begin{equation}
\mathcal L = \mathcal L_0 - m^2( \underbrace{ (\cos \theta) \phi_A + (\sin \theta) \phi_B }_{\equiv \varphi_1 } )^2 + f(m,\theta) ( \underbrace{-(\sin \theta) \phi_A + (\cos \theta) \phi_B}_{\equiv \varphi_2} )^2. \label{eq:phi4}
\end{equation}
For $\theta = 0$ (taking $f(m,0) = 0$), the A d.o.f. are indeed gapped out and decoupled from the critical B d.o.f., whereas for $\theta = \pi/2$ the roles are reversed. We now show how to define $f(m,\theta)$ to keep the system Ising critical for intermediate values of $\theta$.

If we express the Lagrangian in terms of the new fields $\varphi_1$ and $\varphi_2$ which were defined in Eq.~\eqref{eq:phi4}, we obtain
\begin{equation}
\mathcal L = \sum_{i=1,2} \left( (\partial \varphi_i)^2 - \left( \frac{3+\cos(4\theta)}{4} \right) \varphi_i^4 - m_c^2  \varphi_i^2 \right) - m^2\varphi_1^2 + f(m,\theta) \varphi_2^2 - V,
\end{equation}
where the coupling $V$ arises due to the quartic term, $V =\frac{3}{2} \left( 1- \cos(4\theta) \right) \varphi_1^2 \varphi_2^2 + \sin(4\theta) \left( \varphi_1 \varphi_2^3 - \varphi_1^3 \varphi_2 \right)$. If we work in the limit of large $m^2$, then $\varphi_1$ will be pinned to $\varphi_1 = 0$, such that
\begin{equation}
\mathcal L = (\partial \varphi_2)^2 - \left( \frac{3+\cos(4\theta)}{4} \right) \varphi_2^4 - \left( m_c^2 + \frac{3}{2}(1-\cos(4\theta)) \langle \varphi_1^2 \rangle_{m,\theta} - f(m,\theta) \right) \varphi_2^2. \label{eq:phi4b}
\end{equation}
Here $\langle \varphi_1^2 \rangle_{m,\theta}$ depends on the UV (lattice) scale $a$. (Note that there is no effective coupling through $\langle \varphi_1 \rangle$ or $\langle \varphi_1^3 \rangle$ since symmetry forces this to be zero.) The Lagrangian in Eq.~\eqref{eq:phi4b} is at the Ising critical point if we enforce the ratio
\begin{equation}
m_c^2 = \frac{m_c^2 + \frac{3}{2}(1-\cos(4\theta)) \langle \varphi_1^2 \rangle_{m,\theta} - f(m,\theta)}{\frac{3+\cos(4\theta)}{4}} \quad \Rightarrow \quad f(m,\theta) \equiv \sin^2(2\theta) \left( \frac{m_c^2}{2} + \langle \varphi_1^2 \rangle_{m,\theta} \right) .
\end{equation}
Note that when integrating out $\varphi_1$, higher-order corrections can be generated; these can be included, leading to a slight shift in $f(m,\theta)$. The main conceptual point is that we can reach (or better yet, stay on) criticality by tuning a single parameter.

\setcounter{figure}{0}
\section{The curious case of $c=1$ \label{app:gaussian}}

\subsection{The $\mathbb Z_2 \times \mathbb Z_2$-enriched $c=1$ CFTs of codimension two \label{app:core}}

We consider the compact boson CFT and the $c=1$ orbifold CFT; these are the main component of the moduli space of $c=1$ theories \cite{Dijkgraaf88,Ginsparg88,Ginsparg88b,Kiritsis89}. Focusing on CFTs of codimension two, we show that any two $\mathbb Z_2 \times \mathbb Z_2$-enriched versions of the aforementioned CFTs can be smoothly connected without abruptly having to change the universality class at some intermediate point (i.e., they are connected by a path of symmetric marginal perturbations). More precisely, we characterize all possible symmetry assignments of local operators and nonlocal symmetry fluxes and demonstrate that these form a connected moduli space. For the case of the Ising CFT, we proved that it is sufficient to consider such symmetry assignments (see the bait-and-switch lemma in Appendix~\ref{app:baitandswitch}); while we do not have a proof for the analogous statement for these $c=1$ CFTs, we suspect it to be true.

\begin{figure}
	\includegraphics[scale=1]{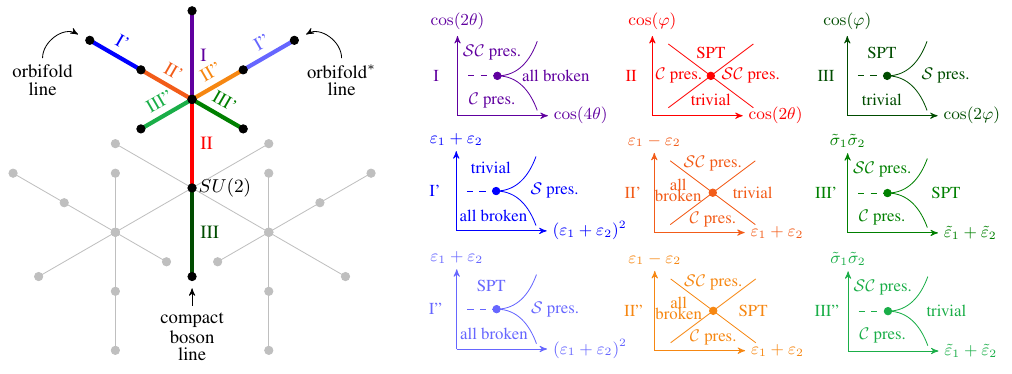}
	\caption{The moduli space of $\mathbb Z_2\times \mathbb Z_2$-enriched $c=1$ CFTs of codimension two (containing the compact boson and its $\mathbb Z_2$-orbifold). The compact boson line appears thrice (twice in gray), corresponding to different identifications of the $\mathbb Z_2 \times \mathbb Z_2$ subgroup generated by charge-conjugation $\mathcal C$ and shift symmetry $\mathcal S$; these are joined at the $SU(2)$-invariant point ($K=1/2$). Each compact boson line sprouts two orbifold lines at the KT point ($K=2$) which are pairwise related by an SPT entangler. Together, there are $3\times 3 \times 3 = 27$ distinct regions which are all smoothly connected; for each region, the nearby phase diagram is sketched.\label{fig:core}}
\end{figure}

Let us first consider the compact boson CFT, with fields $[\partial \varphi(x),\theta(y)] =2\pi i \delta(x-y)$ (see also Section~\ref{subsec:gaussian} of the main text). To achieve the lowest possible codimension (i.e., two) with a $\mathbb Z_2 \times \mathbb Z_2$ symmetry, the symmetry needs to act within the CFT (i.e., it cannot act on only gapped degrees of freedom). The symmetry group of the compact boson (which is generically $U(1)\times U(1) \rtimes \mathbb Z_2$, although our discussion is not limited to generic cases) admits various $\mathbb Z_2 \times \mathbb Z_2$ subgroups. However, there are only two such subgroups which are not related by conjugation; equivalently, by appropriately redefining our fields, the only two $\mathbb Z_2 \times \mathbb Z_2$ subgroups to consider are those generated by $\mathcal C, \mathcal S$ and those generated by $\mathcal C, \tilde{\mathcal S}$, where
\begin{equation}
\mathcal C: \left\{ \begin{array}{ccl} \theta &\to & -\theta \\ \varphi &\to & -\varphi \end{array} \right. \qquad  \qquad
\mathcal S: \left\{ \begin{array}{ccl} \theta &\to & \theta + \pi \\ \varphi &\to & \varphi \end{array} \right. \qquad  \qquad
\tilde{\mathcal S}: \left\{ \begin{array}{ccl} \theta &\to & \theta \\ \varphi &\to & \varphi  + \pi. \end{array} \right.
\end{equation}
We refer to $\mathcal C$ as charge-conjugation and to $\mathcal S,\tilde{\mathcal S}$ as shift symmetries (one could also consider the shift symmetry $\mathcal S \tilde{\mathcal S}$ which shifts both fields, but this is anomalous and thus cannot arise from an on-site $\mathbb Z_2$ symmetry on the lattice).
Moreover, by applying $T$-duality (which corresponds to defining the new fields $\tilde \varphi := \theta$ and $\tilde \theta := \varphi$), one can interchange $\mathcal S$ and $\tilde{\mathcal S}$ at the cost of changing the Luttinger liquid parameter $K \to 1/(4K)$. Since such dual points are connected along the compact boson line (by tuning $K$) \cite{Ginsparg88}, it is sufficient to consider the case where our $\mathbb Z_2 \times \mathbb Z_2$ symmetry is generated by charge-conjugation $\mathcal C$ and shift-symmetry $\mathcal S$.

With regard to symmetry assignments of symmetry fluxes, note that this corresponds to the charges of states in twisted sectors. If two symmetry-enriched CFTs have the same charges for local operators, then the charges of twisted sectors can differ only by a relative global charge\footnote{Otherwise one can argue that there exists a local operator with a different charge in the two CFTs.}. By standard arguments, as in Appendix~\ref{app:chargetococycle}, these charge assignments naturally form a group cocycle in $H^2(G,U(1))$. Hence, we only need to consider stacking the CFT with gapped SPT phases. However, as discussed in the main text (see Section~\ref{subsec:gaussian}), the compact boson CFT is invariant under stacking with the $\mathbb Z_2 \times \mathbb Z_2$ Haldane SPT phase.

In conclusion, a $\mathbb Z_2 \times \mathbb Z_2$-enriched compact boson CFT is unique up to matching the physical symmetries (say, $R_x$, $R_y$, $R_z$) to the continuum symmetry $\mathbb Z_2 \times \mathbb Z_2$ generated by $\mathcal C$ and $\mathcal S$. Since $\mathcal C$ and $\mathcal{CS}$ are related by conjugation (i.e., there is no observable/absolute difference between them), the only real choice is in which of the three symmetries corresponds to $\mathcal S$. This thus gives us three copies of the compact boson line, as shown in Fig.~\ref{fig:core} (two of which are shown in gray). These naturally join at the $SU(2)$-invariant point (where the Luttinger liquid parameter $K=1$), where $R_x$, $R_y$ and $R_z$ are all symmetry-equivalent (see Appendix~\ref{app:SU2} where this is discussed in more detail, as well as the related discussion in the main text in Section~\ref{subsec:gaussian}).

Let us now focus on the cases with only two relevant symmetry-allowed perturbations. Since $e^{i(m\varphi + n \theta)}$ has dimension $m^2 K + n^2/(4K)$, this region is $2/9  < K < 9/2$. Along this line, we show the nearby two-parameter phase diagram in Fig.~\ref{fig:core}. In particular, region I corresponds to $2<K<9/2$, where the relevant operators are $\cos(2\theta)$ and $\cos(4\theta)$; the latter can lead to a first order transition (dashed line in Fig.~\ref{fig:core}). When a nearby phase has spontaneous symmetry breaking, we denote which symmetry is preserved (if any). Region II has $1/2 < K < 2$, containing the free-fermion point at $K=1$; this corresponds to the XY chain in Fig.~\ref{fig:classificationB} (where panel $(a)$ corresponds to the $\cos(2\theta)$ perturbation and panel $(b)$ to the $\cos(\varphi)$ perturbation). Finally, region III has $2/9<K<1/2$, which also contains a dual free-fermion point at $K=1/4$ (however, unlike the $K=1$ above, the symmetry-allowed perturbations are no longer quadratic in the fermionic representation).

The discussion of the orbifold CFT is quite similar. One way of viewing this CFT is as a marginal perturbation of (Ising)$^{2}$, with Ising fields $\sigma_1$ and $\sigma_2$ and marginal perturbation $\varepsilon_1 \varepsilon_2$. Now the symmetry group is $D_8$, the symmetry group of the square, which admits the representation $\langle r,s|r^4=s^2=srsr=1\rangle$ (geometrically: $r$ is rotation and $s$ is reflection) where
\begin{equation}
r: \left\{ \begin{array}{ccr} \sigma_1 &\to & - \sigma_2 \\ \sigma_2 & \to & \sigma_1 \end{array} \right. \qquad  \qquad
s: \left\{ \begin{array}{ccr} \sigma_1 &\to & \sigma_1 \\ \sigma_2 &\to & -\sigma_2 .\end{array} \right. 
\end{equation}
As before, there are two inequivalent choices of a $\mathbb Z_2 \times \mathbb Z_2$ subgroup, generated by $r^2,s$ or $r^2,sr$. And, as before, these two different choices are in fact related by $T$-duality and lie on the same orbifold line. The self-dual point is where the orbifold line meets the compact boson line \cite{Ginsparg88}. E.g., if one passes from II' to III' (in Fig.~\ref{fig:core}), the roles of $s$ and $sr$ are reversed. In particular, coming from one side (II'), we can equate $s = \mathcal C$ (by identifying $\sigma_1 = \cos\theta$ and $\sigma_2 = \sin \theta$, where the latter are the compact boson fields at the KT point) and from the other side, $sr = \mathcal C$ (by identifying $\tilde \sigma_{1,2} = \cos\theta \pm \sin \theta$); we see that in both cases $r^2 = \mathcal S$. Note that either side of the KT point has an (Ising)$^{2}$ point (i.e., in II' and in III'), and for conceptual simplicity, we denote the fields in terms of the respective Ising operators (using tilde variables in III' to avoid confusion); these are related under $T$-duality, in particular: $\tilde \sigma_{1,2} = \sigma_1 \pm \sigma_2$.

As argued above, the nonlocal charges can be toggled by stacking with $G$-SPTs. Stacking with the unique $\mathbb Z_2 \times \mathbb Z_2$-SPT, we obtain the twisted `orbifold*' line in Fig.~\ref{fig:core}. Indeed, note that in the nearby phase diagrams, the roles of trivial and SPT are reversed in I'', II'', III'' relative to I', II', III'. In all cases, these nearby phase diagrams can be obtained by a direct inspection of the symmetric relevant operators, but a shortcut for the two orbifold lines is by noting how these CFTs (and their nearby phase diagrams) can be obtained by gauging charge-conjugation symmetry of the compact boson CFT \cite{Ginsparg88}. Note that as in the case of the compact boson, we only consider the part of the orbifold branch where there are two relevant symmetry-preserving perturbations.

\subsection{\texorpdfstring{$SU(2)$}{SU(2)} symmetry at the self-dual radius \label{app:SU2}}

At the self-dual radius of the compact boson CFT (i.e., $K=1/2$, or $r_c = 1/\sqrt{2}$), we have an emergent $SU(2) \times SU(2)$ symmetry, generated by the following operators with dimensions $(1,0)$ and $(0,1)$, respectively:
\begin{equation}
\begin{array}{lllll}
J^x = \cos(\theta +\varphi) &\quad& J^y = \sin(\theta + \varphi) &\quad& J^z = \frac{1}{2} i\partial_z \left( \theta + \varphi \right) \\
\bar J^x = \cos(\theta - \varphi) &\quad& \bar J^y = \sin(\theta - \varphi) &\quad& \bar J^z = \frac{1}{2} i\partial_{\bar z} \left( \theta - \varphi \right).
\end{array}
\end{equation}
For clarity, we are using a convention such that $\theta+\varphi$ is holomorphic and $\theta-\varphi$ is antiholomorphic at $K=1/2$.
We can define a change of variables which corresponds to a rotation generated by $J^x$, such that $\tilde J^x \equiv J^x$ but $\tilde J^z \equiv J^y$ and $\tilde J^y \equiv - J^z$ (and similarly for the anti-holomorphic sector). Note that this transformation is a symmetry \emph{only} at the self-dual radius (i.e., at that point the theory is identical in these new variables)! We thus find
\begin{equation}
\partial_x \tilde \theta = -2 \cos \theta \sin \varphi, \qquad \partial_x \tilde \varphi = -2 \sin \theta \cos \varphi , \qquad \cos \tilde \theta \cos \tilde \varphi = \cos \theta \cos \varphi, \qquad \sin \tilde \theta \sin \tilde \varphi = \sin \theta \sin \varphi.
\end{equation}

\setcounter{figure}{0}
\section{Relation to the literature}

\begin{figure}[h]
\includegraphics[scale=1]{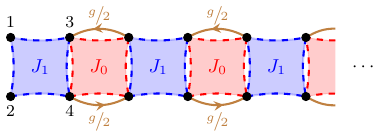}
\caption{A strongly-interacting Majorana model (see Eq.~\eqref{eq:ferm_Ham}) in the DIII class. \label{fig:model}}
\end{figure}

\subsection{``\emph{Quantum Criticality in Topological Insulators and Superconductors: Emergence of Strongly Coupled Majoranas and Supersymmetry}'' (Grover, Vishwanath) \label{app:grover}}

In this subsection, we discuss a model appearing in Ref.~\onlinecite{Grover12}, reinterpreting certain aspects in terms of symmetry-enriched quantum criticality. For that reason, we closely follow the notation used in that paper. To start, let us consider a model which does not directly appear in Ref.~\onlinecite{Grover12}, but which will aid our analysis:
\begin{equation}
H = \sum_{n\geq 1} \left( J_{n \textrm{ mod }2} \; \chi_{2n-1} \chi_{2n} \chi_{2n+1} \chi_{2n+2} - \frac{g}{2} \; i \chi_{4n-1} \chi_{4n+1} + \frac{g}{2}  \; i \chi_{4n} \chi_{4n+2} \right). \label{eq:ferm_Ham}
\end{equation}
The Hamiltonian is pictorially represented in Fig.~\ref{fig:model}. It has an anti-unitary symmetry $T_-$ defined by $(\chi_{2n-1},\chi_{2n} ) \stackrel{T_-}{\mapsto} (\chi_{2n},-\chi_{2n-1})$, satisfying $T^2_- = -1$. If we set $J_0 = J_1 = 0$ (but $g \neq 0$), then in Fig.~\ref{fig:model} we directly see that we are in a topological phase with two zero-energy Majorana edge modes at a \emph{single} edge, $\chi_1$ and $\chi_2$, which $T_-$ prevents us from gapping out. This is the only non-trivial class in DIII. Oppositely, if we make $J_0$ and $J_1$ positive and large, then the ground state wants to occupy vertical bonds, spontaneously breaking $T_-$.

Note that---similar to the SSH model---the above model is only non-trivial if we fix our convention of the physical unit cells to be $(1234)(5678)\cdots$, i.e. the blue boxes in Fig.~\ref{fig:model}.

\subsubsection{Deriving an effective (almost-)spin model (as $J_0 \to +\infty$)}

Note that the $J_0$ term is an integral of motion. In fact, it is already minimized in the above phase, so we can adiabatically ramp up the term $J_0$ such that $J_0 \gg \max\{ |J_1|,|g| \}$. We can then work in the subspace $\mathcal H_0$ where for all $n$, we have $\chi_{4n-1}\chi_{4n}\chi_{4n+1}\chi_{4n+2} = -1$. In this subspace, we can define the following \emph{bosonic} operators:
\begin{align}
\sigma^z_n &:= i\chi_{4n-1}\chi_{4n} =i \chi_{4n+1}\chi_{4n+2}, \label{eq:Z} \\
\sigma^x_n &:= i\chi_{4n-1}\chi_{4n+1} = -i \chi_{4n}\chi_{4n+2}. \label{eq:X}
\end{align}
In this sector, we then obtain
\begin{align}
H_\textrm{eff}
&= \sum_{n\geq 1} \left( J_1 \chi_{4n-3}\chi_{4n-2}\chi_{4n-1} \chi_{4n} - g \; i \chi_{4n-1}\chi_{4n+1} \right) \\
&= - J_1\; \sigma^z_1  \; i \chi_1 \chi_2- \sum_{n \geq 1} \left( J_1 \sigma^z_n \sigma^z_{n+1} + g \sigma^x_n \right). \label{eq:bos_Ham}
\end{align}
The model in Eq.~\eqref{eq:bos_Ham} is the one appearing Ref.~\onlinecite{Grover12} (Eq.~(1) of that work). The anti-unitary symmetries acts the following way on the bosonic operators: $(\sigma^x_n,\sigma^y_n,\sigma^z_n) \stackrel{T_-}{\mapsto} (\sigma^x_n,\sigma^y_n,-\sigma^z_n)$. In other words, for the spin variables, we can write $T_{-,\textrm{eff}} = \left( \prod \sigma^x_n \right)K$. Note that this implies $T_{-,\textrm{eff}}^2 = +1$.

We see that the anti-unitary symmetry in $H_\textrm{eff}$ is spontaneously broken for $|J_1|>|g|$, whereas for $|J_1|<|g|$ we have two protected Majorana modes (per edge).

\subsubsection{Symmetry fluxes}

Let us first obtain the symmetry flux for fermionic parity symmetry $P = \prod_n P_n$ where $P_n = \gamma_{4n-3} \gamma_{4n-2} \gamma_{4n-1} \gamma_{4n}$ is the fermionic parity per unit cell (i.e., per blue box in Fig.~\ref{fig:model}). Since we work in the limit where $J_0 \to + \infty$, we have that \emph{the product of red boxes} has long-range order. We conclude that the symmetry flux of $P$ is $\mathcal S^P_n = \cdots P_{n-2} P_{n-1} (i\gamma_{4n-3} \gamma_{4n-2})$ (note that the $i$ is there to ensure that $\left\langle \left( \mathcal S^P_n \right)^2 \right\rangle >0$). We observe that $T_- \mathcal S^P T_- = - \mathcal S^P$, i.e., the symmetry flux of $P$ is odd under $T$. This gives us a discrete label which is moreover topologically non-trivial.

By the reciprocal nature of charges of symmetry fluxes, we expect that the symmetry flux of $T$ should be odd under $P$. However, in this work we have not studied the notion of symmetry fluxes for anti-unitary symmetries. Nevertheless, a closely related statement is that we expect that $P \mu P = - \mu$. To make this precise, we can consider a lattice $\mathbb Z_2$ symmetry which we can relate to the unitary $\mathbb Z_2$ symmetry of the Ising CFT. This is of course $\prod_n \sigma^x_n$, or in the original fermionic model, the fermionic parity symmetry of a single leg, $P_\textrm{odd} = \prod_n (\chi_{4n-1} \chi_{4n+1})$. Indeed, for $J_0 \to + \infty$, one can derive that $P_\textrm{odd} = \prod_n \sigma^x_n$ (at least without boundaries). Similar to $P$, the symmetry flux of $P_\textrm{odd}$ has an extra Majorana mode: $\mathcal S^{P_\textrm{odd}} = \cdots (P_\textrm{odd})_{n-2} (P_\textrm{odd})_{n-1} \chi_{4n-1}$. Hence, we conclude that the symmetry flux of $P_\textrm{odd}$ is odd under fermionic parity symmetry. Equivalently, the symmetry flux of $\prod_n \sigma^x_n$ is $\cdots \sigma^x_{n-2} \sigma^x_{n-1} \chi_{4n-1}$; we thus conclude that $P \mu P = - \mu$. 

If we do not enforce the lattice $\mathbb Z_2$ symmetry $\prod_n \sigma^x_n$, then it becomes subtle to claim that $P \mu P = -\mu$ is well-defined. However, in this particular case, we expect that due to $T_-$ symmetry being enforced, this invariant remains well-defined even in the absence of explicit $\prod_n \sigma^x_n$ symmetry; the methods developed in this paper do not allow us to prove this.

\subsubsection{Edge modes}

Since $\mu$ is charged under $P$, we know that by the general arguments in Section~\ref{sec:edge}, the Ising CFT will have a degeneracy with open boundaries. This is in fact easy to see in Eq.~\eqref{eq:ferm_Ham} and Eq.~\eqref{eq:bos_Ham}. In both cases, we see that $i \chi_1 \chi_2$ and $\chi_1 \sigma^x_2 \sigma^x_3 \cdots$ are symmetries of the Hamiltonian, whereas they mutually anticommute. The half-infinite geometry thus has a twofold degeneracy labeled by the occupation $i \chi_1 \chi_2 = \pm 1$. Remarkably, this degeneracy is protected by the bulk CFT being enriched non-trivially by $P$ and $T_-$. Even with generic symmetry-preserving perturbations, the system will have a global twofold degeneracy whose splitting is exponentially small in system size.

\subsection{``\emph{Gapless symmetry-protected topological phase of fermions in one dimension}'' (Keselman, Berg) \label{app:Berg}}

Let us recall the model in Eq.~\eqref{eq:Berg} of the main text:
\begin{equation}
H = - \sum_{n,\sigma} \left( c^\dagger_{n,\sigma} c^{\vphantom \dagger}_{n+1,\sigma} + h.c. \right) + U \sum_n \left( - c_{n,\uparrow}^\dagger c_{n,\downarrow} c_{n+1,\uparrow}  c_{n+1,\downarrow}^\dagger  + h.c. + n_{i,\uparrow} n_{i+1,\downarrow} + n_{i,\downarrow} n_{i+1,\uparrow}  \right). \label{eq:Berg2}
\end{equation}
If we relabel the indices as $2j-1 \equiv (j,\uparrow)$ and $2j \equiv (j,\downarrow)$, then one can straightforwardly apply the Jordan-Wigner transformation to obtain the following spin-$1/2$ chain:
\begin{equation}
H = \frac{1}{2} \sum_i \left( X_{i-1} Z_i X_{i+1} +  Y_{i-1} Z_i Y_{i+1} \right) + U \sum_i \left( S^+_{2i-1} S^-_{2i} S^-_{2i+1} S^+_{2i+2} + h.c. + n_{2i-1} n_{2i+2} + n_{2i} n_{2i+1} \right), \label{eq:Berg3}
\end{equation}
where $S^\pm = (X_i \pm i Y_i)/2$ and $n_i = (1+Z_i)/2$. In this bosonic language, we have the symmetries $P = \prod_i P_i$ (where $P_i = Z_{2i-1} Z_{2i}$) and $T = \prod_j e^{i \frac{\pi}{4} \left( S^x_{2j-1} S^y_{2j} - S^y_{2j-1} S^x_{2j} \right)} K$. Note that $T^2 = P$.

\begin{figure}
\includegraphics[scale=1]{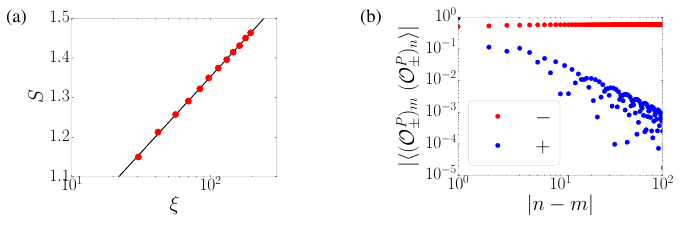}
\caption{Spinful fermions with triplet-pairing: the model of Ref.~\onlinecite{Keselman15} as shown in Eq.~\eqref{eq:Berg2} (or Eq.~\eqref{eq:Berg3}) with $U=-0.9$. (a) Confirmation that the bulk (which has central charge $c=2$ for $U=0$) has flown to central charge $c=1$. The black line is a fit $S \sim \frac{c_\textrm{fit}}{6} \ln \xi$ with $c_\textrm{fit} \approx 0.994$. (b) Correlation function for $\mathcal O^P_\pm$ (where $T \mathcal O^P_\pm T = \pm \mathcal O^P_\pm$). We find that the $T$-odd string operator has long-range order; this is thus the unique symmetry flux of $P$. \label{fig:Berg}}
\end{figure}

We will consider the two string operators $(\mathcal O^P_{\pm})_i \equiv \cdots Z_{2i-3} Z_{2i-2}\left( Z_{2i-1} \pm Z_{2i} \right)$. Note that $T \mathcal O^P_\pm T^{-1} = \pm \mathcal O^P_\pm$. If $U=0$, the bulk is a $c=2$ CFT and both operators have the same scaling dimension $\Delta = 1/2$. For $U \neq 0$, the bulk gaps out the spin sector (of the original spinful fermions), stabilizing a $c=1$ CFT. We confirm this in Fig.~\ref{fig:Berg}(a) for $U=-0.9$. Moreover, we find that $\mathcal O^P_+$ remains critical, whereas $\mathcal O^P_-$ now has long-range order; this is shown in Fig.~\ref{fig:Berg}(b). We conclude that the system has flown to a $c=1$ CFT which is non-trivially enriched by $P$ and $T$; in particular, the (unique) symmetry-flux of $P$ is odd under $T$.
	
\subsection{``Gapless Symmetry-Protected Topological Order'' (Scaffidi, Parker, Vasseur) \label{app:Scaffidi}}

We consider the cluster model with an Ising term on one of the two sublattices \cite{Scaffidi17}:
\begin{equation}
H = - \sum_n \left( Z_{A,n} X_{B,n} Z_{A,n+1} + Z_{B,n-1} X_{A,n} Z_{B,n} + J Z_{A,n} Z_{A,n+1} \right) . \label{eq:Scaffidi_app} 
\end{equation}
This model has a $\mathbb Z_2 \times \mathbb Z_2$ symmetry generated by $P_A = \prod_n X_{A,n}$ and $P_B = \prod_n X_{B,n}$.
To obtain its phase diagram, note that the unitary $U =\prod_n (CZ)_{A,n;B,n} (CZ)_{B,n;A,n+1}$ maps $H$ to two decoupled chains. In particular, if we define
\begin{equation}
\left\{ \begin{array}{rcl}
\tilde X_{A,n} &:= U X_{A,n} U^\dagger &=  Z_{B,n-1} X_{A,n} Z_{B,n} \\
\tilde X_{B,n} &:= UX_{B,n} U^\dagger &= Z_{A,n} X_{B,n} Z_{A,n+1} \\
\tilde Z_{A,n} &:= U Z_{A,n} U^\dagger &= Z_{A,n} \\
\tilde Z_{B,n} &:= U Z_{B,n} U^\dagger &= Z_{B,n}
\end{array} \right. 
\end{equation}
then for \emph{periodic} boundary conditions, Eq.~\eqref{eq:Scaffidi_app} becomes an Ising chain on $A$ sites and a fixed-point paramagnet on $B$ sites:
\begin{equation}
H = - \sum_n \tilde X_{B,n} - \sum_n \left( \tilde X_{A,n} + J \tilde Z_{A,n} \tilde Z_{A,n+1} \right). \label{eq:Scaffidi_dual_app}
\end{equation}
Hence, from this we infer that Eq.~\eqref{eq:Scaffidi_app} is in the SPT cluster phase for $|J|<1$, spontaneously breaks $P_A$ for $|J|>1$ and is at an Ising critical point for $|J|=1$. In Ref.~\cite{Scaffidi17}, it was pointed out that for Eq.~\eqref{eq:Scaffidi_app} with \emph{open} boundary conditions, the edge modes of the SPT phase (partially) survive at the Ising critical point (even upon including arbitrary $\mathbb Z_2 \times \mathbb Z_2$-symmetric perturbations).

Using the concepts introduced in Section~\ref{sec:symflux}, we can identify a non-trivial bulk topological invariant explaining the aforementioned edge modes. It is straightforward to obtain the symmetry flux associated to $P_B$: we see that Eq.~\eqref{eq:Scaffidi_dual_app} has long-range order in $\prod_{m\leq n} \tilde X_{B,m} = \left( \prod_{m\leq n} X_{B,m} \right)Z_{A,n+1}$, which is odd under $P_A$. This thus gives us a topological invariant whenever $P_A$ is preserved\footnote{This charge is not well-defined in the symmetry-breaking phase. E.g., if $J=+\infty$, then $Z_{A,n} = \pm 1$, so both $\left( \prod_{m\leq n} X_{B,m} \right)Z_{A,n+1}$ and $\prod_{m\leq n} X_{B,m}$ have long-range order. }: i.e., in the gapped SPT phase ($|J|<1$) or at the critical point ($|J|=1$). The fact that it has long-range order is due to the symmetry being gapped; it is thus also instructive to consider the other symmetry, which is more representative of the purely-gapless case. In particular, at criticality ($|J|=1$), the Ising chain Eq.~\eqref{eq:Scaffidi_dual_app} has an algebraically-decaying disorder parameter $\mu \sim \prod_{m \leq n} \tilde X_{A,n} = \left(\prod_{m \leq n} X_{A,n} \right) Z_{B,n}$ with scaling dimension $\Delta = 1/8$ (e.g., see the discussion in Section~\ref{sec:overview}). We thus see that the Ising critical point of Eq.~\eqref{eq:Scaffidi_app} has a disorder operator $\mu$ which is odd under $P_B$, encoding the non-trivial topological invariant.

\setcounter{figure}{0}
\section{Details about the construction of 2D examples \label{app:2D}}

\begin{figure}
	\includegraphics[scale=0.5]{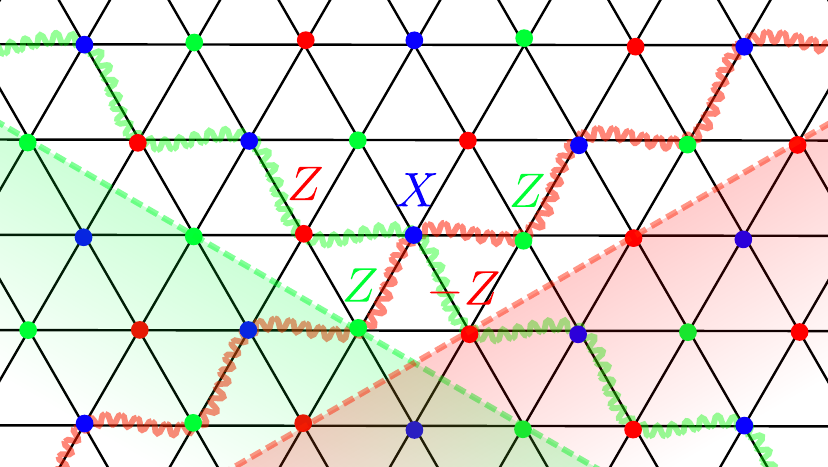}
	\caption{If one start with the trivial paramagnet on the triangular lattice, and then SPT-entangles using $U_\textrm{2D}$, twists by $P_A$ along the green dashed line, and SPT-entangles back, then the resulting model is a trivial paramagnet \emph{except} for the sites along the green wiggly line, where it is in a 1D cluster model. Similarly for twisting by $P_B$ along the red dashed line. If one twists by both symmetries at the same time, then the blue site near the defect intersection is mapped to the five-site operator as shown, with a minus sign due to the $Z$ operator on the green wiggly line anticommuting with the $P_B$ twist. After a global unitary transformation ($V = U_\textrm{0D} U_\textrm{1D} U_\textrm{2D}$ in the text), the system can be mapped back to the original trivial paramagnet, at the cost of flipping the sign of $P_C$. \label{fig:appTL}}
\end{figure}

Let $H$ be a spin-$1/2$ model on the triangular lattice which is symmetric with respect to the $\mathbb Z_2 \times \mathbb Z_2 \times \mathbb Z_2$ symmetry generated by $P_A = \prod_{\bm n} X_{A,\bm n}$, $P_B = \prod_{\bm n} X_{B,\bm n}$ and $P_C = \prod_{\bm n} X_{C,\bm n}$ (here $A,B,C$ denote the three sublattices which are respectively colored green, red and blue in Fig.~\ref{fig:appTL}). Denote $U_\textrm{2D} = \prod_\triangle CCZ$ which applies the $CCZ$ gate on every triangle of the triangular lattice, and define $H' = U H U$ (NB: all unitary operators considered in this section will square to unity so we will not distinguish $U$ from $U^\dagger$). We will now show that the twisted sectors of $H$ and $H'$, twisted by $P_A$ and $P_B$, are unitarily equivalent in such a way that eigenstates with well-defined $P_C$ have opposite quantum numbers.

\subsection{Warm-up: the fixed-point case \label{subsec:simple}}

It is instructive to first consider the simpler case where $H = -\sum_{\bm r} X_{\bm r}$ (here $\bm r$ is short-hand for $(\lambda,\bm n)$ with $\lambda=A,B,C$). In this case, $H' = U HU$ corresponds to the Yoshida model which is a $\mathbb Z_2^3$ SPT phase \cite{Yoshida16,Yoshida17}. Using the property that $(CCZ)_{1,2,3} X_1 (CCZ)_{1,2,3} = X_1 (CZ)_{2,3}$, we see that $U X_{\bm r} U = X_{\bm r} \prod_{\textrm{hex}(\bm r)} CZ$ where the product of $CZ$ gates runs along the hexagon surrounding the site $\bm r$; see Fig.~\ref{fig:TL}(b).

Since each term in $H$ commutes with $P_A$ and $P_B$, twisting has no effect on the trivial Hamiltonian, i.e., $H_{\textrm{twist}} = H$. In particular, since the ground state satisfies $X_{\bm r} = 1$, the ground state parity in the twisted sector is $P_C = \prod_{\bm n}X_{C,\bm n} = 1$. (This signifies that we are in a trivial $\mathbb Z_2^3$ SPT phase.)

In contrast, the terms in the Yoshida model $H'$ are affected whenever they overlap with the defect line. Let us first twist by $P_A$, which corresponds to conjugating the Hamiltonian term with a product of $X_{A,\bm n}$ operators if $\bm n$ lies below or on the green dashed line in Fig.~\ref{fig:appTL}. Using the fact that $X_{1}(CZ)_{1,2} X_1 = (CZ)_{1,2} Z_2$, we see that Hamiltonian terms that intersect the defect line will acquire extra factors of $Z_{B,\bm n}$ or $Z_{C,\bm n}$ after twisting with $P_A$. The net result can be phrased as follows: if after twisting with $P_A$ we conjugate the term again with $U_\textrm{2D}$ (i.e., removing the $\prod CZ$ around the hexagon), we end up with a trivial $X_{\bm r}$ \emph{except} along the green wiggly line in Fig.~\ref{fig:appTL} where the Hamiltonian is the 1D cluster chain.

Twisting by $P_B$ is analogous, with one exception: the blue site at the intersection of the two twist lines (indicated in Fig.~\ref{fig:appTL}) also obtains a minus sign since only one of the two $Z_{B,\bm n}$ operators is acted on by the $P_B$-twist (and $Z_{B,\bm n}$ of course anticommutes with $X_{B,\bm n}$). In summary, twisting by $P_A$ and $P_B$ for the Yoshida' model is equivalent to conjugating by $U_\textrm{1D} U_\textrm{0D}$ where $U_{\textrm{1D}} = \prod CZ$ with the product running along the green and red wiggly lines in Fig.~\ref{fig:appTL} and where $U_\textrm{0D} = Z_{C,\bm{n}_0}$ with $\bm{n}_0$ being the (blue) site at the intersection of the two defect lines. More precisely, $H_\textrm{twist}' = U_\textrm{1D} U_\textrm{0D} H' U_\textrm{1D} U_\textrm{0D} = VHV$ with $V = U_{\textrm{2D}} U_{\textrm{1D}} U_{\textrm{0D}}$. We thus obtain that the ground state of $H_\textrm{twist}'$ satisfies $V X_{C,\bm n} V = 1$, such that \begin{align}
P_C &= \prod_{\bm n} X_{C,\bm n} = - \prod_{\bm n} U_\textrm{0D} X_{C,\bm n} U_\textrm{0D} = - \prod_{\bm n} U_\textrm{0D} \bigg( \prod_{\bm m \in l \cup l'} Z_{B,\bm m}^2 \bigg) X_{C,\bm n}  \bigg( \prod_{\bm m \in l \cup l'} Z_{B,\bm m}^2 \bigg) U_\textrm{0D} \\
&= - \prod_{\bm n} U_\textrm{0D}  U_\textrm{1D} X_{C,\bm n}  U_\textrm{1D} U_\textrm{0D} 
= - \prod_{\bm n} U_\textrm{0D}  U_\textrm{1D} \bigg( \prod_{\triangle} (CCZ)^2 \bigg) X_{C,\bm n}\bigg( \prod_{\triangle} (CCZ)^2 \bigg)  U_\textrm{1D} U_\textrm{0D} 
 = - \prod_{\bm n} V X_{C,\bm n} V = -1.
\end{align}
Here we used the aforementioned properties that $(CZ)_{1,2}X_{1}(CZ)_{1,2} =  X_1 Z_2$ and $(CCZ)_{1,2,3} X_1 (CCZ)_{1,2,3} = X_1 (CZ)_{2,3}$, and $l$ and $l'$ denote the sites along the green and red wiggly lines in Fig.~\ref{fig:appTL}.

\subsection{The general case}

In the simple case above, we saw that the twisted sector of $H'$ (i.e., $H_\textrm{twist}'$) is in fact unitarily equivalent to $H'$ (which is in turn unitarily equivalent to $H$). The above derivation directly extends to Hamiltonians which only contain products of $X_{\bm r}$ operators. However, $H'$ and $H'_\textrm{twist}$ are no longer unitarily equivalent when there are also terms containing $Z_{\bm r}$ (or $Y_{\bm r}$, but this need not be discussed separately since one can rewrite this as the product $iX_{\bm r} Z_{\bm r}$). Fortunately, it is still true that $H_\textrm{twist}$ and $H_\textrm{twist}'$ are unitarily equivalent. The idea is simple: for any $X_{\bm r}$ that appears, we apply the previous unitary mapping, whereas for any $Z_{\bm r}$ that appears, the twisting by $P_A$ and $P_B$ is the same for both $H$ and $H'$.

Let us now confirm that the details work out. Without loss of generality, we can write\footnote{Note that $\alpha$ is just a label and need not be interpreted as a spatial index.} $H = \sum h_\alpha$ where each $h_\alpha$ is a product of Pauli operators: $h_\alpha = a_\alpha \prod_{\bm r \in S_{\alpha,x}} X_{\bm r} \prod_{\bm r \in S_{\alpha,z}} Z_{\bm r}$ where $a_\alpha$ is a number and $S_{\alpha,\gamma=x,z}$ are unspecified sets of indices (note that $S_{\alpha,x}\cap S_{\alpha,z}$ could be nonzero, corresponding to a Pauli-$Y$ operator; hence, hermiticity requires $a_\alpha \in i^{|S_{\alpha,x}\cap S_{\alpha,z}|}\mathbb R$). 

The twisted sector of $H$ can then be written as $H_\textrm{twist} = \sum_\alpha h_{\textrm{twist},\alpha}$ with $h_{\textrm{twist},\alpha} = a_\alpha \prod_{\bm r \in S_{\alpha,x}} X_{\bm r} \; \mathcal T ( \prod_{\bm r \in S_{\alpha,z}} Z_{\bm r} )$. Here $\mathcal T(\cdot)$ simply denote the twisting operation; the result of twisting the $ Z_{\bm r}$ operators could be written out more explicitly but we will not need such detailed expressions to establish the unitary equivalence.

We now turn to $H' = UHU = \sum_\alpha h_\alpha'$ with $h_\alpha' = a_\alpha \prod_{\bm r \in S_{\alpha,x}} \left( U X_{\bm r} U \right) \prod_{\bm r \in S_{\bm r,z}} Z_{\bm r}$. Twisting this gives $H'_\textrm{twist} = \sum_\alpha h_{\textrm{twist},\alpha}'$ and $h_{\textrm{twist},\alpha}' = a_\alpha \big( \prod_{\bm r \in S_{\alpha,x}} \mathcal T \left( U X_{\bm r} U \right)\big) \mathcal T(\prod_{\bm r \in S_{\alpha,z}} Z_{\bm r})$. From the discussion in Sec.~\ref{subsec:simple} we learned that $\mathcal T \left( U X_{\bm r} U \right)\big) = V X_{\bm r} V $ with $V = U_\textrm{2D} U_\textrm{1D} U_\textrm{0D}$. Hence, if we define the new variables under the unitary mapping, $\tilde X = V X_{\bm r} V$ and $\tilde Z = V Z_{\bm r} V = Z_{\bm r}$, then we see that the twisted sector of $H$ and $H'$ are unitarily equivalent! However,
\begin{equation}
\tilde P_C = \prod_{\bm n} \tilde X_{C,\bm n} = \prod_{\bm n} V X_{C,\bm n} V = - \prod_{\bm n} X_{C,\bm n} = - P_C.
\end{equation}
In other words, the unitary mapping between $H_\textrm{twist}$ and $H'_\textrm{twist}$ toggles the eigenvalue of $P_C$ (for each eigenstate).

\end{document}